\documentclass[11pt,a4paper]{article}
\usepackage{xingyang,amsmath,amssymb,slashed,url,bm,textgreek,upgreek}
\usepackage[table]{xcolor}
\usepackage{bbm,array,amsfonts,wrapfig,lscape,float,mathtools,multirow,longtable}
\usepackage{graphicx}
\usepackage{epstopdf}
\usepackage{braket}
\usepackage{newtxtext,newtxmath}
\usepackage{empheq}
\newcommand*\widefbox[1]{\fbox{\hspace{2em}#1\hspace{2em}}}

\usepackage{amsthm}
\usepackage{tikz}
\usetikzlibrary{arrows.meta,decorations.pathmorphing,calc}
\def\be{\begin{equation}}
\def\ee{\end{equation}}

\def\hat{\widehat}

\def\tilde{\widetilde}

\def\[{\bigl [}

\def\]{\bigr ]}

\def\tilde{\widetilde}

\font\teneurm=eurm10 \font\seveneurm=eurm7  \font\fiveeurm=eurm5
\newfam\eurmfam
\textfont\eurmfam=\teneurm \scriptfont\eurmfam=\seveneurm
\scriptscriptfont\eurmfam=\fiveeurm

\font\teneusm=eusm10 \font\seveneusm=eusm7 \font\fiveeusm=eusm5
\newfam\eusmfam
\textfont\eusmfam=\teneusm \scriptfont\eusmfam=\seveneusm
\scriptscriptfont\eusmfam=\fiveeusm

\font\tencmmib=cmmib10 \skewchar\tencmmib='177
\font\sevencmmib=cmmib7 \skewchar\sevencmmib='177
\font\fivecmmib=cmmib5 \skewchar\fivecmmib='177

\usepackage{eso-pic}
\newcommand{\reportnum}[2]{
  \AddToShipoutPictureBG*{%
    \AtPageUpperLeft{%
      \hspace{0.75\paperwidth}%
      \raisebox{#1\baselineskip}{%
        \makebox[0pt][l]{\textnormal{#2}}
  }}}%
}

\title{SymTFT Entanglement and Holographic (Non)-Factorization}

\author[a]{Ethan Torres}
\author[b]{and Xingyang Yu}

\affiliation[a]{Theoretical Physics Department, CERN, 1211 Geneva 23, Switzerland}
\affiliation[b]{Physics Department, Robeson Hall, Virginia Tech,
 Blacksburg, VA 24061, USA}

\emailAdd{ethan.martin.torres@cern.ch}
\emailAdd{xingyangy@vt.edu}

    \abstract{Given two otherwise decoupled $D$-dimensional CFTs which possess a common (finite) symmetry subcategory, one can consider entangled boundary states of their $(D+1)$-dimensional SymTFTs. This roughly corresponds to performing a gauging of the tensor product of two CFTs, and we call this phenomena ``SymTFT entanglement" (or ``S-entanglement" for short). In the case when these CFTs have semiclassical holographic duals, the S-entanglement relates the bulk gauge charges between two otherwise disconnected AdS spacetimes as we highlight in several top-down examples. We show that taking partial traces of such S-entangled states leads to a streamlined approach to preparing ensemble-averaged CFTs in string theory. This ensemble averaging coincides with that generated by $\alpha$-states in the baby universe Hilbert space, and we propose a symmetry-enriched generalization of this Hilbert space via generalized global symmetries. We quantify how this symmetry-governed averaging violates holographic factorization and leads to the emergence of bulk global symmetries. We also consider the eternal (two-sided) AdS black hole geometries, where our SymTFT entanglement considerations imply that there exist refinements of the usual theromofield double state preparation of the system. We show that one may prepare the system in such a way that the total CFT data does not factorize into left and right copies. As anticipated by Marolf and Wall \cite{Marolf:2012xe}, we highlight that such considerations are necessary to define the gauge charges of eternal black holes, and in certain cases, can imply that there exist extended bulk objects stretching across the wormhole which cannot be expressed in terms of a product of left and right CFT operators.
}

\begin{document}
\reportnum{-8}{\quad \quad \quad \quad CERN-TH-2025-194}
\maketitle
\flushbottom



\section{Introduction}
\begin{figure}[h]
    \centering
\includegraphics[width=12cm, trim = {0cm 2cm 0cm 0cm}]{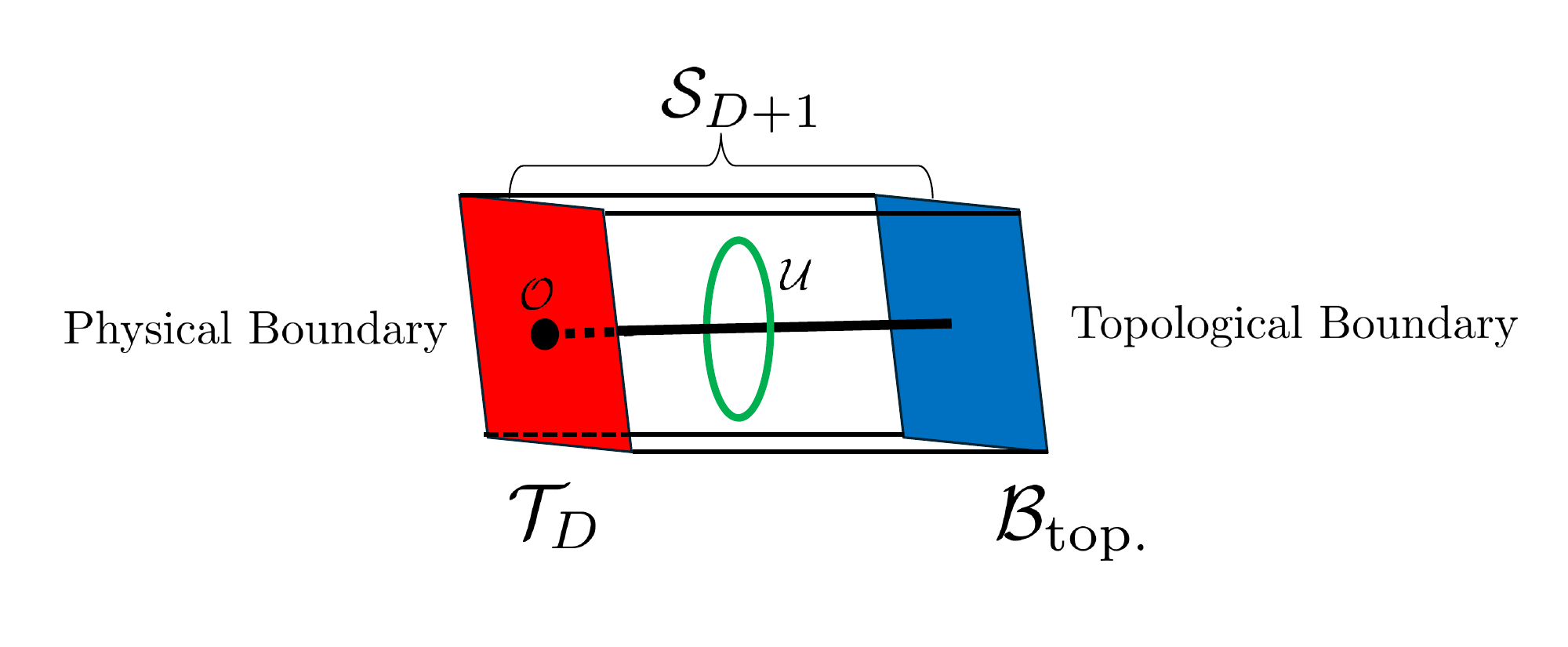}
    \caption{Illustration of a basic SymTFT ``sandwich" construction whereby a $(D+1)$-dimensional TFT has a physical boundary, where typically gapless degrees of freedom are localized, and a topological boundary. Compactifying the interval leads to an absolute $D$-dimensional QFT, which one can denote $\mathcal{T}^{(\mathcal{B}_{\mathrm{top.}})}_D$, $\mathcal{U}$ is then a topological symmetry operator and $\mathcal{O}$ denotes a charged operator.}
    \label{fig:basicsandwich}
\end{figure}

In the study of generalized symmetries of quantum field theories (QFTs) (for reviews see \cite{Costa:2024wks,Shao:2023gho, Schafer-Nameki:2023jdn}) Symmetry Topological Field theories (SymTFTs) has emerged as a central concept. This is due to the fact that SymTFTs are both conceptually and technically useful in understanding symmetry structures. For instance, they are useful in calculating the fusion rules of topological operators and especially in understanding the families of theories related to each other by gauging. See Figure \ref{fig:basicsandwich}, for a basic illustration of a SymTFT ``sandwich" construction. The utility of SymTFTs is especially helpful in more complicated field theory settings such as when the QFTs of interest are placed on manifolds with boundaries, see for instance \cite{Cvetic:2024dzu, GarciaEtxebarria:2024jfv, Cordova:2024iti, Choi:2024tri, Bhardwaj:2024igy, Das:2024qdx}, or when the QFT does not possess a well-defined partition function \cite{Lawrie:2023tdz,Heckman:2025lmw}.

In the context of string theory, SymTFT constructions naturally arise when engineering QFTs decoupled from gravity. These are often constructed from branes and/or geometric singularities which are localized on some codimension-$k$ sublocus of the string theory target space geometry. SymTFT constructions are then realized, roughly speaking, by treating a radial direction away from the codimension-$k$ sublocus as the interval direction in the SymTFT sandwich. Typically\footnote{This in particular assumes that no cycles in such a scaling limit have a finite volume and that the radial shells are smooth. Otherwise, the SymTFT may no longer be topological as it contains a free abelian or non-abelian gauge theory respectively \cite{Cvetic:2024dzu}. The examples considered in this paper will not be met with such technicalities.}, the SymTFT is obtained by first taking a infinite scaling of this radial direction and restricting to topological subsector of the compactification supergravity on the radial shells surrounding the sublocus \cite{Apruzzi:2021nmk}. In such realizations, topological symmetry operators typically arise as branes (wrapping cycles) on the radial shells, while objects charged under the symmetry operators intersect the sublocus \cite{Apruzzi:2022rei, GarciaEtxebarria:2022vzq, Heckman:2022muc,Heckman:2022xgu}.

\begin{figure}[h]
    \centering
\includegraphics[width=12cm, trim = {0cm 2cm 0cm 0cm}]{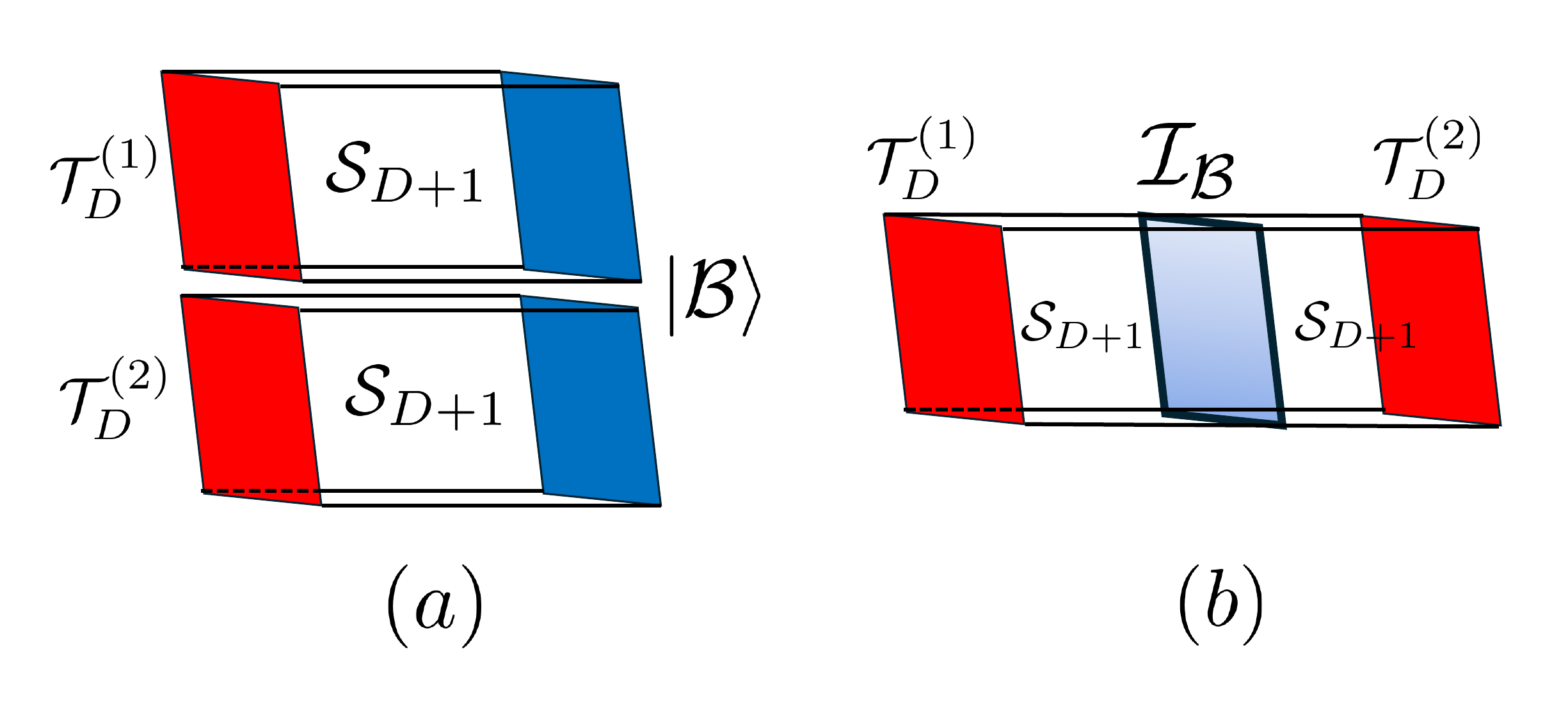}
    \caption{Two equivalent ways of presenting a disconnected SymTFT construction. In $(a)$, the worldvolume of the SymTFT $\mathcal{S}_{D+1}$ is disconnected and the topological boundary condition is specified by a possibly entangled state $\ket{\mathcal{B}}$. This is the convention we will most commonly take in this work. In (b), the data topological coupling caused by the entanglement in $\ket{\mathcal{B}}$ is given by a topological interface $\mathcal{I}_{\mathcal{B}}$. When $\mathcal{I}_{\mathcal{B}}$ being trivial implies $\ket{\mathcal{B}}$ is maximally entangled.}
    \label{fig:twoversions}
\end{figure}

In this work, we study generalizations of the usual SymTFT setup of Figure \ref{fig:basicsandwich} where the usual (Euclidean) worldvolume for the SymTFT, $M_D\times I$, for some $D$-dimensional manifold $M_D$ and some interval $I$, is replaced as:
\begin{equation}\label{eq:multipleI}
    M_D\times I \rightarrow M_D\times \left( I\coprod I\coprod \dots I\right).
\end{equation}
In other words, we consider multiple disconnected sandwich constructions labeled by $i=1,...,K$ and with multiple (potentially different) gapless boundary theories $\mathcal{T}_D^{(i)}$. Our main interest is in studying the role of entangled states in a $K$-fold tensor product Hilbert space for SymTFTs, which is necessary to understand the possible choices of topological boundaries in the setting of \eqref{eq:multipleI}. We will call such boundary states \textit{SymTFT entangled} (or the phonetically friendlier \textit{S-entangled}) which means that the total boundary state cannot be written as a direct product: $\ket{\mathcal{B}}\neq \ket{\mathcal{B}_1}\otimes \ket{\mathcal{B}_2} ...\otimes \ket{\mathcal{B}_K}$. We will see that this essentially amounts to a gauging of a common finite\footnote{We restrict ourselves to finite symmetries as the SymTFT construction is currently most firmly established for these cases, but we will motivate interesting consequences for generalizing to continuous symmetries in Section \ref{ssec:CERNVT}.} subsymmetry between the various $\mathcal{T}^{(i)}_D$ theories. We will also see that S-entanglement can be equivalently viewed as coupling multiple theories via a topological interface, see Figure \ref{fig:twoversions}.\footnote{For similar existing constructions in the literature which have a pair of gapless theories of some dimension $D$ as boundaries of some $(D+1)$-dimensional topological theory, including the construction of 2D RCFTs, see e.g. \cite{Fuchs:2002cm, Kapustin:2010if, Moller:2024xtt, Heckman:2025lmw}.}

A straightforward implication of S-entanglement is that it leads to a lack of factorization of field theory data (e.g. partition functions, correlation functions, algebra of genuine line operators, etc.) into the $K$ constituent QFTs. Specifically (for the case $K=2$) we show that:
\begin{empheq}[box=\widefbox]{align*}
  & \textnormal{\textbf{Statement 1:} Given two S-entangled QFTs $\mathcal{T}^{(1)}_D$ and $\mathcal{T}^{(2)}_D$, there exists a $p$ such that}\\
  & \textnormal{the algebra of genuine $p$-dimensional defect operators of the total system does not factorize} \\
   & \textnormal{into the tensor product of genuine $p$-dimensional defect operators of $\mathcal{T}^{(1)}_D$ and $\mathcal{T}^{(2)}_D$.}
\end{empheq}

\begin{figure}[h]
    \centering
\includegraphics[width=12cm, trim = {0cm 2cm 0cm 0cm}]{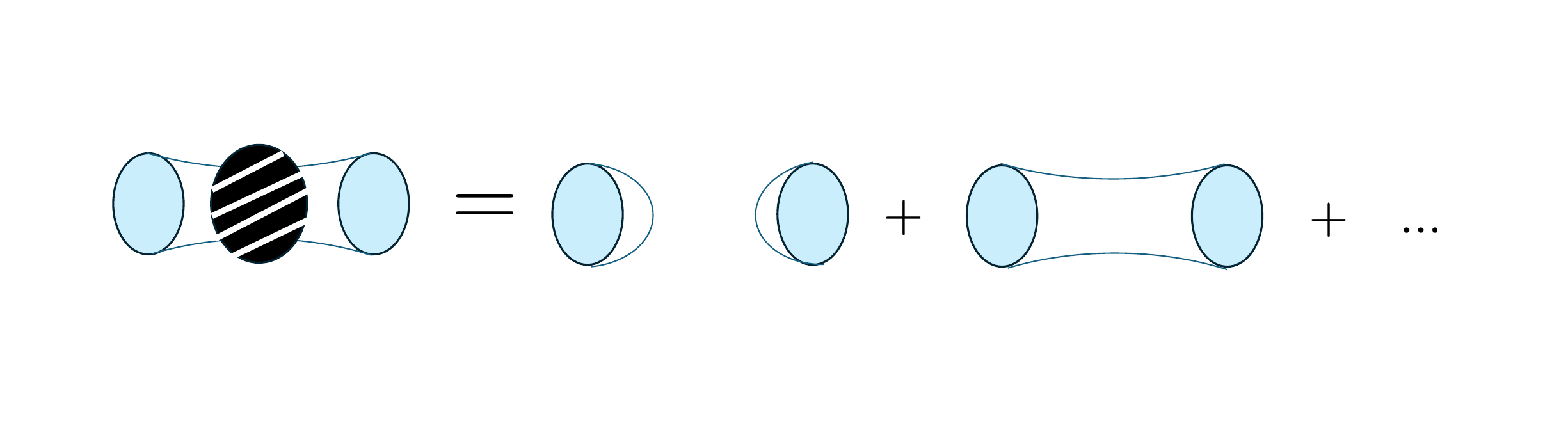}
    \caption{At least naively, a gravitational path integral for the case of two asymptotic boundaries (blue circles) includes both disconnected and connected contributions.}
    \label{fig:dummysum}
\end{figure}

When the QFTs $\mathcal{T}^{(i)}_D$ have semiclassical\footnote{I.e. the low-energy bulk gravitational sector is described by Einstein gravity.} holographic duals, then the factorization of physical data, or lack-thereof, is of central interest in understanding generalizations of the AdS/CFT correspondence to spacetimes with multiple (disconnected) asymptotic boundaries \cite{Witten:1999xp, Maldacena:2004rf}. In particular, what is known as the ``factori\textbf{z}ation puzzle" (US English spelling) arises when one considers a gravitational path integral in the bulk which includes a sum over topologies, see Figure \ref{fig:dummysum}. Including wormhole configurations into this sum seems to lead to a violation of factorization in the partition function, correlators, Hilbert spaces, etc. of the system. While such a sum can be made precise (to varying degrees) in low dimensions, such as in the case of JT gravity or perturbative string worldsheets, Figure \ref{fig:dummysum} is merely a heuristic for $D\geq 4$. In the context of top-down realizations of AdS/CFT in string theory one can easily construct holographic duals of direct product CFTs $\mathcal{T}_1\otimes \mathcal{T}_2$ which are simple two disconnected asymptotically AdS spacetimes\footnote{For example, one can generalize the original argument considered in \cite{Maldacena:1997re} of the duality between 4D $\mathcal{N}=4$ Super Yang-Mills (SYM) and IIB on $\mathrm{AdS}_5\times S^5$ by starting with IIB string on a disconnected target space $\mathbb{R}^{1,9}\coprod \mathbb{R}^{1,9}$. One can consider a large $N_1$ amount of $D3$-branes in the first component and a large $N_2$ amount of $D3$-branes in the second component, and the gravitational backreaction will produce two near-horizon throats which are decoupled. In this language, a key point of our work is to point out the that transverse spaces of the $D3$s, $\mathbb{R}^6\coprod \mathbb{R}^6$, can have boundary conditions which are entangled.}.

When two CFTs are coupled via a discrete gauging, such as 4$D$ $\mathcal{N}=4$ SYM theory with gauge group\footnote{This can be obtained from gauging a diagonal $\mathbb{Z}_N$-valued 1-form symmetry of the $G=SU(N)^2$ theory. Readers perhaps unfamiliar with such statements are referred to the review material in Section \ref{sec:SymTFTentanglement}.} $G=(SU(N)\times SU(N))/\mathbb{Z}_N$, this leads to S-entangled theories and part of the purpose of this work is to give a top-down string construction of such scenarios as well as bulk interpretation of their physics. Roughly speaking, we will find that S-entanglement correlates discrete gauge theory data between two or more disconnected AdS spacetimes. These AdS spacetimes are otherwise not coupled to one another as topological coupling between CFTs does not affect the stress-tensor, by definition.

Such bulk interpretations of S-entanglement is particularly interesting in the context of AdS eternal (two-sided) black holes. The aforementioned correlations between gauge charges clarifies a conceptual puzzle first raised in \cite{Marolf:2012xe} regarding original proposal of Maldacena \cite{Maldacena:2001kr} that these systems are described by a thermofield double (TFD) state in a direct product CFT $\mathcal{T}\otimes \mathcal{T}^\dagger$ where $\dagger$ denotes CPT conjugate. While we leave a review of the puzzle posed by Marolf/Wall to Section \ref{ssec:UCSBCambridge}, we argue that given that the CFT $\mathcal{T}$ has a discrete, internal 0-form global symmetry $H^{(0)}$ then:
\begin{empheq}[box=\widefbox]{align*}
  & \textnormal{\textbf{Statement 2:} A two-sided black hole in AdS which has a single well-defined $H^{(0)}$ gauge charge}\\
  & \textnormal{is dual to a TFD state prepared in a diagonal gauging of a product CFT: $(\mathcal{T}\otimes \mathcal{T}^\dagger)/H^{(0)}$.}
\end{empheq}
If one does not include the diagonal gauging, then we show that the corresponding eternal AdS black hole has two separate $H^{(0)}$ gauge charges, one associated with the left horizon and the other the right horizon. Overall, we argue that the AdS eternal black hole systems can be prepared in a plurality of ways, some of which (including those for which there is only one $H^{(0)}$-charge) have a total Hilbert space which does not factorize $\mathcal{H}\neq \mathcal{H}_L\otimes \mathcal{H}_R$. In fact, by the state-operator correspondence, one can see this as a special case of Statement 1 above when $p=0$. In the literature, the question of factorization for two-side black holes often called the ``factori\textbf{s}ation puzzle" (UK English spelling)\footnote{According to \cite{Penington:2023dql}, this linguistic distinction is due to Henry Maxfield.}, but we will simply use ``factorization" throughout this work for notational uniformity.\footnote{This is despite the fact that this violates the preprint requirements of one of the author's organization \cite{cernpreprint}.}

Another application of S-entanglement arises when we focus on just one of the constituent theories $\mathcal{T}^{(i)}_D$. This requires one to take a partial trace of the pure state density matrix $\ket{\mathcal{B}}\bra{\mathcal{B}}$, thus, one is led to consider the generalization of the SymTFT construction whereby the topological boundary state is a \textit{mixed state}. Our top-down string theory examples of S-entanglement therefore also construct top-down examples of ensemble averaging in string theory, which unifies constructions found in \cite{Heckman:2021vzx, Baume:2023kkf}. One set of examples we consider which has particular overlap with previous considerations of ensemble averaging from bottom-up perspectives display S-entanglement of $(-1)$-form symmetries. Since $(-1)$-form global symmetry background fields are simply parameters of a QFT, non-trivial mixed topological boundary states in the SymTFT sector describing $(-1)$-form symmetries precisely lead to ensemble averaging. Additionally, in light of recent work on the connection between (non-)factorization and the dimension of baby universe Hilbert spaces, see for instance \cite{Usatyuk:2024isz} as well as related discussions in \cite{Marolf:2020xie, McNamara:2020uza, Penington:2019kki, Engelhardt:2025vsp, Akers:2025ahe, Coleman:1988cy}, we comment on how our top-down non-factorizing AdS models have a dimension of their baby universe Hilbert spaces larger than one and also display features of having bulk global symmetries.\footnote{We emphasize that we of course do not claim to construct in bulk global symmetries in a quantum gravity model as this only arises when one looks at part of the system. Specifically, the bulk picture of our top-down models always take the form of multiple disconnected AdS spacetimes with some entangled boundary conditions at infinity, and our baby universe/bulk global symmetries statements hold only if one takes into account only a subset of these AdS spacetimes.} In particular, we obtain the following quantitative result:
\begin{empheq}[box=\widefbox]{align*}
  & \textnormal{\textbf{Statement 3:} Partial traces of SymTFT-entangled states prepare ensemble-averaged boundary}\\
  & \textnormal{theories, whose effective baby universe Hilbert space dimension is given by the exponential of the}\\
  & \textnormal{von Neumann entropy or, equivalently, the dimension of the Lagrangian algebra of the SymTFT.}
\end{empheq}

This paper is organized as follows. In Section \ref{sec:SymTFTentanglement} we start by reviewing the basics of SymTFT constructions and how it is naturally realizes in AdS/CFT.
We then define S-entanglement, give a holographic interpretation, and explore various examples. This section works entirely in the context of 4D $\mathcal{N}=4$ SYM theories. Section \ref{sec:Otherexamples} discusses other examples of S-entanglement including $6D$ $(2,0)$ SCFTs, $3D$ ABJM theories, and $2D$ symmetric orbifold CFTs. In Section \ref{sec:ohbaby} generalize the SymTFT construction to mixed boundary states as it naturally arises from taking partial traces of S-entangled constructions. We show how this leads to (higher) ensemble averaging in various examples and explain the holographic implications for baby universe states and bulk global symmetries. In Section \ref{sec:eternal}, we apply what we have learned about S-entanglement to eternal AdS black holes and how this enriches the original proposal of \cite{Maldacena:2001kr} and discuss some physical consequences of this. Section \ref{sec:conc} contains our conclusions and future directions. Appendix \ref{app:5point1extended} gives more details on thermofield double state preparation with S-entanglement relevant for AdS$_3$/CFT$_2$, and Appendix \ref{app:theta} gives an example showcasing how a simple generalization of our construction can yield S-entanglement for continuous symmetries and a continuous ensemble averaging.





\emph{Note added:} While the main results in this work were being completed and presented by one of the authors (ET) at \cite{ET:2025talk1, ET:2025talk2}, an independent preprint \cite{Heckman:2025lmw} appeared on the ArXiv which includes examples of what we refer to as S-entanglement. We also mention upcoming work \cite{Jakecite1, Jakecite2} which has some overlapping themes with this paper.


\section{SymTFT Entanglement for 4D $\mathcal{N}=4$ Super Yang-Mills Theories}\label{sec:SymTFTentanglement}

Having outlined the concept of SymTFT entanglement (or S-entanglement for short) in the introduction, we turn now in this section to concretely presenting our proposal for the case when the physical boundary theories are copies of:
\begin{equation}\label{eq:Texample}
    \mathcal{T}:=\left( \textnormal{4D $\mathcal{N}=4$ SYM with gauge algebras $\mathfrak{su}(N)$}\right)
\end{equation}
Note that $\mathcal{T}$ is a relative theory in the sense of \cite{Freed:2012bs} and so only depends on the gauge algebra $\mathfrak{su}(N)$. After first reviewing the SymTFT construction for a single copy of $\mathcal{T}$ and the dictionary with the holographic dual (assuming large $N$), we discuss the field theoretic and holographic ramifications of our proposal for two identical copies of $\mathcal{T}$.






\subsection{Review of SymTFT and Holographic Interpretation for $\mathcal{T}$}\label{ssec:reviewsymhol}

We first begin with a review of the SymTFT picture for the 1-form symmetries of $\mathcal{T}$ as well as its holographic interpretation. Readers familiar with SymTFTs can safely skip the first half of this subsection. Recall that the gauge algebra $\mathfrak{su}(N)$ determines a set of line operators that come in two (irreducible) flavors: Wilson lines and 't Hooft lines. Wilson lines are labeled by representations, $\mathbf{R}$, of $\mathfrak{su}(N)$, and are defined as
\begin{equation}\label{eq:wilsonline}
    W_\mathbf{R}(L):=\mathrm{Tr}_{R}\mathcal{P}\exp\bigg(i\int_{L}A\bigg)
\end{equation}
where $\mathcal{P}$ is the path-ordering and $L$ is the support of the Wilson line. A Wilson line can be thought of as an infinitely massive probe electric particle in the $R$ representation, on the other hand 't Hooft operators are infinitely massive probe magnetic monopoles. Their charge is characterized by representations of the Langlands dual algebra of $\mathfrak{su}(N)$ which is, incidentally, also $\mathfrak{su}(N)$. This means that we can label the minimal charge 't Hooft line, $H_{\mathbf{N}}(L)$, by the fundamental representation $\mathbf{N}$, just as $W_{\mathbf{N}}$ has minimum electric charge. One can also consider dyonic lines by fusing Wilson and 't Hooft lines. Note that while the operator \eqref{eq:wilsonline} is not strictly speaking BPS, we will implicitly use $W_\mathbf{R}(L)$ to also denote its $\frac{1}{2}$-BPS cousin. We do this both because we are mainly worried about the global symmetry charge of such line operators (which is the same for supersymmetric versions of \eqref{eq:wilsonline}) and the amount of supersymmetry in the situations we consider will be clear.

\begin{figure}[h]
    \centering
\includegraphics[width=10cm, trim = {0cm 0cm 0cm 0cm}]{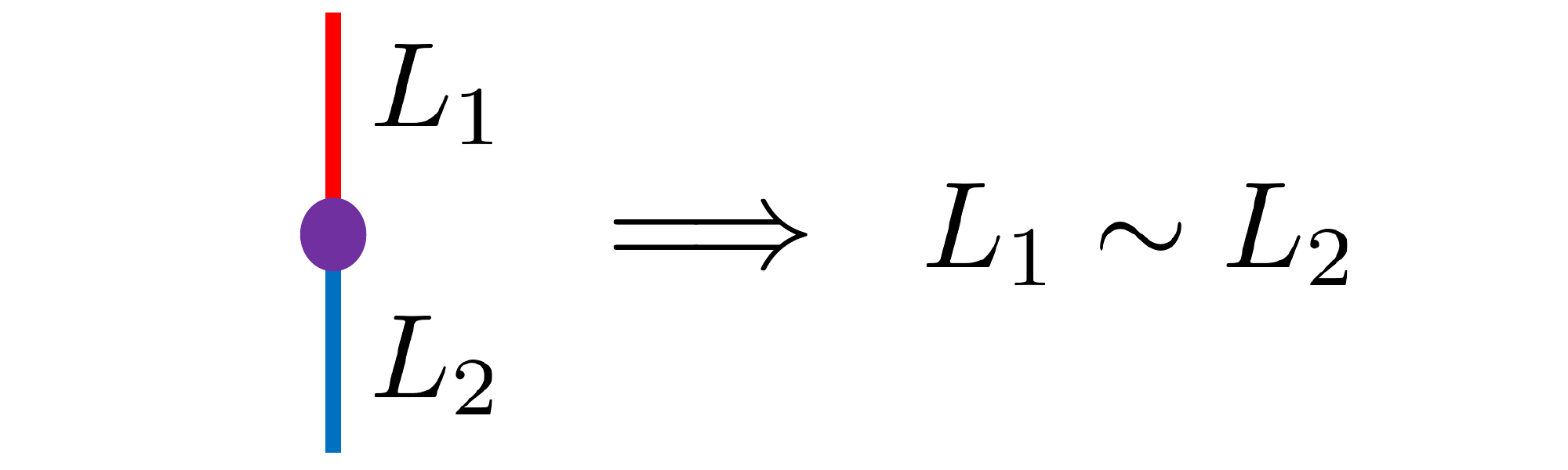}
    \label{fig:simdefinition}
    \caption{}
\end{figure}

Within the set of Wilson and 't Hooft lines, one can consider the equivalence class of line operators where two line operators are identified under an equivalence relation $\sim$ iff there is a local operator separating the two lines, see Figure \ref{fig:simdefinition}. We will call this equivalence class the defect group of line operators \cite{Aharony:2013hda, DelZotto:2015isa} (see also Appendix A of \cite{Lee:2021crt} for a gentle introduction), which we will denote by $\mathbb{D}^{(1)}$ with the superscript anticipating that these are charged under 1-form symmetries. For pure 4D gauge theories (possibly coupled to adjoint matter like in $\mathcal{N}=4$ SYM) with gauge algebra $\mathfrak{g}$ this is given by\footnote{In the definition of $\mathbb{D}(\mathfrak{g})^{(1)}$, we consider both genuine line operators, as well as line operators which may have topological surface operators attached.}
\begin{equation}
    \mathbb{D}(\mathfrak{g})^{(1)}:=\{\textnormal{{Line operators}}\}/\sim \; = Z(G_{s.c.})^{(1)}\oplus Z(G_{s.c.})^{(1)}
\end{equation}
where $G_{s.c.}$ is the unique simply-connected Lie group associated with $\mathfrak{g}$, and $Z(G_{s.c.})$ denotes its center. The group operation of $\mathbb{D}(\mathfrak{g})^{(1)}$ follows from fusion of line operators. For $\mathfrak{g}=\mathfrak{su}(N)$ we have
\begin{equation}\label{eq:defectgroupsun}
    \mathbb{D}(\mathfrak{su}(N))^{(1)}=(\mathbb{Z}_N)^{(1)}_e\oplus (\mathbb{Z}_N)^{(1)}_m
\end{equation}
which are generated by $W_\mathbf{N}$ and $H_{\mathbf{N}}$ respectively, which explains the subscripts. This implies that $N$ copies of each can end on local operators. For example, the fact that $N$ copies of $W_\mathbf{N}$ can end on a creation operator for a vector boson (such that the combined configuration is gauge invariant) follows from the fact that
\begin{equation}
    \mathbf{Adj}\subset \mathbf{N}^{\otimes N}.
\end{equation}
One can define 2d topological surface operators $U(\Sigma_2)$ and $V(\Sigma_2)$ which measure these equivalence classes:
\begin{align}
    U(\Sigma_2)W_\mathbf{N}(L)=e^{2\pi i\mathrm{Link}(\Sigma_2,L)/N}W_\mathbf{N}(L)U(\Sigma_2) \\
    V(\Sigma_2)H_\mathbf{N}(L)=e^{2\pi i\mathrm{Link}(\Sigma_2,L)/N}H_\mathbf{N}(L)V(\Sigma_2)
\end{align}
where $\mathrm{Link}(-,-)$ denotes the integer-valued link pairing of 1-cycles and 2-cycles in 4D. Together $U$ and $V$ generate a 1-form symmetry group $(\mathbb{Z}_N)^{(1)}_e\oplus (\mathbb{Z}_N)^{(1)}_m$. Due to the non-trivial Dirac pairing\footnote{Technically we mean the Dirac pairing modulo $1$. As an example, let us take $N=2$ and turn on a vev for an adjoint scalar which Higges the gauge group to $U(1)$. In the normalization where the W-bosons have electric $\pm 2$ under this $U(1)$, the Wilson line operators decompose as $W_\mathbf{2}\rightarrow W_{+1}+W_{-1}$ where the subscript is the $U(1)$-charge. Meanwhile, the 't Hooft line operators decompose as $H_{\mathbf{2}}\rightarrow H_{+1/2}+H_{-1/2}$ where $H_{\pm 1/2}$ is defined by the condition that $\int_{S^2}F_{U(1)}=\pm 1/2$ for an $S^2$ linking the 't Hooft operator. This Higgsing picture makes clear that $W_\mathbf{2}$ and $H_\mathbf{2}$ have Dirac pairing $\frac{1}{2} \; \mathrm{mod}\; 1$.} between electric and magnetic particles associated with the representation $\mathbf{N}$, there is a non-trivial 't Hooft anomaly between $(\mathbb{Z}_N)^{(1)}_e$ and $(\mathbb{Z}_N)^{(1)}_m$ which is captured by the 5D topological action
\begin{equation}\label{eq:sunsymtftaction}
   S_{5D}= \frac{2\pi}{N}\int_{X_5} b_2\cup
    \delta c_2
\end{equation}
via inflow as we take the 4D worldvolume of $\mathcal{T}$, $M_4$, to be such that $M_4\subset \partial X_5$. Here $c_2$ and $b_2$ are $\mathbb{Z}_N$-valued 2-form gauge potentials. As a disclaimer, note that the aforementioned 't Hooft anomaly is not, strictly speaking, a 't Hooft anomaly that an absolute QFT possess. Rather, this is mixed 't Hooft anomaly for a relative theory in the sense that it is inconsistent for both $(\mathbb{Z}_N)^{(1)}_e$ and $(\mathbb{Z}_N)^{(1)}_m$ to be gauged. More formally, a 't Hooft anomaly for a discrete symmetry of an absolute 4D QFT can be captured by an invertible 5D TFT, while the TFT \eqref{eq:sunsymtftaction} capturing this slightly more general notion of 't Hooft anomaly is non-invertible. 

We will regard the 5D TFT with action \eqref{eq:sunsymtftaction} as the\footnote{As mentioned in the introductions, what we call SymTFT throughout most of this work captures discrete internal symmetries.} SymTFT for $\mathcal{T}$ where we take the 5D TFT worldvolume to be $X_5=M_4\times I$ with the interval $I\equiv [0,1]$ parametrized by a coordinate $y$. The CFT $\mathcal{T}$ is localized at $y=0$ while $y=1$ is a topological boundary condition whose purpose will be made clear below. The topological surface operators of this 5D TFT are given by
\begin{align}
     U(\Sigma_2)=\exp\bigg(\frac{2\pi i}{N}\int_{\Sigma_2}c_2\bigg) \\
      V(\Sigma_2)=\exp\bigg(\frac{2\pi i}{N}\int_{\Sigma_2}b_2\bigg) \label{eq:defV}
\end{align}
which satisfy the algebraic relations
\begin{align}
    U(\Sigma_2)V(\Sigma'_2)&=e^{\frac{2\pi i }{N}\mathrm{Link}(\Sigma_2,\Sigma'_2)}V(\Sigma'_2)U(\Sigma_2) \label{eq:5dlink}\\
    U(\Sigma_2)^N&=1, \quad \quad V(\Sigma_2)^N=1
\end{align}
where $\mathrm{Link}(-,-)$ denotes the integer-valued link pairing of 2-cycles in $X_5$ and $\sim$ denotes homological equivalence. Given some $y_0\in [0,1]$ the dictionary between SymTFT topological operators and symmetry operators of $\mathcal{T}$ is
\begin{align}
    U(\Sigma_2\times \{y_0\}) \Leftrightarrow \quad \textnormal{$(\mathbb{Z}_N)^{(1)}_e$ Symmetry Operator}\\
     V(\Sigma_2\times \{y_0\}) \Leftrightarrow \quad \textnormal{$(\mathbb{Z}_N)^{(1)}_m$ Symmetry Operator}.
\end{align}
Moreover, Wilson / 't Hooft line operators with support $L\subset M_4\times \{y=0\}$ are respectively attached to $V(L\times I)$ and $U(L\times I)$. This implies that the 5D link pairing \eqref{eq:5dlink} reproduces the usual action of symmetry operators on charged line operators. In other words, after reducing on $I$ we have for instance
\begin{equation}\label{eq:4dsymaction}
    U(\Sigma_2)W_{\mathbf{N}}(L)=e^{ \frac{2\pi i }{N}\mathrm{Link}(L,\Sigma_2)}W_{\mathbf{N}}(L)U(\Sigma_2)
\end{equation}
and similarly for action of $V$ on $H_\mathbf{N}$ Note that the link pairing between 1-cycles and 2-cycles in \eqref{eq:4dsymaction} descends from the 2-cycle link pairing in 5D.

Note that when the theory $\mathcal{T}$ is placed on a Euclidean 4-manifold $M_4$ such that $|H_2(M_4,\mathbb{Z}_N)|\neq 0$, it is well-known that the relative theory $\mathcal{T}$ does not have an unambiguous partition function, but should be described by a partition vector \cite{Witten:1998wy, Freed:2012bs}. In SymTFT language is simply due to the fact that the quantization of the action \eqref{eq:sunsymtftaction} along $I$ assigns a Hilbert space of dimension $N\times |H_2(M_4,\mathbb{Z}_N)|$ to $M_4$. One can obtain a well-defined partition function by selecting topological boundary conditions at $y=1$. Such boundary data is equivalent to selecting a polarization\footnote{Choosing a polarization specifies the absolute theory only up to the addition of local counterterms or SPT phases. One way to refine this choice is to consider a \emph{polarization pair}, namely two Lagrangian algebras in the SymTFT, with one determining the global form of the gauge group and the other selecting the corresponding counterterm/SPT stacking \cite{Lawrie:2023tdz}.} for the Heisenberg algebra in \eqref{eq:5dlink} (see for instance \cite{Belov:2006jd} for more details). Upon selecting such a boundary state $\ket{\mathcal{B}_{\mathrm{top.}}}\in \mathcal{H}_{\mathcal{S}}(M_4)$ then reducing on $I$ yields an absolute 4D QFT, $\mathcal{T}^{(\mathcal{B}_{\mathrm{top.}})}$, whose partition function is schematically
\begin{equation}
    Z_{\mathcal{B}_{\mathrm{top.}}}(M_4)=\langle \mathcal{T}| \mathcal{B}_{\mathrm{top.}} \rangle
\end{equation}
where $\ket{\mathcal{T}}$ denotes the non-topological boundary state with the localized $\mathcal{N}=4$ gauge degrees of freedom. We will highlight below how these choice of boundary conditions determine which line operators are genuine or non-genuine (i.e. those which require a two-dimensional surface operator to be attached to them)\footnote{In general, given a boundary state of a SymTFT for the symmetry category $\mathcal{C}$, it corresponds to a topological boundary if and only if it is associated to an object $m$ of the (higher) module category $\mathcal{M}$ over $\mathcal{C}$. Physically, $\mathcal{M}$ gives rise to ``representations'' of $\mathcal{C}$, which needs the states form a closed set under the action of topological operators generating the symmetry $\mathcal{C}$. For fusion category symmetry in 2D QFTs, this is completely understood in rigorous mathematical language \cite{etingof2015tensor} (see also \cite{Bartsch:2022ytj,Bartsch:2022mpm,Bhardwaj:2023ayw,Cordova:2024iti,Yu:2025iqf} for a partial list of recent physical discussions).}.


For example, let us take the case of $N=2$ and $H_2(M_4,\mathbb{Z}_2)=\mathbb{Z}_2$ for simplicity. Then along the $y=1$ boundary, we can for instance choose Dirichlet boundary conditions for $b_2$ and a Neumann boundary conditions for $c_2$
\begin{equation}
    b_2|_{y=1} \; \mathrm{fixed}, \quad d_yc_2=0
    \end{equation}
where $d_y:=\frac{\partial}{\partial y} dy$ is the exterior derivative transverse to $M_4$.

\begin{figure}[h]
    \centering
\includegraphics[width=12cm, trim = {0cm 2cm 0cm 3cm}]{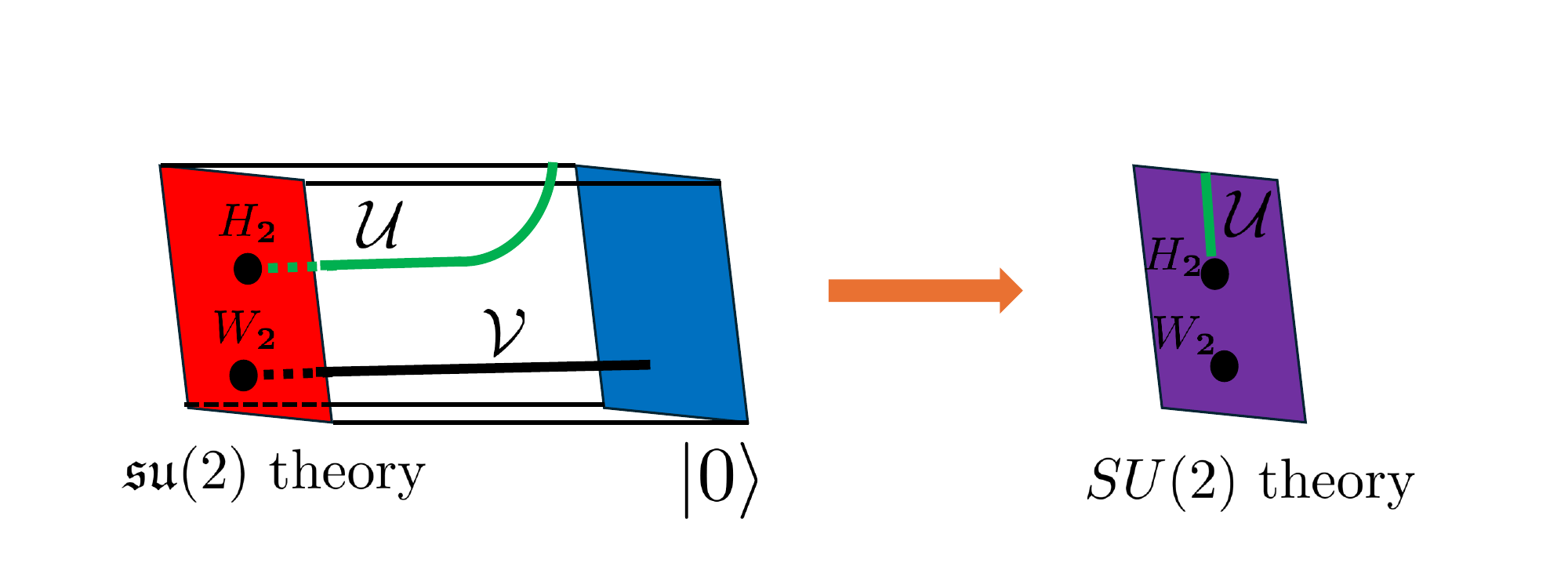}
    \label{fig:closingsandwich}
    \caption{Left: SymTFT setup for 4D $\mathcal{N}=4$ $\mathfrak{su}(2)$ theory with topological boundary condition $\ket{0}$. Right: Closing the sandwich fixes the gauge group to $SU(2)$ where minimal charge Wilson/'t Hooft line $W_{\mathbf{2}}$/$H_{\mathbf{2}}$ is genuine/non-genuine.}
\end{figure}

This amounts to choosing the global topology of the 4D gauge group to be $SU(2)$ which follows from our above definitions as the Dirichlet value $b_2|_{y=1}=B_2$ (well-defined up to shifting by exact 2-forms and large gauge transformations) is the background field for the $(\mathbb{Z}_2)^{(1)}_e$ global symmetry. Notice that $c_2$ being dynamical along $y=1$ is necessary for $U(\Sigma_2)$ to be a non-trivial operator in the theory while $V(\Sigma_2)$ is simply a c-number times the identity surface operator. Additionally, such boundary conditions also have implications for $U$ and $V$ with support along the interval direction. For instance, we can allow $V(L\times I)$ to end on the topological boundary at $y=1$, but are forbidden from ending $U(L\times I)$ on it in a gauge invariant manner.\footnote{There is a slight abuse of notation here since the worldvolume of such $U$ operators would be as depicted in Figure \ref{fig:closingsandwich} rather than $L\times I$. } Meanwhile both $U$ and $V$ can end on the non-topological boundary at $y=0$ as 't Hooft operators or Wilson operators, respectively, which are non-trivial in the defect group equivalence class. Upon closing the SymTFT sandwich, this implies that such Wilson lines will be genuine line operators (not attached to higher-dimensional operators), which such 't Hooft lines must be attached to the symmetry operators $U$, see Figure \ref{fig:closingsandwich} for a summary.

Highlighting more general boundary choices, we can take a Dirichlet boundary condition for $c_2$, where $c_2|_{y=1}=C_2$ is the background field for the $(\mathbb{Z}_2)^{(1)}_m$ global symmetry, and a Neumann boundary condition for $b_2$ amounts to choosing $SO(3)$ as the gauge group. Of course when $|H_2(M_4,\mathbb{Z}_2)|>1$, there are more general choices one can make but we will always choose polarizations which treat each of the generators of $H_2(M_4,\mathbb{Z}_2)$ on equal footing for simplicity. Additionally, even when $|H_2(M_4,\mathbb{Z}_2)|=1$, there is yet another polarization choice which amounts to picking Dirichlet condition for $b_2+c_2$. This results in a $SO(3)_-$ gauge theory where the subindex labels the nontrivial discrete theta-angle \cite{Aharony:2013hda}. The $(\mathbb{Z}_2)^{(1)}$ global symmetry in this case acts non-trivially on dyonic line operators $W_{\mathbf{2}}H_{\mathbf{2}}$. The case without the discrete theta angle will be denoted $G=SO(3)_+$ below.

We now take a concrete look at the Hilbert space the SymTFT assigns to $M_4$ in this context with $N=2$ and $b_2=1$, and what it means to choose various polarizations.
This can be presented as a two-dimenisonal space with qubit basis vectors $\ket{0}$ and $\ket{1}$ which we take to be eigenvectors of $U$:
\begin{align}
  &  V\ket{0}=\ket{0}, \quad \quad V\ket{1}=-\ket{1} \\
   & U\ket{0}=\ket{1}, \quad \quad U\ket{1}=\ket{0}. \label{eq:uket}
\end{align}
$V$ and $U$ are thus also known as ``clock'' and ``shift'' operators, respectively. These eigenstates are relevant for the theory with the global form $SU(2)$ because from \eqref{eq:defV}, we see that the $\ket{k}$ state is associated with a background $\int_{\gamma_2} B_2=k \; \mathrm{mod}\; 2$ where $\langle \gamma_2\rangle=\mathbb{Z}_2=H_2(M_4,\mathbb{Z}_2)$. Similarly from \eqref{eq:uket}, the operator $U$ acts as the symmetry operator since its presence changes the background field. For the choice of $G=SO(3)_+$, one could make similar statements for the basis of vectors which are eigenstates of $V$ instead, but we will choose to stay in the $U$-diagonalized basis. This means that the $V$-eigenstates are
\begin{equation}
    \frac{1}{\sqrt{2}}(\ket{0}\pm \ket{1})
\end{equation}
where the $\pm=(-1)^\ell$ coefficient is determined by the value of $\int_{\gamma}C_2=\ell \; \mathrm{mod}\; 2$. Meanwhile the $G=SO(3)_-$ polarization boundary states must be eigenstates of the operator
\begin{equation}
    W\equiv iVU=\exp\left(i\int_{\Sigma_2}(b_2+c_2)\right)
\end{equation}
which are $\frac{1}{\sqrt{2}}(\ket{0}\pm i\ket{1})$ where the $\pm=(-1)^m$ coefficient is determined by the value of $\int_{\gamma}(B_2+C_2)=m \; \mathrm{mod}\; 2$. The results are summarized in Table \ref{tab:qubits for SYM global forms}. For latter purposes, we find it helpful to introduce the notation $\mathcal{V}^{\textnormal{gen. lines}}_{\ket{\mathcal{B}}}$ for the algebra of genuine line operators for the theory associated with SymTFT boundary state $\ket{\mathcal{B}}$. For example, $W_{\mathbf{2}}\in \mathcal{V}^{\textnormal{gen. lines}}_{\ket{0}}$, while $H_{\mathbf{2}}\notin \mathcal{V}^{\textnormal{gen. lines}}_{\ket{0}}$.
\begin{table}[h]
    \centering
    \begin{tabular}{|c|c|c|}
    \hline
     Global Form/Polarization   & Qubit Boundary States & Min. Charge Genuine Lines\\
     \hline
      $SU(2)$   & $|0\rangle$ and $\ket{1}$ & $W_{\mathbf{2}}$\\
      \hline
      $SO(3)_+$   & $\frac{1}{\sqrt{2}}\left( |0\rangle\pm |1\rangle\right)$ & $H_{\mathbf{2}}$\\
      \hline
      $SO(3)_-$   & $\frac{1}{\sqrt{2}}\left(|0\rangle\pm i|1\rangle\right)$ & $W_{\mathbf{2}}H_{\mathbf{2}}$\\
      \hline
    \end{tabular}
    \caption{Qubit states in the Hilbert space of the SymTFT for global forms of 4D $\mathfrak{su}(N)$ SYM, and the genuine line operators with minimal 1-form global symmetry charge.}
    \label{tab:qubits for SYM global forms}
\end{table}

One can now recover the fact that the $G=SU(2)$ and $G=SO(3)_+$ theories are related by gauging their 1-form global symmetries by the fact that their topological boundary states are discrete Fourier transforms of each other. For instance denoting $Z_{SU(2)}(k)\equiv Z_{SU(2)}(\int_{\gamma_2}B_2=k)= \langle \mathcal{T}|k\rangle$ then
\begin{equation}\label{eq:so3plus}
    Z_{SO(3)_+}(C_2)=\frac{1}{\sqrt{2}}\left(Z_{SU(2)}(0)+e^{\pi i \int_{\gamma_2}C_2}Z_{SU(2)}(1)\right).
\end{equation}
Indeed, one perspective of the fact that 't Hooft operators are non-genuine for the $SU(2)$ theory is that are charged under a $(\mathbb{Z}_N)^{(1)}$ \textit{gauge} symmetry, which is the same 1-form gauging that would reverse the process of \eqref{eq:so3plus} to take one from $SO(3)_+$ to $SU(2)$.

For general $N$, the set of possible polarizations one can take are heavily dependent on number-theoretic properties of $N$, but we can always consider the choices $G=SU(N)$, and $G=PSU(N):=SU(N)/\mathbb{Z}_N$ which are related to each other by gauging their respective $(\mathbb{Z}_N)^{(1)}$ global symmetries. Assuming $H_2(M_4,\mathbb{Z}_N)=\mathbb{Z}_N$, their SymTFT boundary states (depending on the 1-form symmetry background fields) are given by:
\begin{equation}
    \begin{split}
        G&=SU(N): \quad \ket{k} \; \implies \; \int B_2=k \; \mathrm{mod}\; N \\
    G&=PSU(N): \quad \frac{1}{\sqrt{N}}\sum^{N-1}_{k=0}e^{2\pi i k \ell/N}\ket{k} \; \implies \; \int C_2=\ell \; \mathrm{mod}\; N.
    \end{split}
\end{equation}
As an example of more intermediate cases, the consider $\mathfrak{su}(2N)$ SYM theories. These additionally include the cases (ignoring discrete theta angles)
\begin{equation*}
    \begin{split}
        G&=SU(2N)/\mathbb{Z}_2: \quad \frac{1}{\sqrt{2}}\left( \ket{k}+(-1)^{\ell}\ket{N+k}\right) \; \implies \; \int C_2=\ell \; \mathrm{mod}\; 2, \; \int B_2=k \; \mathrm{mod}\; N \\
        G&=SU(2N)/\mathbb{Z}_{N}: \quad \frac{1}{\sqrt{N}}\sum^{N-1}_{k'=0}e^{2\pi i k'\ell/N}\ket{2k'+k}\; \implies \; \int C_2=\ell  \; \mathrm{mod}\; N, \; \int B_2=k \; \mathrm{mod}\; 2
    \end{split}
\end{equation*}
Such cases have two background fields because there exist both electric and magnetic 1-form symmetries: the former has a $\mathbb{Z}^{(1)}_{N,e}\times \mathbb{Z}^{(1)}_{2,m}$ global symmetry, while the latter has $\mathbb{Z}^{(1)}_{2,e}\times \mathbb{Z}^{(1)}_{N,m}$.

Before moving onto the holographic interpretation of this SymTFT story we comment on the case of taking $M_4=\mathbb{R}^4$. One might conclude that because the SymTFT Hilbert space assigned to $\mathbb{R}^4$ is trivial and that there are no meaningful notions of global form of the gauge group in $\mathbb{R}^4$. However we know that this latter statement is incorrect as it was first shown in \cite{Aharony:2013hda} that gauge theories in flat space which differ by global forms of gauge theories can indeed have different phase structures. In the SymTFT setting, one can address this technicality by formally compactifies the Euclidean worldvolume and taking the volume to infinite size. Since we already mentioned above that we only consider polarizations that treat cycles uniformly, this technicality will not play any role in this work.

\begin{figure}[h]
    \centering
\includegraphics[width=12cm, trim = {0cm 2cm 0cm 1cm}]{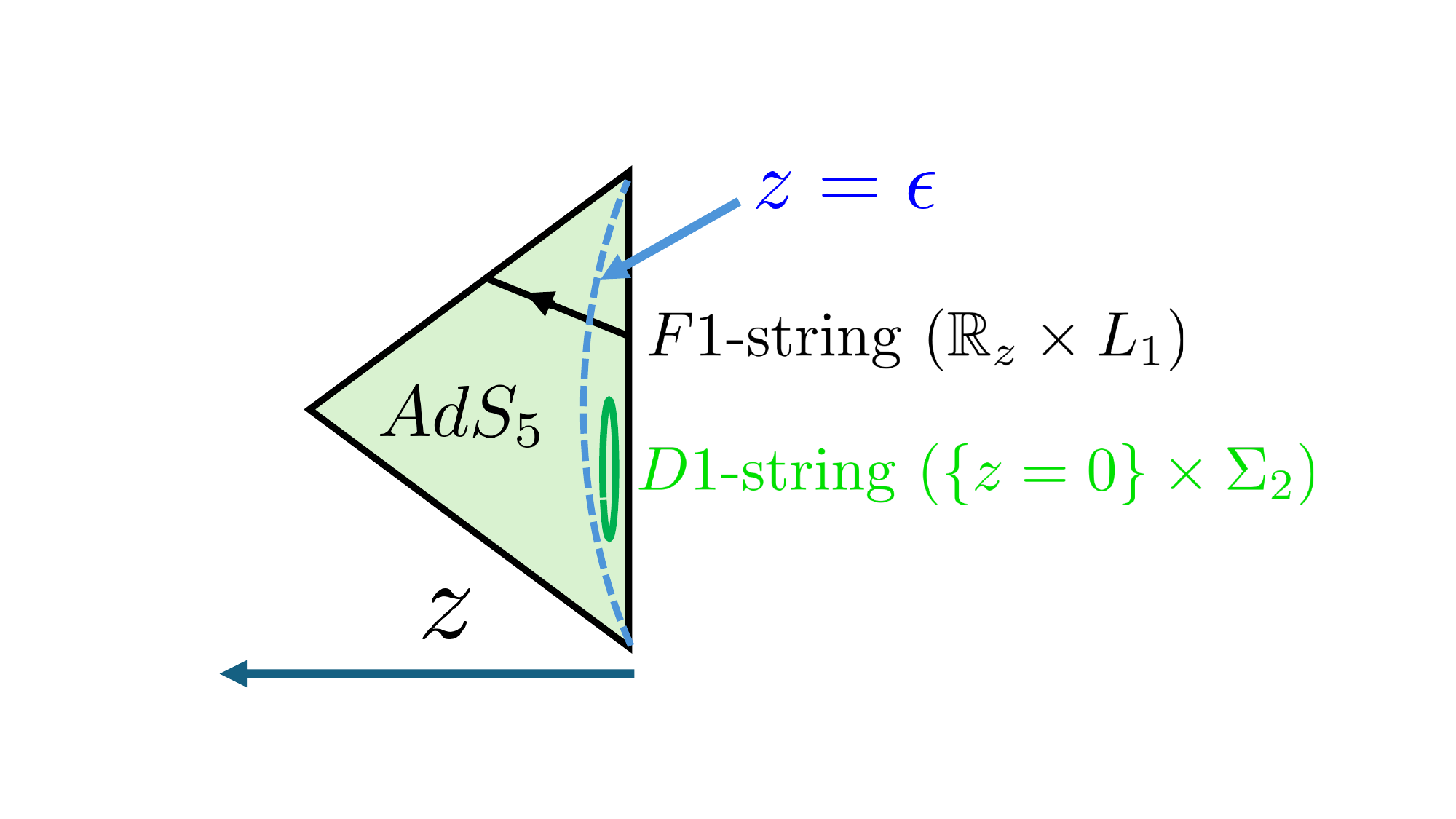}
    \label{fig:adssliver}
    \caption{Poincar\'e patch of AdS$_5$ spacetime. The standard SymTFT sandwich in a limit $\epsilon\rightarrow 0$ where the spacetime $z>\epsilon$ is captured by the physical boundary condition (i.e. $\mathfrak{su}(N)$ $\mathcal{N}=4$ SYM), and the $z=0$ boundary corresponds to the topological boundary condition which implicitly take to be $\ket{0}$. The region $[0,\epsilon]$ as the SymTFT interval. Note that we do not show the temporal directions of $L_1$ or $\Sigma_2$. }
\end{figure}

\paragraph{Holographic Interpretation}
In the large $N$ limit, there exists a dictionary between the SymTFT picture and the holographically dual IIB theory on $\mathrm{AdS}_5\times S^5$ as first discussed for similar theories in \cite{Apruzzi:2022rei, GarciaEtxebarria:2022vzq} (see also \cite{Heckman:2024oot} for a helpful introduction, and the earlier \cite{Witten:1998wy}). The basic idea starts by reducing the IIB supergravity action on the internal $S^5$ with $N$ units of five-form flux, the effective action includes a topological term $N\int_{\mathrm{AdS}_5} B_2\wedge dC_2$ as it descends from the 10D term $\int C_4\wedge H_3\wedge F_3$. We see that this 5D term is identical to that of \eqref{eq:sunsymtftaction} up to a normalization, and one obtains bona fide discretely valued fields as in \eqref{eq:sunsymtftaction} if the kinetic terms of $B_2$ and $C_2$ vanish. This is indeed the case in an infinitesimal neighborhood of the boundary as the field fluctuations in this neighborhood have an infinitely large redshift with respect to the bulk degress-of-freedom. That is the sense in which we obtain a topological field theory close to the boundary where the boundary conditions of AdS space are identified with the topological boundary of the SymTFT picture, see Figure \ref{fig:adssliver}. Meanwhile the bulk physics, as captured by the gapless degrees of freedom of the CFT, are of course identified with the physical boundary.

To state the SymTFT-AdS spacetime dictionary precisely, let us consider a Poincar\'e patch of the AdS$_5$ spacetime which has metric
\begin{equation}
    ds^2=R^2_{AdS}\left( \frac{dz^2+\eta^{4D}_{\mu\nu}dx^{\mu\nu}}{z^2}\right).
\end{equation}
Equivalently, this means we either take the boundary theory to be $\mathbb{R}^{3,1}$ or $\mathbb{R}^4$ depending on the signature of the bulk spacetime. We then have the following dictionary for the lines generating the defect group \eqref{eq:defectgroupsun}:
\begin{align}
    W_{\mathbf{N}}(L_1) \quad \Leftrightarrow \quad \textnormal{$F1$-string on $\mathbb{R}_z\times L_1$} \label{eq:wdictionary}\\
     T_{\mathbf{N}}(L_1) \quad \Leftrightarrow \quad \textnormal{$D1$-string on $\mathbb{R}_z\times L_1$} \label{eq:tdictionary}
\end{align}
In Figure \ref{fig:adssliver}, we took the boundary condition $\ket{0}$ which yields a $SU(N)$ SYM theory, which in the spacetime picture implies that the $F1$ string is free to end on the boundary $z=0$, while $D1$-strings are forbidden from ending. The fact that $N$ $F1$-strings can end on a $D5$ brane wrapping the internal $S^5$, is mirrored in the field theory by the fact that $N$ Wilson lines can end on a local operator (often called a ``baryon vertex") \cite{Witten:1998xy}\footnote{See also \cite{Bergman:2025isp} for a recent discussion on these baryonic operators from a SymTFT perspective.}. While the statements \eqref{eq:wdictionary} and \eqref{eq:tdictionary} have long been appreciated in holography (see also \cite{Maldacena:1998im}), it was relatively recently shown that the topological operators generating the electric and magnetic 1-form symmetries can also be realized from branes \cite{Apruzzi:2022rei, GarciaEtxebarria:2022vzq}:
\begin{align}
     V(\Sigma_2) \quad \Leftrightarrow \quad \textnormal{$D1$-string on $\{ 0\leq z_0\leq \epsilon \}\times\Sigma_2$} \label{eq:vdictionary} \\
       U(\Sigma_2) \quad \Leftrightarrow \quad \textnormal{$F1$-string on $\{ 0\leq z_0\leq \epsilon \}\times\Sigma_2$} \label{eq:udictionary}
\end{align}
These objects become topological in the limit $\epsilon$ in the sense their fluctuations are red-shifted away from affecting bulk observables. These statements can of course be generalized from a flat space CFT worldvolume to a general 4D manifold where a coordinate $z$ can be defined in the negatively curved bulk near the boundary.

These statements all have parallels in the dictionary between SymTFT and geometric/brane engineering of field theories \cite{Heckman:2022muc, Heckman:2022xgu}. Relevant to the case at hand is the system of $N$ parallel $D3$ branes, where the radial coordinate of the transverse $\mathbb{R}^6$ is related as $r\sim 1/z$ to the Poincar\'e patch coordinate above. Wilson/'t Hooft lines are engineered from $F1$/$D1$ branes stretching from the $D3$-brane stack out to $r=\infty$, while topological operators are engineered from $F1$/$D1$-branes with worldvolume support at $r=\infty$. Note that an overall $U(1)$ center-of-mass gauge theory factor may or may not be present depending on what boundary conditions we take along the angular $S^5$ at $r=\infty$. These boundary conditions are more general than the topological ones considered earlier in this section, and the corresponding holographic version of the boundary conditions are spelled out in Appendix A of \cite{Maldacena:2001ss}.

Finally, we end with a disclaimer that we are, strictly speaking, not studying the ``full SymTFT" of 4D $\mathcal{N}=4$ SYM but only restricting to the 1-form symmetry operators and their charged objects. As mentioned in \cite{Apruzzi:2022rei, GarciaEtxebarria:2022vzq, Heckman:2022muc}, the Wess-Zumino terms on branes can enhance the algebraic structure of these topological operators built from branes to be non-invertible. These technicalities will not play much role in this work since our main focus on invertible symmetries will factor through, but we leave a more thorough treatment for future work.


\subsection{SymTFT Entanglement Between Two Copies of $\mathcal{T}$}\label{ssec:symtftentanglementtwocopies}

We turn now to studying the behavior of two copies of the SymTFT sandwich constructions considered in the previous subsection. Note that if we take two copies of the same physical boundary QFT $\mathcal{T}$ in \eqref{eq:Texample} which we denote $\mathcal{T}_1$ and $\mathcal{T}_2$, then of course a priori we could choose two separate topological boundary conditions, $\mathcal{B}_1$ and $\mathcal{B}_2$, to obtain two absolute QFTs $\mathcal{T}^{(\mathcal{B}_1)}_{1}$ and $\mathcal{T}^{(\mathcal{B}_2)}_{2}$. After compactifying the intervals of both sandwiches, the system is equivalent to two decoupled copies of $\mathcal{T}$ on disconnected 4-manifolds\footnote{Notice that it is equivalent to present the SymTFT picture for this composite system as the tensor product of
SymTFTs $\mathcal{S}_{\mathcal{T}}\otimes \mathcal{S}_{\mathcal{T}}$ placed on $M_{D}\times I$ or, as the SymTFT $\mathcal{S}_\mathcal{T}$ placed on the disconnected manifold $(M_{D}\times I) \coprod (M_{D}\times I)$. This equivalence follows from the Atiyah-Segal axioms where TFTs can be regarded as functors between symmetric monoidal categories, and the fact that one can take tensor products of functors. We opt for the latter of these presentations throughout our work.}. By the axioms of QFT, this is equivalent to a tensor product CFT $\mathcal{T}^{(\mathcal{B}_1)}_{1}\otimes \mathcal{T}^{(\mathcal{B}_2)}_{2}$ where by definition, the Hilbert space assigned to some spatial slice $M_{D-1}$ takes the factorized form
\begin{equation}
    \mathcal{H}_{\mathcal{T}_1\otimes \mathcal{T}_2}(M_{D-1}):=\mathcal{H}_{\mathcal{T}_1}(M_{D-1})\otimes \mathcal{H}_{\mathcal{T}_2}(M_{D-1}).
\end{equation}
Similarly, partition functions also factorize $Z_{\mathcal{T}_1\otimes \mathcal{T}_2}(M_D)=Z_{\mathcal{T}_1}(M_D) \cdot Z_{\mathcal{T}_2}(M_D)$, and data such as the algebra of genuine line operators also factorizes: $\mathcal{V}^{(\mathrm{gen. lines})}_{\ket{\mathcal{B}_1\mathcal{B}_2}}=\mathcal{V}^{(\mathrm{gen. lines})}_{\ket{\mathcal{B}_1}}\otimes \mathcal{V}^{(\mathrm{gen. lines})}_{\ket{\mathcal{B}_2}}$. For example, let us take $\ket{\mathcal{B}_1}=\ket{0}$ and $\ket{\mathcal{B}_2}=\frac{1}{\sqrt{N}}\sum^{N-1}_{k=0}\ket{k}$, then our CFT is simply $\mathcal{N}=4$ SYM with gauge group $SU(N)\times PSU(N)$.

The key point of this section is that there is no reason to restrict topological boundary states to be of the factorized form $\ket{\mathcal{B}_1}\ket{\mathcal{B}_2}$ for this two-sandwich SymTFT system. Indeed, the Hilbert space the 5D SymTFT $\mathcal{S}_{5D}$ (with action \eqref{eq:sunsymtftaction}) assigns to the disconnected topological boundary $M_{D}\coprod M_{D}$ a tensor product Hilbert space
\begin{equation}
    \mathcal{H}_{\mathcal{S}}\left(M_{D}\coprod M_{D}\right)=\mathcal{H}_{\mathcal{S}}(M_{D})\otimes \mathcal{H}_{\mathcal{S}}(M_D)
\end{equation}
and generic states of this are entangled. To obtain a topological boundary, one is free to consider entangled states which are representations of the Heisenberg algebra
\begin{equation}
    V_IU_J=\exp\left(\frac{2\pi i}{N} \delta_{IJ}\right) U_J V_I, \;\;  I,J=1,2.
\end{equation}

\begin{figure}[h]
    \centering
\includegraphics[width=12cm, trim = {0cm 5cm 0cm 4cm}]{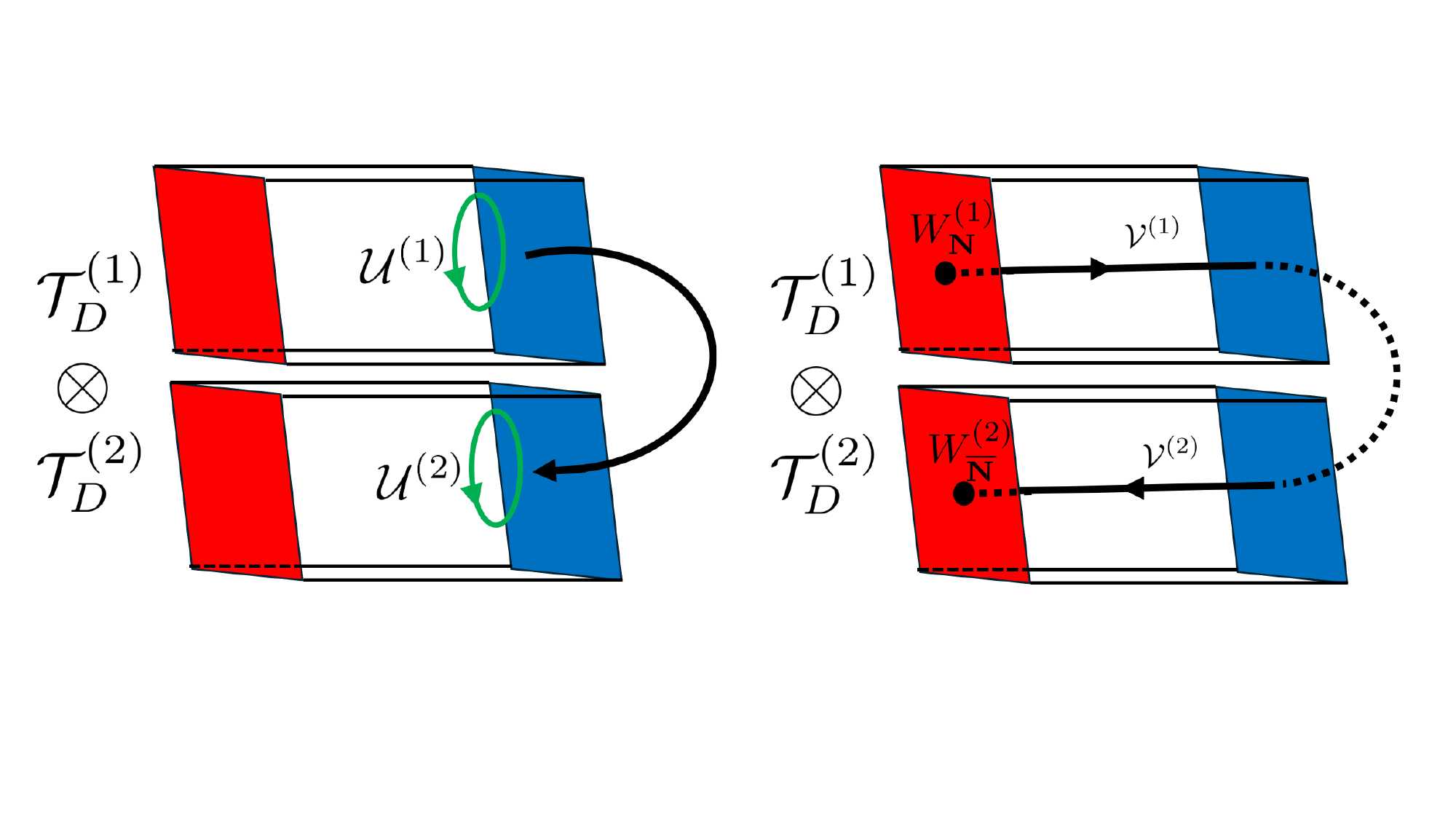}
    \label{fig:portal}
    \caption{Teleportation of operators due to the entangled boundary conditions relevant for the gauge group structure $SU(N)^2/\mathbb{Z}_N(1,-1)$.}
\end{figure}

Perhaps the simplest example of such an entangled boundary state is
\begin{equation}\label{eq:entangledsimpexample}
    \ket{\mathcal{B}}=\frac{1}{\sqrt{N}}\sum^{N-1}_{k=0}\ket{k}\ket{-k}.
\end{equation}
Upon reducing along the interval direction of the SymTFT, this will cause the Euclidean partition functions on general 4-manifolds to no longer factorize:
\begin{equation}
    Z=\big(\bra{\mathcal{T}_1}\bra{\mathcal{T}_2}\big)\ket{\mathcal{B}}=\frac{1}{\sqrt{N}}\sum^{N-1}_{k=0}\left( Z_{SU(N)}(k)\right)^2.
\end{equation}
To better understand this, notice that the state \eqref{eq:entangledsimpexample} satisfies the following relations
\begin{equation}\label{eq:teleportation}
    U_1\ket{\mathcal{B}}=U_2\ket{\mathcal{B}}, \quad \quad V_1\ket{\mathcal{B}}=V^{-1}_2\ket{\mathcal{B}}.
\end{equation}
In the SymTFT setup, this means that we are free to ``teleport" symmetry operators $U_1(\Sigma_2)$ and $V_1(\Sigma_2)$ from the first interval to the second by moving them to the topological boundary, see Figure \ref{fig:portal}. At the level of symmetry operators, this indicates that there is a correspondence between the electric and magnetic 1-form symmetries between the two intervals, and from the definition of $U$ and $V$ operators is equivalent to enforcing the boundary conditions
\begin{align}
  \mathrm{(Dirichlet)}:&  \; \;  c^{(1)}_2|_{y=1}-c^{(2)}_2|_{y=1}=0, \quad b^{(1)}_2|_{y=1}+b^{(2)}_2|_{y=1}=0\label{eq:entbcs1}\\
    \mathrm{(Neumann)}:& \; \;  d_yc^{(1)}_2|_{y=1}+d_yc^{(2)}_2|_{y=1}=0, \quad d_yb^{(1)}_2|_{y=1}-d_yb^{(2)}_2|_{y=1}=0\label{eq:entbcs2}
\end{align}
for the 5D gauge fields of the SymTFT.

At the level of line operators, notice first that if we rewrite \eqref{eq:teleportation} as
\begin{equation}\label{eq:teleport2}
    U^{-1}_2U_1 \ket{\mathcal{B}}= V_2V_1 \ket{\mathcal{B}}= \ket{\mathcal{B}},
\end{equation}
then the fact that $\ket{\mathcal{B}}$ is an eigenvector of $U^{-1}_2U_1$ and $V_2V_1$ means that these operators can terminate on the topological boundary as implied by the orientation of the arrows in Figure \ref{fig:portal}. Such a configuration however again appears as a form of teleportation: an operator $U_1$ with support along the first interval can thread through the topological boundary to become $U_2$ (oriented in the opposite direction). In fact, because $\ket{\mathcal{B}}$ is \textit{not} an eigenvector under $U_{I}$ nor $V_{I}$, this means that one \textit{cannot} end $U_1$ on the topological boundary, without also including $U_2$ ending on its topological boundary. This allows us to conclude that the set of genuine line operators for this system does not factorize, explicitly realizing Statement 1 in the Introduction. In particular, let $(\mathbf{R}_2,\mathbf{R}_2)$ denote a representation of $\mathfrak{su}(N)\times \mathfrak{su}(N)$, then the above S-entanglement implies, for example,
\begin{align}
   &  W_{(\mathbf{N},\overline{\mathbf{N}})}, \; \mathrm{and} \; H_{(\mathbf{N},\mathbf{N})} \; \in \mathcal{V}^{(\mathrm{gen. lines})}_{\ket{\mathcal{B}}} \\
   W_{(\mathbf{N},\mathbf{1})}, &\;  W_{(\mathbf{1}, \overline{\mathbf{N}})}, \;
    H_{(\mathbf{N},\mathbf{1})}, \; \mathrm{and}\; H_{(\mathbf{1}, \overline{\mathbf{N}})}\;  \notin \mathcal{V}^{(\mathrm{gen. lines})}_{\ket{\mathcal{B}}}.
\end{align}
Such a list of genuine/non-genuine line operators are identical to that of SYM theory with gauge group
\begin{equation}\label{eq:quotientgaugegroup}
    G=\frac{SU(N)\times SU(N)}{\mathbb{Z}_N(1,-1)}
\end{equation}
where the quotient identifies $(\zeta_N \mathbf{1}_{N\times N}, \mathbf{1}_{N\times N})\sim (\mathbf{1}_{N\times N}, \zeta_N\mathbf{1}_{N\times N})$ in matrix notation. One can see that this is indeed the case from the fact that if one starts with $SU(N)^2$ SYM, then gauge the subgroup of the electric 1-form global symmetry $\mathbb{Z}^{(1)}_N\times \mathbb{Z}^{(1)}_N$ generated by $(1,-1) \; \mathrm{mod} \; N$, this yields the gauge group \eqref{eq:quotientgaugegroup}. This is similar what we saw in the previous subsection where the $PSU(N)$ SYM theory can be obtained from $SU(N)$ by gauging the electric 1-form symmetry. The theory with quotient gauge group \eqref{eq:quotientgaugegroup} will have a $\mathbb{Z}^{(1)}_{N,m}$ magnetic symmetry due to the fact that $\pi_1(G)=\mathbb{Z}_{N}$, and there will additionally still be a $\mathbb{Z}^{(1)}_{N,e}$ electric symmetry since $Z(G)=\mathbb{Z}_N$.

One can also consider turning on background fields for the $\mathbb{Z}^{(1)}_{N,e}\times \mathbb{Z}^{(1)}_{N,m}$ global symmetries which corresponds to modifying \eqref{eq:teleport2} to
\begin{equation}\label{eq:teleport3}
    U^{-1}_2U_1\ket{\mathcal{B}}=e^{2\pi i \ell/N}\ket{\mathcal{B}}, \quad  V_2V_1\ket{\mathcal{B}}=e^{2\pi i \ell'/N}\ket{\mathcal{B}}
\end{equation}
which is satisfied by the states
\begin{equation}\label{eq:backgrounds1minus1}
     \ket{\mathcal{B}}=\frac{1}{\sqrt{N}}\sum_{k=0}e^{-2\pi i k\ell/N}\ket{k+\ell'}\ket{-k}.
\end{equation}
The eigenvalues in \eqref{eq:teleport3} correspond to the Dirichlet values for the background fields, i.e.
\begin{equation}
    \int_{\gamma_2} c^{(1)}_2|_{y=1}-c^{(2)}_2|_{y=1}=\ell, \quad \int_{\gamma_2}b^{(1)}_2|_{y=1}+b^{(2)}_2|_{y=1}=\ell'
\end{equation}
for all 2-cycles $\gamma_2$ in $M_4$.

Alternatively, we can consider the gauge group
\begin{equation}
    G=\frac{SU(N)\times SU(N)}{\mathbb{Z}_N(1,1)}
\end{equation}
where the $\mathbb{Z}_N$ quotient identifies $(\zeta_N \mathbf{1}_{N\times N}, \mathbf{1}_{N\times N})\sim (\mathbf{1}_{N\times N}, \zeta^{-1}_N\mathbf{1}_{N\times N})$. Some examples of genuine/non-genuine line operators in this case are
\begin{align}
   &  W_{(\mathbf{N},\mathbf{N})}, \; \mathrm{and} \; H_{(\mathbf{N},\overline{\mathbf{N}})} \; \in \mathcal{V}^{(\mathrm{gen. lines})}_{\ket{\mathcal{B}}} \\
   W_{(\mathbf{N},\mathbf{1})}, &\;  W_{(\mathbf{1}, \overline{\mathbf{N}})}, \;
    H_{(\mathbf{N},\mathbf{1})}, \; \mathrm{and}\; H_{(\mathbf{1}, \overline{\mathbf{N}})}\;  \notin \mathcal{V}^{(\mathrm{gen. lines})}_{\ket{\mathcal{B}}}.
\end{align}
Once again there is $\mathbb{Z}^{[(1)}_{N,e}\times \mathbb{Z}^{(1)}_{N,m}$ global symmetry group, and the analog of the conditions \eqref{eq:teleport3} are
\begin{equation}\label{eq:teleport4}
    U_2U_1\ket{\mathcal{B}}=e^{2\pi i \ell/N}\ket{\mathcal{B}}, \quad  V^{-1}_2V_1\ket{\mathcal{B}}=e^{2\pi i \ell'/N}\ket{\mathcal{B}}
\end{equation}
and the relevant boundary states are
\begin{equation}
     \frac{1}{\sqrt{N}}\sum_{k=0}e^{-2\pi i k\ell/N}\ket{k+\ell'}\ket{k}.
\end{equation}

\paragraph{Interface Picture}
As mentioned in the introduction, see Figure \ref{fig:twoversions}, one can equivalently present two disconnected SymTFT sandwiches (with or without S-entanglement) as a single interval configuration with physical boundary conditions on both sides and a topological interface in the middle. Indeed, even staring at Figure \ref{fig:portal} suggests that the teleportation of the symmetry operators/defect operators could be interpreted as a simple translation along the $y$-direction after gluing the two intervals.

Such a claim follows from a simple application of the Atiyah-Segal axioms to the BF-like theory \eqref{eq:sunsymtftaction}. If we quantize the TFT in the $y$-direction, then the vector space of all possible topological interfaces localized at some $y=y_0$ is equivalent to the vector space that the TFT assigns to $M_4\coprod \overline{M}_4$. This means that an interface $\mathcal{I}_{\mathcal{B}}$ acts as a matrix
\begin{equation}
    \mathcal{I}_{\mathcal{B}} \; \leftrightarrow \; \sum_{ij}c_{ij}\ket{i}\bra{j}
\end{equation}
when translating states along the $y$-direction. We have labeled by a subscript $\mathcal{B}$ because one can obtain this matrix from a state $\ket{\mathcal{B}}=\sum_{ij}c_{ij}\ket{i}\ket{j}$ flipping the second ``ket" for a "bra". Take for instance the $SU(N)^2/\mathbb{Z}_N(1,-1)$ case, then the corresponding interface is the identity (up to an overall normalization):
\begin{equation}
    \ket{\mathcal{B}}=\frac{1}{\sqrt{N}}\sum^{N-1}_{k=0}\ket{k}\ket{k}\; \rightarrow \; \mathcal{I}_{B}=\frac{1}{\sqrt{N}}\mathbf{1}_{N\times N}.
\end{equation}
To complete the dictionary between the two-sandwich picture and the interface picture, we must apply the identity $\langle \mathcal{T}| k\rangle=\langle k|\mathcal{T}\rangle^*=\langle k|\mathcal{T}^{\dagger}\rangle$ to match the partition functions/observables to match. Here $\ket{\mathcal{T}^{\dagger}}$ is the boundary state for the orientation-reversal of the theory $\mathcal{T}$ which will differ to two ways from $\mathcal{T}$. The first is that the complexified gauge coupling $\tau=4\pi i/g_{SYM}^2+\theta/2\pi$ will transform as $\tau\mapsto -\tau^*$, while the second is that the 't Hooft line 1-form charges will flip sign. The latter is due to the fact that because magnetic fields are pseudo-vectors so magnetic charges are pseudo-scalars. Indeed this allows us to make sense of the fact that in Figure \ref{fig:portal}, that orientation-reversal properties of the $U$ and $V$ operators are different\footnote{This relation should be viewed at the level of \textit{forms} where for instance in terms of components where, if $\mu$ and $\nu$ are denote directions along $M_4$, then $(b_{2})_{\mu\nu}\rightarrow -(b_2)_{\mu\nu}$ while $(b_{2})_{y\nu}\rightarrow +(b_2)_{y\nu}$.}:
\begin{equation}
    \textnormal{$y\rightarrow -y$   $\quad$ $\implies$ $\quad$  $b_2\rightarrow -b_2$, \; $c_2\rightarrow +c_2$}.
\end{equation}
This is a symmetry of the action \eqref{eq:sunsymtftaction} and upon acting with S-duality permutes to a transformation under which $b_2$ is even and $c_2$ is odd. This is the perhaps familiar fact that $P$ and $CP$ are exchanged under S-duality in $\mathcal{N}=4$ SYM.

If we consider the more general boundary states with 1-form background fields turned on \eqref{eq:backgrounds1minus1}, then the corresponding interface will act on the $U$ and $V$ operators crossing it by phases. Meanwhile, the interface for the $SU(N)^2/\mathbb{Z}_N(1,1)$ case will flip the charges of the $U$ and $V$ operators crossing it, so will have the opposite behavior in Figure \ref{fig:portal}. Finally, for the case $G=SU(N)^2$ which has no S-entanglement, the interface allows $U$ operators to end on it, from either the left or the right independently which means that no operators are allowed to pass through.

\paragraph{Bipartite SymTFT Entanglement Generalities}
While we have only considered S-entanglement between identical theories in this section, we emphasize that such entanglement may always exist so long as two theories $\mathcal{T}_1$ and $\mathcal{T}_2$ share a common symmetry subcategory. Let $\mathcal{S}$ be the SymTFT corresponding to this subcategory, then by assumption, there exist two physical boundary conditions $\mathcal{T}_1$ and $\mathcal{T}_2$. For a general $\mathcal{S}$, classifying the set of topological interfaces may be highly non-trivial in contrast to the 5D BF theory above. However, because the identity interface always exists, one can always consider a maximal (sub)symmetry entanglement between $\mathcal{T}_1$ and $\mathcal{T}_2$. We also touch on such scenarios in Section \ref{ssec:CERNVT}. As a simple example, consider 4D $\mathcal{N}=4$ SYM with gauge group $(SU(2N)\times SU(6))/\mathbb{Z}_2$. This entangles the $\mathbb{Z}_2$ subgroup of the $\mathbb{Z}_{2N}$ 1-form symmetry of the $\mathfrak{su}(2N)$ theory with the $\mathbb{Z}_2$ subgroup of the $\mathbb{Z}_{6}$ 1-form symmetry of the $\mathfrak{su}(6)$ theory. In the SymTFT picture, the operators $U^N$ and $V^N$  from the left side are allowed to pass through the interface and become $U^3$ and $V^3$ respectively.

\subsection{Holographic Interpretation}\label{ssec:HolInt}
\begin{figure}[h]
    \centering
\includegraphics[width=12cm, trim = {0cm 1cm 0cm 1cm}]{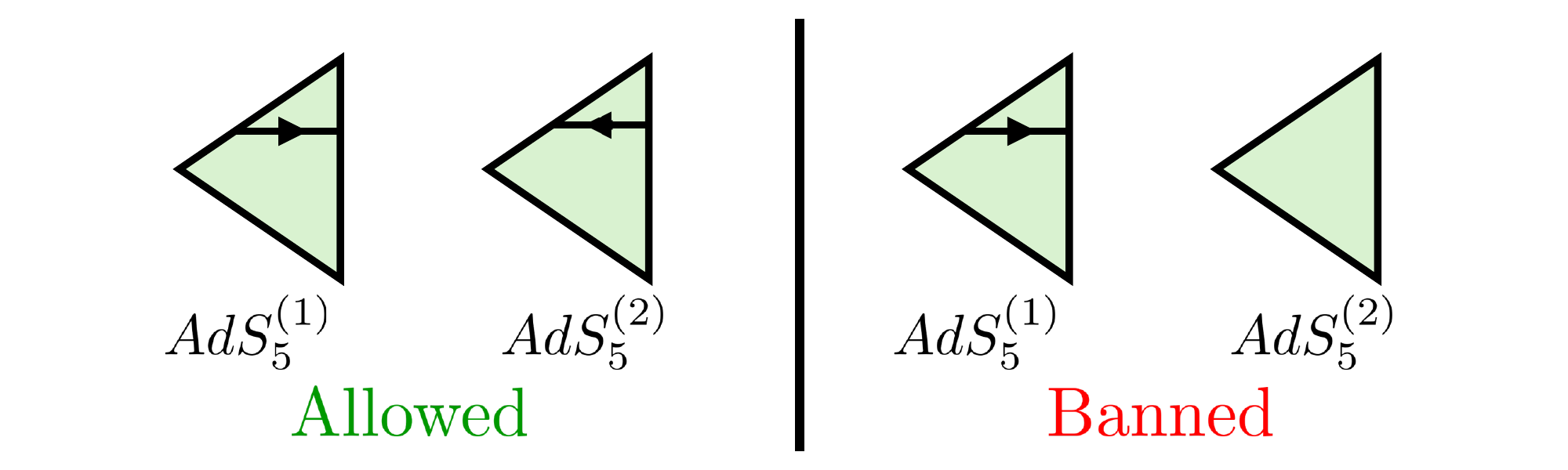}
    \label{fig:allowedbanned}
    \caption{Examples of allowed and banned $F1$/$D1$ string configurations for the bulk dual of 4D SYM theory with gauge group $SU(N)^2/\mathbb{Z}_N(1,-1])$.}
\end{figure}

We now address the holographic interpretation of S-entanglement which, given the SymTFT / holographic dictionary given in Section \ref{ssec:reviewsymhol}, is fairly straightforward. Consider a configuration consisting of two disconnected copies of the $\mathrm{AdS}_5\times S^5$ spacetimes depicted in Figure \ref{fig:adssliver}. As we learned from Section \ref{ssec:reviewsymhol}, the SymTFT fields $b_2$ and $c_2$ can be respectively identified with the reductions of the IIB 2-form potentials $B_2$ and $C_2$ on the $S^5$ (with all legs along the AdS$_5$ directions). S-entanglement between the two conformal boundaries just means that the boundary conditions for these 2-form potentials are correlated in some way. For instance from \eqref{eq:entbcs1} and \eqref{eq:entbcs2}, we see that the boundary conditions relevant for the $G=SU(N)^2/\mathbb{Z}_N(1,-1)$ case are
\begin{align}\label{eq:holentbcs1}
  \mathrm{(Dirichlet)}:&  \; \;  C^{(1)}_2|_{z=0}-C^{(2)}_2|_{z=0}=0, \quad B^{(1)}_2|_{z=0}+B^{(2)}_2|_{z=0}=0 \\
   \mathrm{(Neumann)}:& \; \;  d_zC^{(1)}_2|_{z=0}+d_zC^{(2)}_2|_{z=0}=0, \quad d_zB^{(1)}_2|_{z=0}-d_zB^{(2)}_2|_{z=0}=0 \label{eq:holentbcs2}
\end{align}
where $d_z:=\frac{\partial}{\partial z}dz$ is the exterior derivative along the AdS$_5$ radial direction. Combining the holographic dictionary of Section \ref{ssec:reviewsymhol} with the SymTFT conclusions of Section \ref{ssec:symtftentanglementtwocopies} implies that $F1$ and $D1$ strings can teleport from one universe to the other by connecting through their conformal boundaries. Moreover, as we saw from the previous section, it is not possible to place a $U^{(1)}(L\times [0,1])$ or $V^{(1)}(L\times [0,1])$ operator along the first SymTFT interval without including a corresponding $U^{(2)}(L\times [0,1])$ or $V^{(2)}(L\times [0,1])$ in the second interval. In bulk language, this leads us to conclude that allowed configurations of $F1$/$D1$-strings are such that an $F1$/$D1$ string can only end on the conformal boundary of AdS$^{(1)}_5$, if there is a corresponding $F1$/$D1$ string ending on the on the conformal boundary of AdS$^{(2)}_5$, see Figure \ref{fig:allowedbanned}. All other configurations are inconsistent with the boundary conditions \eqref{eq:holentbcs1}.

Another bulk interpretation of such a phenomena is that there exists some sort of wormhole connecting the disconnected AdS spacetimes at their conformal boundaries. This wormhole is topological in the sense that the precise positions of the branes ending on the boundary are not dynamically correlated. In dual CFT language, this is equivalent to the fact that we are free to make a deformation
\begin{equation}
    W_{(\mathbf{N},\overline{\mathbf{N}})}(L)\rightarrow W_{(\mathbf{N},\mathbf{1})}(L)\cdot V(\Sigma)\cdot W_{(\mathbf{1},\overline{\mathbf{N}})}(L')
\end{equation}
where $\partial \Sigma=L\coprod \overline{L'}$. This topological nature is ultimately due to the fact the stress-tensors of the two CFT copies commute with each other.




\paragraph{Comment on $D3$ Brane Picture}
From the perspective of the above AdS$_5$ background before considering the backreaction and near-horizon limit, we simply have a string background consisting of two disconnected 10D Minkowski spacetimes, each containing a stack of $N$ $D3$-branes. String theory on disconnected spacetimes is hardly exotic as the perturbative worldsheet CFT picture just consists of a sigma model with a disconnected target space. The 2D CFT Hilbert space just factorizes as a connected sum. Altogether, this just means that the bulk fields along the asymptotic boundaries $(S^5)^{(i=1,2)}_\perp$ surrounding the $D3$-stacks are coupled as in \eqref{eq:holentbcs1}, but with $z=0$ replaced with $r=\infty$ the $\mathbb{R}^6_{\perp}$ radial coordinate.




\section{Other Examples of SymTFT Entanglement}\label{sec:Otherexamples}
In this section we extend the construction of S-entanglement to holographic setups in dimensions other than four. Our aim here is not to analyze these cases with the same level of detail as in the example of 4D $\mathcal{N}=4$ SYM, but rather to illustrate the generality of the framework. A more systematic treatment of specific models will be deferred to future work.

\subsection{D=6: Absolute SCFTs via SymTFT Entanglement}

Six-dimensional SCFTs are intrinsically relative QFTs, in the sense that their associated 7D (Sym)TFTs\footnote{Strictly speaking, ``SymTFT'' is an abuse of terminology here, since the absence of topological boundary conditions prevents the separation of symmetry operators from charged defects in the TFT.} generally do not admit topological boundary conditions. As a result, the corresponding Heisenberg algebra of topological operators admits no polarization, and hence the 6D theory lacks a well-defined global structure.

Nevertheless, it has been shown in \cite{Gukov:2020btk, Lawrie:2023tdz} that direct sums of such relative 6D SCFTs can sometimes admit polarizations, thereby yielding absolute theories with well-defined global structure. We reinterpret this phenomenon as an instance of S-entanglement.

Consider the 6D $\mathcal{N} = (2,0)$ theory of type $A_4$, i.e., with Lie algebra $\mathfrak{g} = \mathfrak{su}(5)$. Its associated 7D TFT is a Chern–Simons theory at level $k = 5$:
\begin{equation}
    \frac{5}{4\pi}\int_{M_6\times I} c_3 \wedge dc_3.
\end{equation}
The topological line operators
\begin{equation}
    U_n = \exp \left(in \int_{\Sigma_3} c_3 \right), \quad n \in \mathbb{Z}_5,
\end{equation}
generate a Heisenberg algebra
\begin{equation}
    U_m U_n = U_n U_m \exp\left( \frac{2\pi i m n}{5} \right),
\end{equation}
whose defect group $\mathbb{D}^{(2)} = \mathbb{Z}_5$ does not admit a Lagrangian subgroup. As a result, the theory is intrinsically relative \cite{Franco:2024mxa}, and no polarization exists for constructing a Hilbert space basis on a general 6-manifold $M_6$.

However, consider the decoupled pair of theories $A_4 \oplus A_4$. The associated bulk 7D theory consists of two Chern–Simons gauge fields $c_3, c_3'$, and the corresponding topological operators take the form
\begin{equation}
    U_{n,n'} = \exp \left( i \int_{\Sigma_3} n c_3 + n' c_3' \right), \quad n,n' \in \mathbb{Z}_5.
\end{equation}
The resulting Heisenberg algebra
\begin{equation}
    U_{m,m'} U_{n,n'} = U_{n,n'} U_{m,m'} \exp\left( \frac{2\pi i (mn + m'n')}{5} \right)
\end{equation}
now has defect group $\mathbb{Z}_5 \oplus \mathbb{Z}_5$, which does admit Lagrangian subgroups. For example, the subgroup generated by $(1,2) \in \{(a,b)|a,b=0,1,2,3,4 \} \cong \mathbb{Z}_5 \oplus \mathbb{Z}_5$ gives rise to a set of mutually commuting operators:
\begin{equation} \label{eq:lagrangian-algebra}
    \{ V_0, V_1, V_2, V_3, V_4 \} = \{1, U_{1,2}, U_{2,4}, U_{3,1}, U_{4,3} \}.
\end{equation}
These operators define a polarization of the tensor product Hilbert space $\mathcal{H} \otimes \mathcal{H}$, which becomes a five-dimensional qudit space with basis vectors $|a\rangle$ labeled by
\begin{equation}
    V_n |a\rangle = \exp\left( \frac{2\pi i n a}{5} \right) |a\rangle, \quad a \in \mathbb{Z}_5.
\end{equation}

The resulting absolute theory, defined by this polarization, is denoted by $((SU(5) \times SU(5))/\mathbb{Z}_5)_+$. Another Lagrangian subgroup, generated by $(2,1)$, yields an orthogonal basis $\{|b\rangle\}$ corresponding to the global form $((SU(5) \times SU(5))/\mathbb{Z}_5)_-$. We summarize:
\begin{equation}
\begin{aligned}
    |a\rangle &\longleftrightarrow ((SU(5)\times SU(5))/\mathbb{Z}_5)_+, \\
    |b\rangle &\longleftrightarrow ((SU(5)\times SU(5))/\mathbb{Z}_5)_-.
\end{aligned}
\end{equation}

\begin{figure}[h]
    \centering
    \includegraphics[width=14cm]{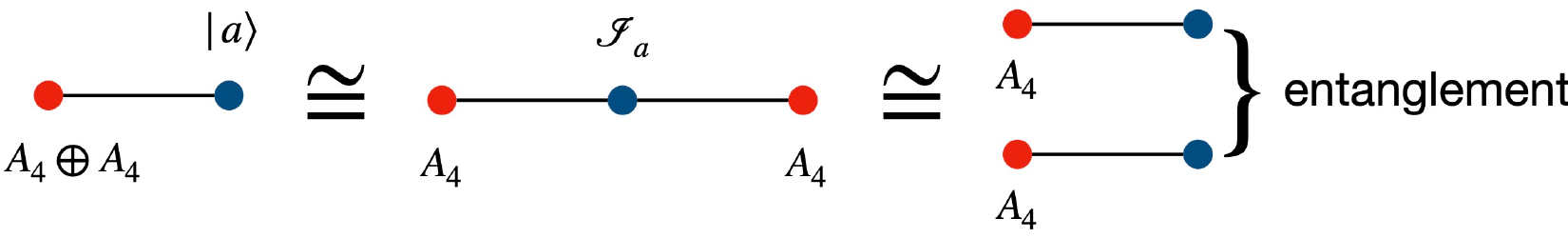}
    \caption{Schematic depiction of S-entanglement for two copies of the 6D $A_4$ SCFT. Each copy corresponds to a 7D Chern–Simons theory with a relative boundary condition on the bottom edge (black) and a symmetry boundary condition on the top edge (colored). The interface (middle) glues the two systems via a Lagrangian algebra in the combined Heisenberg algebra, yielding a topological boundary condition for the folded theory. This construction produces a symmetry-entangled state in $\mathcal{H} \otimes \mathcal{H}$, which corresponds to an absolute theory with global form $((SU(5) \times SU(5))/\mathbb{Z}_5)_{+}$.}
    \label{fig:6d_symtft_entanglement}
\end{figure}

These states $|a\rangle$ and $|b\rangle$ are not tensor product of states in $\mathcal{H}$. Rather, they are entangled states, and it is this entanglement that renders the combined system absolute. Figure~\ref{fig:6d_symtft_entanglement} gives a schematic representation of this construction.
In this sense, S-entanglement provides a concrete mechanism for realizing global structure in intrinsically relative 6D theories.

See also \cite{Heckman:2025lmw} for a recent related interpretation via Wigner wavefunctions.

\subsection{D=3: ABJM theories and Mixed-Form Symmetry Entanglement}

In contrast to the examples in 4D and 6D, where we only consider intermediate defect groups \cite{Lawrie:2023tdz}, the three-dimensional ABJM theory \cite{Aharony:2008ug} exhibits a particularly rich symmetry structure: its associated SymTFT involves multiple background fields coupling to both 0-form and 1-form global symmetries. This enables novel forms of S-entanglement that intertwine different types of symmetry sectors.

\begin{figure}[h]
    \centering
    \includegraphics[width=0.45\textwidth]{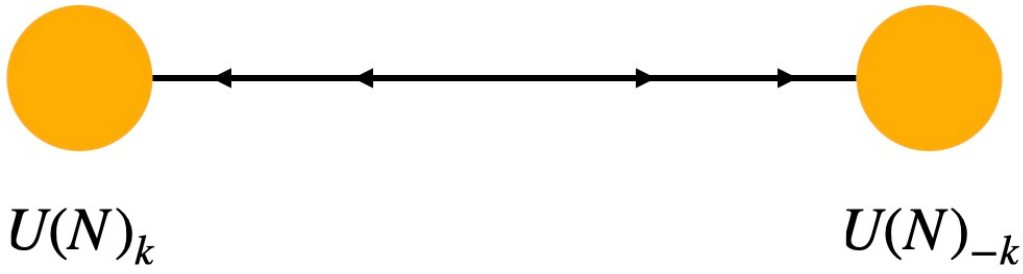}
    \caption{The ABJM theory is a Chern–Simons–matter theory with gauge group $U(N)_k \times U(N)_{-k}$ and bifundamental matter fields. It describes $N$ M2-branes probing a $\mathbb{C}^4/\mathbb{Z}_k$ singularity.}
    \label{fig:abjm-quiver}
\end{figure}

To set the stage, we first review the topological boundary conditions in the ABJM SymTFT and the corresponding global structures of the theory \cite{Bergman:2020ifi}. These boundary conditions select distinct polarizations of the bulk operator algebra, determining which global symmetries are genuine, which defects can end, and which global form of the gauge group is realized. In the second part of this section, we will construct symmetry-entangled states by gluing two such boundary conditions diagonally, and interpret the resulting global structures as entangled state of the SymTFT.

\paragraph{A Review of SymTFTs for ABJM theories.
}
We begin by a lighting review of the structure of the SymTFT associated with the ABJM theory and refer the reader to \cite{Bergman:2020ifi, vanBeest:2022fss} for more details. This will allow us to precisely define the meaning of boundary conditions, and clarify which global symmetries are genuine in each case.

The ABJM theory with gauge group $U(N)_k \times U(N)_{-k}$ admits three interrelated global symmetries:
\begin{itemize}
    \item A magnetic $U(1)_m^{(0)}$ symmetry, under which monopole operators are charged;
    \item A baryonic $U(1)_B^{(0)}$ symmetry, under which baryonic operators built by bi-fundamental matter are charged;
    \item A discrete 1-form symmetry, either $\mathbb{Z}_k^{(1)}$ or $\mathbb{Z}_N^{(1)}$, depending on the global form.
\end{itemize}
However, not all three symmetries can be simultaneously genuine: they are related by the SymTFT action \cite{Bergman:2020ifi}:
\begin{equation}\label{eq: symtft of ABJM}
    \frac{1}{2\pi} \int_{M_4} B_2 \wedge d(k C_1 + N A_1),
\end{equation}
where $C_1$, $A_1$, and $B_2$ are background gauge fields for $U(1)_B$, $U(1)_m$, and the 1-form symmetry, respectively.

To define a well-posed 3D QFT, we must impose topological boundary conditions (on asymptotic boundary in string theory setup) on these background fields. For simplicity, below we describe three canonical boundary conditions, corresponding to three standard global structures.

\begin{itemize}
    \item \textbf{Dirichlet for $A_1$ and $C_1$, Neumann for $B_2$.} This boundary condition allows $D0$-branes (electrically charged under $A_1$) and $D4$-branes (charged under $C_1$) to end on the boundary, but forbids $F1$ strings from ending. As a result, we have a
$$
U(1)^{(0)}\times \mathbb{Z}^{(0)}_{\text{gcd}\{N, k\}}
$$
global symmetry. This corresponds to the global form
\begin{equation}
    \frac{U(N)_k \times U(N)_{-k}}{\mathbb{Z}_k} = \frac{SU(N)_k \times SU(N)_{-k}}{\mathbb{Z}_N},
\end{equation}
where Wilson lines are not genuine operators.
   \item \textbf{Dirichlet for $A_1$ and $B_2$, Neumann for $C_1$.} This boundary condition allows $F1$ strings and $D0$-branes to end, but not $D4$-branes. Consequently we have $$
U(1)_m^{(0)}\times \mathbb{Z}_k^{(0)}
$$
global symmetry.  The associated global form is
\begin{equation}\label{eq: un un ABJM}
    U(N)_k \times U(N)_{-k}.
\end{equation}
   \item \textbf{Dirichlet for $C_1$ and $B_2$, Neumann for $A_1$.} This allows $D4$-branes and $F1$ strings to end, but not $D0$-branes. Thus we have
$$
U(1)^{(0)}_B \times \mathbb{Z}_N^{(1)}
$$
global symmetry. The global form in this case is
\begin{equation}
    SU(N)_k \times SU(N)_{-k}.
\end{equation}
\end{itemize}

In each of these three cases, the boundary condition selects a maximal commuting subalgebra of the SymTFT operator algebra, effectively specifying a polarization. These polarizations correspond to physical choices of global structure for the theory. More exotic, ``intermediate'' boundary conditions may exist (e.g., partial gauging or discrete quotients), but we focus on the above three for clarity and concreteness.

In the following, we will construct entangled states in the doubled SymTFT Hilbert space by gluing together two copies of the theory along these boundary conditions. Each such gluing defines a different symmetry-entangled interface, leading to distinct combinations of gauged symmetries and resulting global forms.

\paragraph{SymTFT Entanglement for ABJM theories.}

Consider two copies of the ABJM theory and their associated SymTFTs. A simple, unentangled state corresponds to the direct product of two absolute ABJM theories. For instance, the global form
\begin{equation}\label{eq: nonentangled ABJM}
    \frac{U(N)_k \times U(N)_{-k}}{\mathbb{Z}_k}\times \frac{U(N)_k\times U(N)_{-k}}{\mathbb{Z}_k}
\end{equation}
can be interpreted as such an unentangled configuration. Alternatively, this state can be viewed as the SymTFT (\ref{eq: symtft of ABJM}) bounded by two copies of the relative ABJM theory, glued along a topological interface that does not mix the field profiles on either side. Explicitly labeling fields on the two sides with superscripts $(1)$ and $(2)$ respectively\footnote{We hope the reader will not confuse these superscripts with those used elsewhere in the paper to denote the degree of generalized symmetry.}, this configuration corresponds to the boundary condition:
\begin{equation}
\text{Dirichlet for } A_1^{(1)}, ~C_1^{(1)}, ~A_1^{(2)}, ~C_1^{(2)}.
\end{equation}

To construct an \emph{entangled} state in the tensor product Hilbert space of the two SymTFTs, we instead consider a topological interface $\mathcal{I}$ that glues the field profiles across the two sides. A natural choice of gluing condition is:
\begin{equation}\label{eq: abjm gluing condition}
A_1^{(1)}|_{\mathcal{I}} = A_1^{(2)}|_{\mathcal{I}}, \quad C_1^{(1)}|_{\mathcal{I}} = C_1^{(2)}|_{\mathcal{I}}.
\end{equation}
This boundary condition leads to a new absolute global form:
\begin{equation}\label{eq: entangled ABJM global form}
    \frac{U(N)_k \times U(N)_{-k} \times U(N)_k \times U(N)_{-k}}{\mathbb{Z}_k}=\frac{SU(N)_k \times SU(N)_{-k} \times SU(N)_k \times SU(N)_{-k}}{\mathbb{Z}_N}
\end{equation}

The genuine global symmetries associated with this absolute theory --- and hence the entangled state --- can be analyzed as follows. We begin with the unentangled state (\ref{eq: nonentangled ABJM}), which enjoys a global symmetry group:
\begin{equation}
(U(1)_m^{(0)} \times \mathbb{Z}_k^{(1)})^2.
\end{equation}
Upon gauging the diagonal $\mathbb{Z}_k^{(1)}$ symmetry, we arrive at the global form (\ref{eq: entangled ABJM global form}). The gauging process produces a quantum symmetry $\mathbb{Z}_k^{(0)}$, while leaving an ungauged residual $\mathbb{Z}_k^{(1)}$ intact.

However, due to the presence of a mixed anomaly between $U(1)_m^{(1)}$ and $\mathbb{Z}_k^{(1)}$ \cite{Bergman:2020ifi} prior to gauging , the emergent $\mathbb{Z}_k^{(0)}$ does not exist as a standalone symmetry. Instead, it becomes part of an extended $0$-form symmetry group, in which $U(1)^{(0)}\times U(1)^{(0)}$ is nontrivially fibered over a $\mathbb{Z}_{\text{gcd}\{N, k \}}$.

The resulting global symmetry group can be schematically written as:
\begin{equation}
U(1)^{(0)} \times U(1)^{(0)} \times \mathbb{Z}_{\text{gcd}\{ N, k \}}^{(0)} \times \mathbb{Z}_k^{(1)}.
\end{equation}

\begin{figure}[h]
\centering
\includegraphics[width=0.75\textwidth]{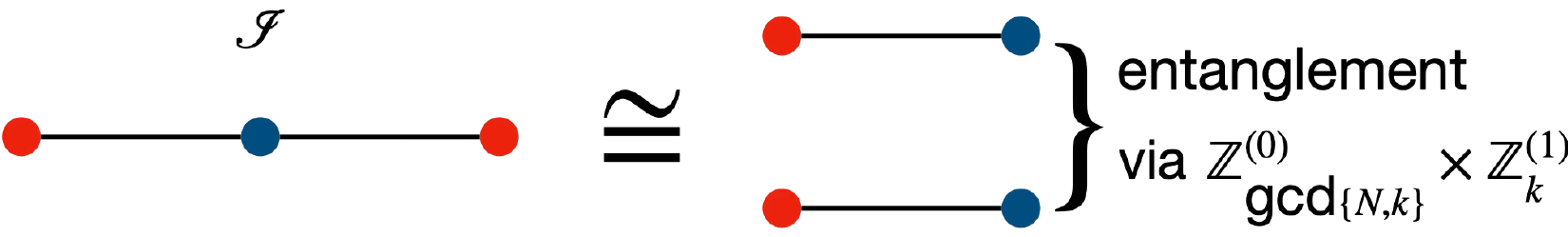}
\caption{Left: Topological interface $\mathcal{I}$ between two relative ABJM boundaries of the SymTFT (\ref{eq: symtft of ABJM}) with the gluing condition (\ref{eq: abjm gluing condition}). Right: Interpretation of the entangled state as a global symmetry quotient.}
\label{fig:abjm-entangled}
\end{figure}

This construction illustrates a mixed-form symmetry entanglement between the two ABJM copies. Specifically
\begin{itemize}
    \item Their unscreened Wilson lines are charged under a common $\mathbb{Z}_k^{(0)}$ symmetry.
    \item There exist $\mathbb{Z}_{\text{gcd}(N, k)}^{(0)}$ symmetry operators that can transit freely through both SymTFT bulks.
\end{itemize}
This structure is schematically illustrated in Figure~\ref{fig:abjm-entangled}.

We close this subsection by noting that similar mixed-form symmetry entanglement structures can be constructed for 7D SYMs and 5D SCFTs (see, e.g., \cite{Apruzzi:2021phx, Baume:2023kkf, Cvetic:2024dzu}), both of which admit M-theory approach to their SymTFTs. In both cases, one may consider gauging the diagonal 1-form symmetry of two copies of electrically polarized theories. The resulting absolute theory then corresponds to an entangled state, correlated by the residual 1-form symmetry as well as the emergent quantum symmetry, which is 2-form for the 5D case and 4-form for the 7D case. Since the SymTFTs in these examples involve only a single BF term (up to possible cubic anomaly terms), the structure is considerably cleaner than in the ABJM case, and we will therefore refrain from further elaboration here.

\subsection{D=2: SymTFT Entanglement for Non-invertible Symmetries}

\label{subsec:noninver symentangle}

The previous examples focused on invertible symmetries—either 0-form or higher-form—that act as group-like global transformations. A more general class of symmetries, however, are non-invertible symmetries. In 2D QFTs, such symmetries are particularly ubiquitous, and are now well understood to be captured by fusion categories (at least for finite symmetries).

This makes 2D a natural setting in which to generalize the notion of S-entanglement beyond invertible cases. In this subsection, we construct entangled states for theories with non-invertible symmetry by using the associated 3D SymTFT built from a modular tensor category (MTC), and coupling two such SymTFTs via a topological interface defined by a Lagrangian algebra.

Since our general theme is the role of symmetry and entanglement in holography, we focus on a well-studied AdS$_3$/CFT$_2$ dual pair where such non-invertible symmetries arise naturally. In particular, type IIB string theory on $\mathrm{AdS}_3 \times S^3 \times M_4$ in the tensionless limit (with $k=1$ units of NS-NS flux and $M_4 = T^4$ or $K3$) is dual to a 2D CFT given by the symmetric orbifold
\begin{equation}
    \frac{M_4^{\otimes N}}{S_N},
\end{equation}
describing $N$ fundamental strings (see, e.g, \cite{Gaberdiel:2017oqg,Giribet:2018ada,Gaberdiel:2018rqv,Eberhardt:2018ouy,Eberhardt:2019ywk}). This theory has a non-invertible global symmetry governed by the fusion category $\mathrm{Rep}(S_N)$, which can be regarded as the quantum symmetry under the $S_N$ orbifold \cite{Bhardwaj:2017xup}. This holographic non-invertible symmetry has been recently discussed in \cite{Heckman:2024obe,Gutperle:2024vyp,Knighton:2024noc}. We will see in the following that this setup provides a controlled and physically meaningful setting in which to explore S-entanglement for non-invertible symmetries.

\paragraph{SymTFT for Rep($S_N$)}

The SymTFT associated with this non-invertible symmetry is the 3D Turaev–Viro theory based on the Drinfeld center $\mathcal{Z}(\mathrm{Rep}(S_N))$ (see, e.g., \cite{Turaev:1992hq,Barrett:1993ab,Kirillov:2010nh,Turaev:2010pp}). Since Rep$(S_N)$ and $S_N$ are connected by gauging in 2D QFTs (mathematically speaking, the fusion category Rep$(S_N)$ and Vec$_{S_N}$ are Morita-equivalent), one can instead use $\mathcal{Z}(\mathrm{Vec}_{S_N})$ as the SymTFT. The simple line operators in this theory are labeled by pairs
$$
([g], \pi_g),
$$
where $[g]$ is a conjugacy class in $S_N$ and $\pi_g$ is an irreducible representation of the centralizer $Z_g$.

There are two canonical choices of Lagrangian algebra in $\mathcal{Z}(\mathrm{Vec}_{S_N})$ (see, e.g., \cite{davydov2016lagrangianalgebrasgrouptheoreticalbraided})
\begin{itemize}
    \item The \emph{electric} boundary condition, corresponding to a 2D QFT with $\mathrm{Vec}_{S_N}$ symmetry, is given by the Lagrangian algebra
    \begin{equation}
        \mathcal{L}_{\text{electric}} = \bigoplus_{\pi \in \mathrm{Irrep}(S_N)} d_\pi \cdot ([1], \pi),
    \end{equation}
    where $d_\pi$ is the dimension of the representation $\pi$.

    \item The \emph{magnetic} boundary condition, corresponding to a 2D QFT with $\mathrm{Rep}(S_N)$ symmetry, is defined by
    \begin{equation}
        \mathcal{L}_{\text{magnetic}} = \bigoplus_{[g]} ([g], \mathbf{1}),
    \end{equation}
    where the sum runs over all conjugacy classes in $S_N$, and $\mathbf{1}$ denotes the trivial representation of the stabilizer $Z_g$.
\end{itemize}

From the SymTFT point of view, the holographic CFT, as a symmetric orbifold theory $\mathrm{Sym}^N(M_4)$, is realized by placing a magnetic boundary condition in for $\mathcal{Z}(\mathrm{Vec}_{S_N})$. In this language, if we have the following product theory
\begin{equation}
    \left( \frac{M_4^{\otimes N}}{S_N} \right) \otimes \left( \frac{M_4^{\otimes N}}{S_N} \right),
\end{equation}
it will correspond to two copies of the SymTFT, each with magnetic boundary condition, and no entanglement between them.

\begin{figure}[h]
    \centering
    \includegraphics[width=0.9\textwidth]{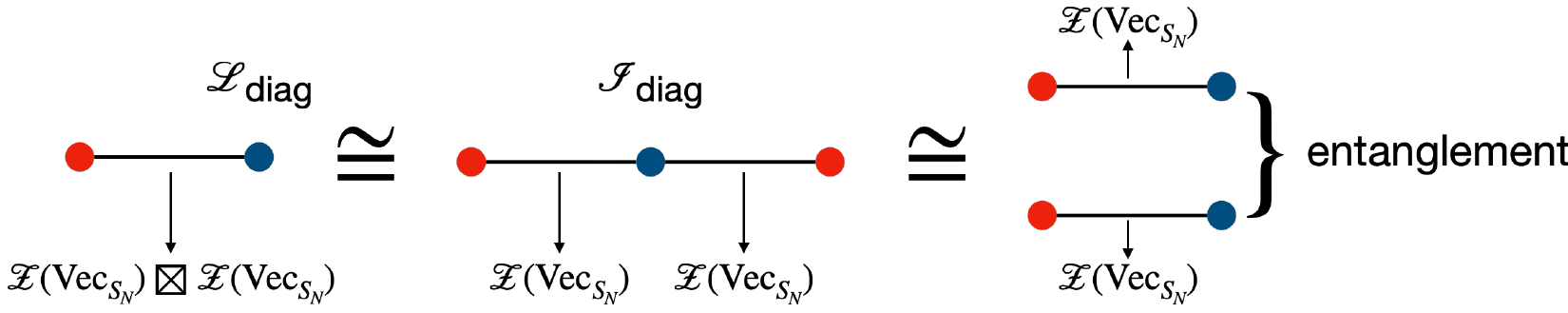}
    \caption{
        A topological interface in $\mathcal{Z}(\mathrm{Vec}_{S_N})$ defined by the diagonal Lagrangian algebra
        $\mathcal{L}_{\text{diag}} = \bigoplus_{[g]} ([g], \mathbf{1}) \otimes ([g], \mathbf{1})$ of $\mathcal{Z}(\mathrm{Vec}_{S_N}) \boxtimes \mathcal{Z}(\mathrm{Vec}_{S_N})$ under the folding trick.
        Folding this yields an entangled state in the tensor product Hilbert space.
        Each individual factor corresponds to a local CFT, while the entangled state corresponds to a diagonally orbifold theory  $(M_4^{\otimes N} \otimes M_4^{\otimes N})/S_N$.
    }
    \label{fig:s_n_entangle}
\end{figure}

\paragraph{SymTFT Entanglement} We now define a genuine entangled state in the doubled Hilbert space by coupling these two boundary conditions via a topological interface. Let us consider two SymTFTs, each associated with $\mathcal{Z}(\mathrm{Vec}_{S_N})$, and couple them through a topological interface corresponding to a \emph{diagonal} Lagrangian algebra in the product category $\mathcal{Z}(\mathrm{Vec}_{S_N}) \boxtimes \mathcal{Z}(\mathrm{Vec}_{S_N})$:
\begin{equation}
    \mathcal{L}_{\text{diag}} = \bigoplus_{[g]} ([g], \mathbf{1}) \otimes ([g], \mathbf{1}).
\end{equation}
This diagonal algebra ensures that for each conjugacy class $[g]$, the same simple object in the Drinfeld center is inserted in both copies. Folding this interface yields a state $|\Psi\rangle$ in the tensor product Hilbert space $\mathcal{H} \otimes \mathcal{H}$, which is pure globally but a mixed state from the perspective of either copy alone. It realizes a symmetry-entangled state between two CFTs with $\mathrm{Rep}(S_N)$ symmetry. See Figure \ref{fig:s_n_entangle} for an illustration.

The 2D interpretation of this entangled state is a single CFT with global structure given by the \emph{diagonal} $S_N$ orbifolding. A natural candidate for such a theory is the symmetric orbifold
\begin{equation}
    \frac{M_4^{\otimes N} \otimes M_4^{\otimes N}}{S_N},
\end{equation}
where the diagonal $S_N$ permutation symmetry acts simultaneously on the two $M_4^{\otimes N}$ factors. This theory cannot be decomposed as a product of two symmetric orbifolds, and its Hilbert space is entangled across the two copies. Thus, the SymTFT framework naturally extends the notion of symmetry-entangled states to non-invertible symmetries.

\section{Baby Universes, Higher Ensemble Averaging and Bulk Global Symmetries}\label{sec:ohbaby}

Having discussed S-entanglements for (S)CFTs in various dimensions admitting a top-down construction, we now explore how taking partial traces of S-entangled SymTFT states makes contact with holographic ensemble averaging and the dimension of baby universe Hilbert spaces. We first review how to build $\alpha$-states in the context of the gravitational path integral, following \cite{Coleman:1988cy, Marolf:2020xie}, and their resulting statistical averaging for the ensemble of CFTs. From a generalized symmetry perspective, this can be viewed as an entanglement for $(-1)$-form symmetries of CFTs, where $\alpha$ serves as the background field labeling elements in the theory ensemble. From the S-entanglement perspective, it is then natural to generalize the baby universe Hilbert space involving S-entanglement for any $p$-form, possibly categorical, global symmetries. We then show how this higher-generalization of baby universe Hilbert space can lead to ensemble averaging of theories over not just coupling constants, but allowing an ensemble whose element theories are differed by higher-form background fields. After this, we show that the baby universe Hilbert space dimension can be expressed in terms of S-entanglement data (Statement 3 from the Introduction). 

We then comment on how one recovers factorization in the context of UV-complete quantum gravity by including the full information of the combined S-entangled system.
This amounts to making the baby universe Hilbert space one-dimensional, following the baby universe conjecture \cite{McNamara:2020uza}. From a bulk perspective, this amounts to breaking potential global symmetries which, for invertible $(-1)$-form symmetries, aligns with the statement in \cite{McNamara:2020uza} that the one-dimensional baby universe Hilbert space can be viewed as a corollary of the Cobordism Conjecture \cite{McNamara:2019rup}. However, given that the global symmetries we are considering can be categorical, we expect the trivialization of topological sectors responsible for the one-dimensional baby universe Hilbert space to be beyond the cobordism classification.

\subsection{Baby Universes and $\alpha$-States}
\label{subsec:mm picture}
\begin{figure}[h]
    \centering
    \includegraphics[width=8cm]{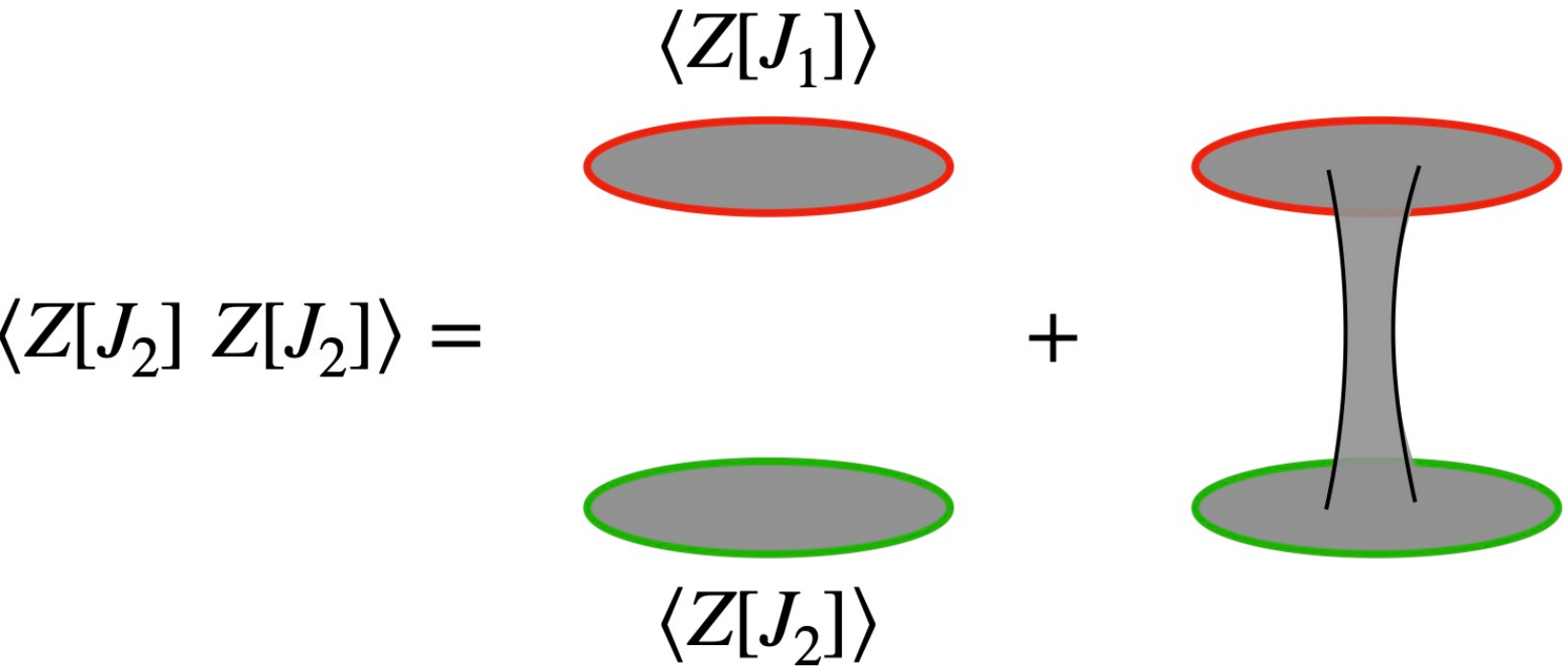}
    \caption{The non-factorization of partition functions of gravitational path integral with spacetime wormholes. The red and green lines label two components of the conformal boundary. From the CFT point of view, these two disconnected boundary components correspond to partition functions $Z[J_1]$ and $Z[J_2]$ with associted sources $J_1$ and $J_2$. However, the full gravitational path integral $\langle Z[J_1]Z[J_2] \rangle$ by summing over bulk topologies include an extra configuration where the red and green asymptotic boundaries are connected via a spacetime wormhole.}
    \label{fig:facpuzzle}
\end{figure}

We start with a lightning review of how the ensemble averaging shows up from $\alpha$-states in baby universe Hilbert space, following \cite{Marolf:2020xie} (see also \cite{Heckman:2021vzx}).  Consider a (semi-classical) bulk theory with a set of fields $\Phi$ (including the metric, scalars, higher-form gauge fields etc) whose asymptotic boundary has $n$ disconnected components. Associate to each component a boundary condition $\Phi \sim J_i$, the gravitational path integral can then be defined schematically as
\begin{equation}\label{eq:GPI}
    \langle Z[J_1]\cdots Z[J_n] \rangle=\int_{\Phi\sim J}\mathcal{D}\Phi~e^{-S[\Phi]}.
\end{equation}
The connected bulk geometry implies the non-factorization of the above path integral, e.g., for $n=2$.
\begin{equation}
    \langle Z[J_1]Z[J_2] \rangle \neq \langle Z[J_1] \rangle \langle Z[J_2] \rangle.
\end{equation}
See Figure \ref{fig:facpuzzle} for a rough depiction.
This non-factorization property of the gravitational path integral is closed related to baby universes, which we review as follows.

\begin{figure}[h]
    \centering
    \includegraphics[width=8cm]{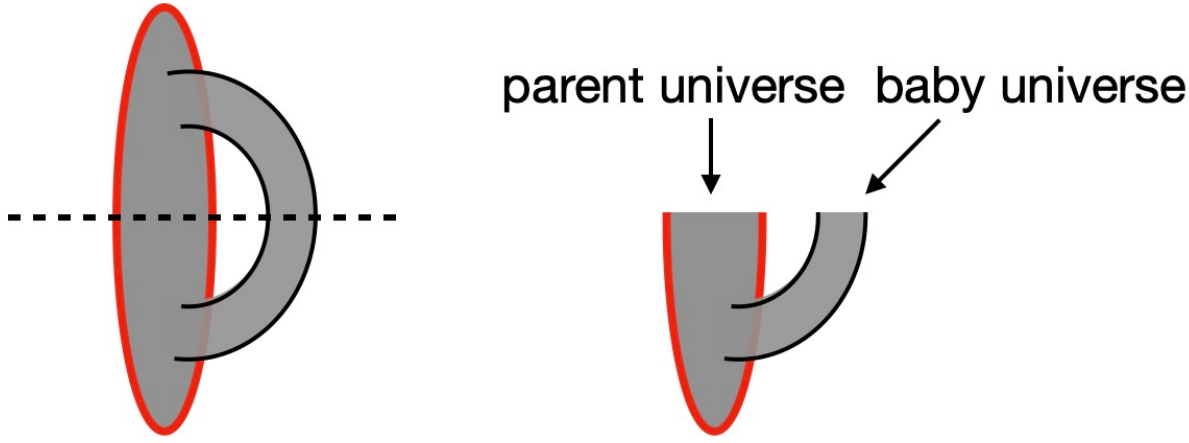}
    \caption{A baby universe is detached from the parent universe. Cutting open the gravitational path integral (\ref{eq:GPI}), the baby universe does not intersect with the parent universe along the spatial slice.}
    \label{fig:buslice}
\end{figure}

Consider cutting open the gravitational path integral at $\tau=0$ so that the resulting spatial slices including sectors not associated
on this slice with any of the asymptotically AdS boundaries, but which is instead
associated with spatially compact universes. These closed universes are referred to as baby universes in \cite{Marolf:2020xie} (following \cite{Coleman:1988cy, Giddings:1988cx, Giddings:1988wv}). See Figure \ref{fig:buslice} for an illustration.
One can then associate a Hilbert space $\mathcal{H}_{\text{BU}}$ consisting of all intermediate states separating the ``past'' ($\tau<0$) and ``future'' ($\tau>0$), corresponding to all possible configurations including baby universes at $\tau=0$. Now given a boundary condition with $n$ components $\{J_1, \cdots J_n \}$, there is an associated state
\begin{equation}\label{eq:general state in HBU}
    |Z[J_1]\cdots Z[J_n]\rangle \in \mathcal{H}_{\text{BU}}.
\end{equation}
Among all states in $\mathcal{H}_{\text{BU}}$, a special one is the Hartle-Hawking state $|\text{HH}\rangle$ corresponding to the configuration with no boundary. The so-called “cosmological partition function” is given by its norm:
\begin{equation}\label{eq: cos partition function}
    \langle \text{HH} | \text{HH} \rangle=\int_{\text{no boundary}}\mathcal{D}\Phi~e^{-S[\Phi]}.
\end{equation}

The operators acting on the $\mathcal{H}_{\text{BU}}$ can be introduced to change the number of boundary components from $n$ to $n+1$. Denoting the operator as $\hat{Z[J]}$ with the boundary condition $\Phi \sim J$, its action reads
\begin{equation}\label{eq:Z operator acting on HH}
    \hat{Z[J]}|Z[J_1]\cdots Z[J_n]\rangle =|Z[J]Z[J_1]\cdots Z[J_n]\rangle.
\end{equation}
The Hartle-Hawking state can then be regarded as the ``ground state'', upon which the ``creation operators'' $\hat{Z[J]}$ can act and  obtain the state (\ref{eq:general state in HBU}) in $
\mathcal{H}_{\text{BU}}$ associated to a general boundary condition with multiple boundary components:
\begin{equation}
    \hat{Z[J_1]}\cdots \hat{Z[J_n]}|\text{HH}\rangle =|Z[J_1]\cdots Z[J_n]\rangle.
\end{equation}
The gravitational path integral (\ref{eq:GPI}) is then be interpreted as the vacuum expectation value
\begin{equation}\label{eq:GPI as vev}
    \int_{\Phi \sim J}\mathcal{D}\Phi ~e^{-S[\Phi]}=\langle \text{HH} | \hat{Z}[J_1]\cdots \hat{Z}[J_n]| \text{HH} \rangle
\end{equation}

For a general state (\ref{eq:general state in HBU}) in $\mathcal{H}_{\text{BU}}$ one can freely exchange the order of all $Z[J]$'s, due to the fact that it will associate to the same state. Together with (\ref{eq:Z operator acting on HH}), this leads to a key property of the set of operators $\hat{Z[J]}$:
\begin{equation}
  |Z[J_1]Z[J_2]\rangle=|Z[J_2]Z[J_1]\rangle \Rightarrow  [\hat{Z[J_1]}, \hat{Z[J_2]} ]=0.
\end{equation}
Therefore, all $\hat{Z[J]}$ operators share eigenvectors, building a orthogonal basis for $\mathcal{H}_{\text{BU}}$, known as $\alpha$-states \cite{Coleman:1988cy, Marolf:2020xie}:
\begin{equation}
    \hat{Z}[J]|\alpha \rangle =
    Z_\alpha[J]|\alpha \rangle, ~\forall J.
\end{equation}
That is to say, the eigenvalue for an $\alpha$-state is a partition function for an asymptotic AdS boundary with a single component.

The gravitational path integral (\ref{eq:GPI}), written in terms of the vev (\ref{eq:GPI as vev}), can now be obtained via projections onto $\alpha$-states
\begin{equation}\label{eq:MM averaging}
\begin{split}
      \langle Z[J_1]\cdots Z[J_n] \rangle &=\sum_{\alpha_0, \cdots, \alpha_n}\langle \text{HH}|\alpha_0 \rangle \langle \alpha_0|Z[J_1]|\alpha_1\rangle \cdots \langle \alpha_{n-1}|Z[J_n]|\alpha_n\rangle \langle \alpha_n| \text{HH}\rangle\\
    &=\mathfrak{N}\sum_\alpha p_\alpha Z_\alpha[J_1]\cdots Z_\alpha[J_n]
\end{split}
\end{equation}
where $\mathfrak{N}=\langle \text{HH} \rangle$ is the norm given by the ``cosmological partition function'' (\ref{eq: cos partition function}), and the probability $p_\alpha$ is given by the innet product in $\mathcal{H}_{\text{BU}}$
\begin{equation}
    p_\alpha=\frac{|\langle \text{HH}|\alpha \rangle |^2}{\mathfrak{N}}.
\end{equation}
This shows that a gravitational path integral involving spacetime wormholes enjoys an ensemble averaging interpretation. Element theories in the ensemble are labeled by $\alpha$ parameters, and $p_\alpha$ specifies the probability for drawing that theory from the ensemble\footnote{It is also possible to consider continuous ensemble averaging, exemplified by, e.g., random matrix models dual to the JT gravity \cite{Saad:2019lba,Stanford:2019vob} and averaging over conformal manifolds of 2D CFTs \cite{Maloney:2020nni}. The according modification of the above discussion is to realize that $|\alpha \rangle$ is now not normalizable by itself, but requires a delta-function normalization. The resulting ensemble will be equipped with an probability distribution $p(\alpha)$, whose averaging reads, schematically, $\langle Z[J_1]\cdots Z[J_n] \rangle =\int d\alpha p(\alpha)Z_\alpha[J_1]\cdots Z_\alpha[J_n]$.}.

\subsection{Ensemble Averaging from SymTFT Entanglement}
\label{ssec:ensembleavg}
We have seen that the $\alpha$ parameter plays a key role in labeling states in the baby universe Hilbert space, effectively indexing different elements in an ensemble of theories. However, in a UV-complete theory of gravity—such as string theory—it is generally believed that there are no free parameters: all couplings and discrete labels are fixed by the vacuum expectation values of dynamical modulus fields. This suggests that only a single theory is physically realized, and the baby universe Hilbert space should collapse to a one-dimensional space~\cite{McNamara:2020uza}.

This raises a natural tension between two perspectives. On the one hand, semiclassical gravitational path integrals with wormholes point to ensemble averaging over $\alpha$-parameters. On the other hand, top-down constructions from string theory typically produce a specific boundary theory with fixed parameters. Reconciling these two pictures remains a subtle problem.

One proposal addressing this issue was put forward in~\cite{Heckman:2021vzx}, where ensemble-averaged SCFTs are engineered on branes, and a distinguished sector of operators is selected to reconstruct a single AdS bulk.\footnote{See also~\cite{Baume:2023kkf} for a follow-up on Schlenker–Witten’s large $N$ averaging proposal~\cite{Schlenker:2022dyo}.}

In this subsection, we describe a complementary perspective based on S-entanglement. The idea is to associate a family of QFTs labeled by a parameter with the Hilbert space of a SymTFT implementing a global $(-1)$-form symmetry. Entanglement within this symmetry sector then gives rise to ensemble-averaged quantities such as $\overline{Z^2}$, in a way that naturally admits a string-theoretic origin.

In favorable cases where the D-dimensional boundary QFT admits a string-theoretic engineering, the corresponding SymTFT for $(-1)$-form symmetries can be understood as a topological limit of the modulus sector of the higher-dimensional bulk theory. Concretely, a modulus field descending from 10D or 11D string theory—such as a gauge coupling, an axion, or a brane separation parameter—gives rise to a propagating dynamical field in (D+1) dimensions. Its topological limit, describing discrete or continuous parameter values modulo gauge redundancy, defines the SymTFT implementing the $(-1)$-form symmetry \cite{Yu:2024jtk} (see also \cite{Najjar:2024vmm,Lin:2025oml,Robbins:2025apg} and Appendix C of \cite{Aloni:2024jpb}). The entanglement structure of this SymTFT encodes the coupling of the boundary theory to this modulus sector and governs the statistical properties of ensemble observables.


This connection provides a UV origin for the ensemble averaging from the S-entanglement: it is not an ad hoc construct, but a limit of a physical modulus theory that propagates in the bulk. It applies equally well to discrete parameters (such as gauge ranks or fluxes) and to (compact) continuous parameters (such as theta-angles or axions), thereby offering a unified language for describing holographic ensemble phenomena.

\subsubsection{SymTFT Hilbert space and averaging mechanism}

Consider a (D+1)-dimensional SymTFT for a $\mathbb{Z}_k$ $(-1)$-form symmetry, with action
\begin{equation}
    S_{\text{SymTFT}} = \frac{k}{2\pi} \int \phi \wedge d f_D,
\end{equation}
where $\phi$ is a 0-form field and $f_D$ is a $D$-form gauge field. This theory contains:

\begin{itemize}
    \item Local operators $U_m = \exp(i m \phi)$, $m=0,1,\cdots, k-1$;
    \item Topological domain wall operators $V_n = \exp\left(i n \int_{M_D} f_D\right)$, $n=0,1,\cdots, k-1$.
\end{itemize}

The Hilbert space $\mathcal{H}$ of the SymTFT admits an orthonormal basis $\{ |a\rangle \}$, labeled by $a = 0, 1, \dots, k-1$, satisfying
\begin{equation}
    V_n |a\rangle = e^{2\pi i n a / k} |a\rangle, \qquad U_m |a\rangle = |a + m \rangle.
\end{equation}
Each basis vector $|a\rangle$ corresponds to a definite value of the background parameter in the associated D-dimensional QFT—for instance, a discrete theta angle or gauge rank modulo $k$.

We now consider an entangled state in the doubled Hilbert space $\mathcal{H}_1 \otimes \mathcal{H}_2$:
\begin{equation}\label{eq:maxentangle (-1) symtft}
    |\psi\rangle = \frac{1}{\sqrt{k}} \sum_{a=0}^{k-1} |a\rangle_1 \otimes |a\rangle_2.
\end{equation}
Acting on $|\psi\rangle$ with a generic operator gives
\begin{equation}
U_{m_1}^{(1)}U_{m_2}^{(2)}V_{n_1}^{(1)}V_{n_2}^{(2)}|\psi\rangle
=
\frac{1}{\sqrt{k}}\sum_{a\in\mathbb Z_k}\omega^{(n_1+n_2)a}
|a+m_1\rangle_1|a+m_2\rangle_2,
\end{equation}
with $\omega=e^{2\pi i /k}$ the $k$-th root of unity.
Shifting the summation variable to $b=a+m_1$, this becomes
\begin{equation}
U_{m_1}^{(1)}U_{m_2}^{(2)}V_{n_1}^{(1)}V_{n_2}^{(2)}|\psi\rangle
=
\omega^{-(n_1+n_2)m_1}\frac{1}{\sqrt{k}}
\sum_{b\in\mathbb Z_k}\omega^{(n_1+n_2)b}
|b\rangle_1|b+(m_2-m_1)\rangle_2.
\end{equation}
This state is proportional to $|\psi\rangle$ only if the relative shift vanishes and the $b$-dependent phase is absent, namely
\begin{equation}
m_2-m_1\equiv 0 \pmod{k},\qquad n_1+n_2\equiv 0 \pmod{k}.
\end{equation}
Thus, setting $m_1=m_2\equiv m$ and $n_2=-n_1\equiv -n$, one finds that the residual overall phase is trivial and
\begin{equation}
\mathcal O_{m,n}|\psi\rangle=|\psi\rangle,
\qquad
\mathcal O_{m,n}\equiv U_m^{(1)}U_m^{(2)}V_n^{(1)}V_{-n}^{(2)},
\qquad m,n\in\mathbb Z_k.
\end{equation}
Therefore $|\psi\rangle$ is a simultaneous $+1$ eigenstate of the $k^2$ operators $\mathcal O_{m,n}$. Since the opposite signs in the two $V$-factors make the Heisenberg phases cancel between the two tensor factors, these operators commute and furnish a Lagrangian subgroup of the doubled $\mathbb Z_k$ Heisenberg algebra, making $\psi\rangle$ a state corresponding to a topological boundary condition.

This state is maximally entangled in the parameter basis. Tracing out the second factor yields a reduced density matrix
\begin{equation}\label{eq: reduced density matrix discrete}
    \rho_1 = \operatorname{Tr}_{\mathcal{H}_2} |\psi\rangle \langle \psi| = \frac{1}{k} \sum_{a=0}^{k-1} |a\rangle \langle a|.
\end{equation}

Now consider a relative QFT whose partition vector  in $\mathcal{H}$ is
\begin{equation}
    |\mathcal{T}\rangle = \sum_{a=0}^{k-1} Z_a |a\rangle,
\end{equation}
where $Z_a$ denotes the partition function of the absolute theory with background label $a$. This vector specifies a physical, non-topological boundary condition for the SymTFT, corresponding to a choice of relative theory \cite{Freed:2012bs}. The partition function for a specific background $b$ of the modulus field $\phi$ is recovered as
\begin{equation}
    \langle b | \mathcal{T} \rangle = Z_b.
\end{equation}

The entangled partition function is then given by pairing\footnote{Given two density matrices $\rho_A$ and $\rho_B$, there is a canonical inner product $\mathrm{Tr}(\rho_A^\dagger \rho_B)$. For the case of pure states, $\mathrm{Tr}(\rho_A^\dagger \rho_B)=|\langle A|B \rangle|^2$} $|\mathcal{T}\rangle$ with $\rho_1$:
\begin{equation}\label{eq:Zsquared}
    |Z_{\text{entangled}}|^2 := \operatorname{Tr}\left( \rho^\dagger_{\mathcal{T}} \rho_1 \right), \quad \text{where } \rho_{\mathcal{T}} := |\mathcal{T}\rangle \langle \mathcal{T}|.
\end{equation}
A straightforward computation yields
\begin{equation}\label{eq:averaged (-1)-form partition function}
    |Z_{\text{entangled}}|^2 = \frac{1}{k} \sum_{a=0}^{k-1} |Z_a|^2.
\end{equation}
This expression exactly matches the ensemble-averaged partition function Eq. (\ref{eq:MM averaging}) obtained by Marolf and Maxfield via baby universe $\alpha$-states with two disconnected AdS boundary components with the probability $p_a=\frac{1}{k}$! In our case, the averaging emerges directly from entanglement under the topological symmetry sector.


\subsubsection{Schlenker-Witten Averaging Over Gauge Ranks in ABJ(M) Theory}
\label{subsec:ABJMaveraging}

An explicit interpretation of ensemble averaging in AdS/CFT was proposed by Schlenker and Witten~\cite{Schlenker:2022dyo}, where the semiclassical gravitational path integral is argued to describe an average over the large-$N$ limit --- namely, over the gauge rank. Motivated by this certain averaging, we model an analogous averaging over the gauge rank $M$ in the ABJ(M) theory $U(N)_k \times U(N+M)_{-k}$ \cite{Aharony:2008gk}, based on SymTFT for the corresponding (-1)-form symmetry.

Recall that the ABJ theory $U(N)_k \times U(N+M)_{-k}$ \cite{Aharony:2008gk} is engineered on $N$ regular M2-branes probing $\mathbb{C}^4/\mathbb{Z}_k$ orbifold singularity, with $M$ fractional M2-branes, i.e. M5-branes wrapping on torsional 3-cycles. This discrete torsion is also $\mathbb{Z}_k$, namely the possible values for $M$ reads $M=0,1,\cdots k-1$.

Promoting $M$ to a background field for a $(-1)$-form symmetry, we ask: what is the corresponding SymTFT? Fortunately, this corresponding theory has been derived in \cite{Yu:2024jtk}\footnote{There is extra topological terms involving the coupling between $b_0$ and a 2-form field $b_2$, which matters little in our current context.}:
\begin{equation}\label{eq:parameter symtft in ABJM}
    k\int_{M_4}b_0 \cup dc_3
\end{equation}
Schematically, $b_0$ and $c_3$ come from $C_3$ and its magnetic dual fields reduced on torsional 3-cycles, respectively\footnote{A careful treatment of this dimensional reduction involves differential cohomology computation. See e.g., \cite{Yu:2023nyn, GarciaEtxebarria:2024fuk}.}. In particular, $b_0$ is a $\mathbb{Z}_k$ field whose background value gives rise to the gauge rank $M$, determining the number of fractional M2-branes. The admitted value of $M$ is upper-bounded by $k$, namely $M=0,1,\cdots, k-1$.

To model averaging over gauge rank, we follow the general procedure outlined above: we consider an entangled state
\begin{equation}
    |\psi\rangle = \frac{1}{\sqrt{k}} \sum_{M}^{k-1} |M\rangle_1 \otimes |M\rangle_2.
\end{equation}
and a relative QFT state
\begin{equation}
    |\mathcal{T}_{\text{ABJM}}\rangle=\sum_{M=0}^{k-1} Z_M|M\rangle
\end{equation}
where $Z_M$ is the partition function of ABJ(M) theory at rank $U(N)_k \times U(N+M)_{-k}$. Entangling with the topological sector as above yields
\begin{equation}
    |Z_{\text{ABJM}}^{\text{avg}}|^2 = \frac{1}{k} \sum_M |Z_M|^2.
\end{equation}

This construction provides a concrete top-down analog of the Marolf–Maxfield picture, where $\alpha$-states responsible for non-factorization are embedded in M-theory as a SymTFT associated to a discrete parameter. In particular, it mimics the averaging over gauge group ranks proposed by Schlenker and Witten. Here, we reproduce the same statistical structure purely from entanglement under the topological sector, encoded in the SymTFT. This offers further support for the idea that ensemble behavior in gravity may be understood as the shadow of symmetry-based entanglement on the boundary.

We emphasize that while the above discussion focused on discrete parameters associated to finite $(-1)$-form symmetries, similar constructions can be extended to compact continuous parameters. A simple illustration is theta-angle averaging in $\mathcal{N}=4$ SYM, modeled using a $U(1)$ SymTFT. We relegate this continuous case to Appendix~\ref{app:theta} to maintain focus on the finite setting.

\subsection{From SymTFT Entanglement Entropy to Baby Universe Hilbert Space Dimension}

In the framework described above, the ensemble average over $k$ discrete element theories arises from the $\mathbb{Z}_k$ $(-1)$-form symmetry implementing S-entanglement. We have chosen an extreme case where the state Eq.(\ref{eq:maxentangle (-1) symtft}) is a maximally entangled state. This entanglement can be measured by the von Neumann entropy computed from the reduced density matrix:
\begin{equation}
    S_{\mathrm{vN}}(\rho_1)=-\text{tr}(\rho_1 \ln \rho_1) = \log k.
\end{equation}
Alternatively, from Eq.(\ref{eq:averaged (-1)-form partition function}), this ensemble averaging can be interpreted as a result of a $k$-dimensional baby universe Hilbert space, spanned by $k$ orthogonal $\alpha$-states, as we reviewed in Section \ref{subsec:mm picture}.

We thus propose that the S-entanglement entropy reflects the number of distinguishable component theories in the ensemble, and corresponds to the dimensionality of the baby universe Hilbert space:
\begin{equation}
    \text{dim}_{\mathcal{H}_\text{BU}}=\exp{(S_{\mathrm{vN}}(\rho_{1}))}
\end{equation}
which in the case of $\mathbb{Z}_k$ SymTFT reads $\text{dim}_{\mathcal{H}_\text{BU}}=k$.

\begin{figure}[h]
    \centering
    \includegraphics[width=12.5cm]{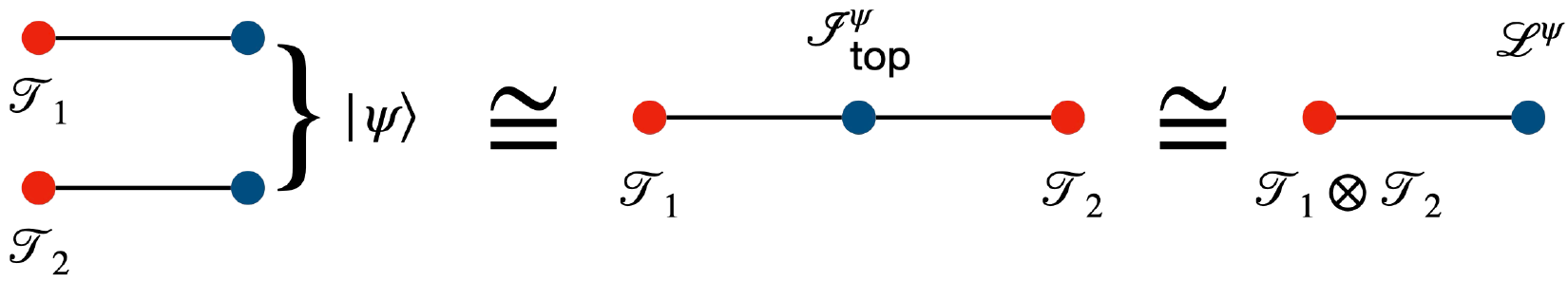}
    \caption{}
    \label{fig:topinterface}
\end{figure}

There is one subtlety to be clarified: For a general entangled state, it is very unlikely that $\exp(S_{\text{vN}}(\rho_1))$ will be an integer. However, we emphasize that the entangled state $|\psi\rangle$ we are considering, corresponds to a polarization of the relative theory $\mathcal{T}_1\otimes \mathcal{T}_2$. Under the SymTFT picture, this entangled state corresponds to not arbitrary but topological interface $\mathcal{I}_{\text{top}}^{\psi}$ within the SymTFT bulk, with two relative boundary states corresponding to $\mathcal{T}_1$ and $\mathcal{T}_2$. This topological interface, under the folding trick, corresponds to a topological boundary $\mathcal{L}^{\psi}$ of the $\text{SymTFT}_1\boxtimes \text{SymTFT}_2$. See Figure \ref{fig:topinterface} for an illustration of these alternative perspectives.

Given the $\mathcal{L}^\phi$ is a topological boundary of the tensor product theory $\text{SymTFT}_1\boxtimes \text{SymTFT}_2$, it defines a maximal commuting set of topological operators. This set of operators diagonalize the Hilbert space of $\text{SymTFT}_1\boxtimes \text{SymTFT}_2$. That is to say, the Hilbert space is spanned by the joint eigenstates of this commuting algebra, whose dimension $k$ counts the number of distinguishable topological sectors supported by the boundary. As a result, any entangled state deviating from the maximally entangled one --- beyond the direct product of polarizations for $\text{SymTFT}_1$ and $\text{SymTFT}_2$ --- would necessarily exhibit a bias toward certain sectors. This would break the maximal commuting structure of the boundary operator algebra and render the boundary condition non-topological.

From the $\mathbb{Z}_k$ SymTFT example, it seems that the dimension of the baby universe Hilbert space is given by
\begin{equation}\label{eq:dimequalsqqrtdim}
\text{dim}_{\mathcal{H}_\text{BU}}=\exp{(S_{\mathrm{vN}}(\rho_{1}))}=|\mathcal{L}^\psi|=\sqrt{\text{dim}_{\text{SymTFT}}}
\end{equation}
where the dimension of SymTFT can be expressed as
\begin{equation}
    \text{dim}_{\text{SymTFT}}=\sum_i |U_i|^2
\end{equation}
is defined as the sum of squares of the quantum dimensions of all topological operators of the SymTFT. The square root corresponds to the dimension of the Hilbert space associated to the maximally commuting subset of these operators (i.e., a Lagrangian subalgebra). This matches the known results: (a) For abelian groups in diverse dimensions, it corresponds to the order of the Lagrangian subgroup defining the polarization, and (b) for non-invertible symmetries in 2D, dimension of the fusion category $\mathcal{C}$ associated to a Lagrangian algebra of the center $\mathcal{Z}(\mathcal{C})$ (i.e., of the SymTFT). 

\begin{figure}[h]
    \centering
    \includegraphics[width=0.7\linewidth]{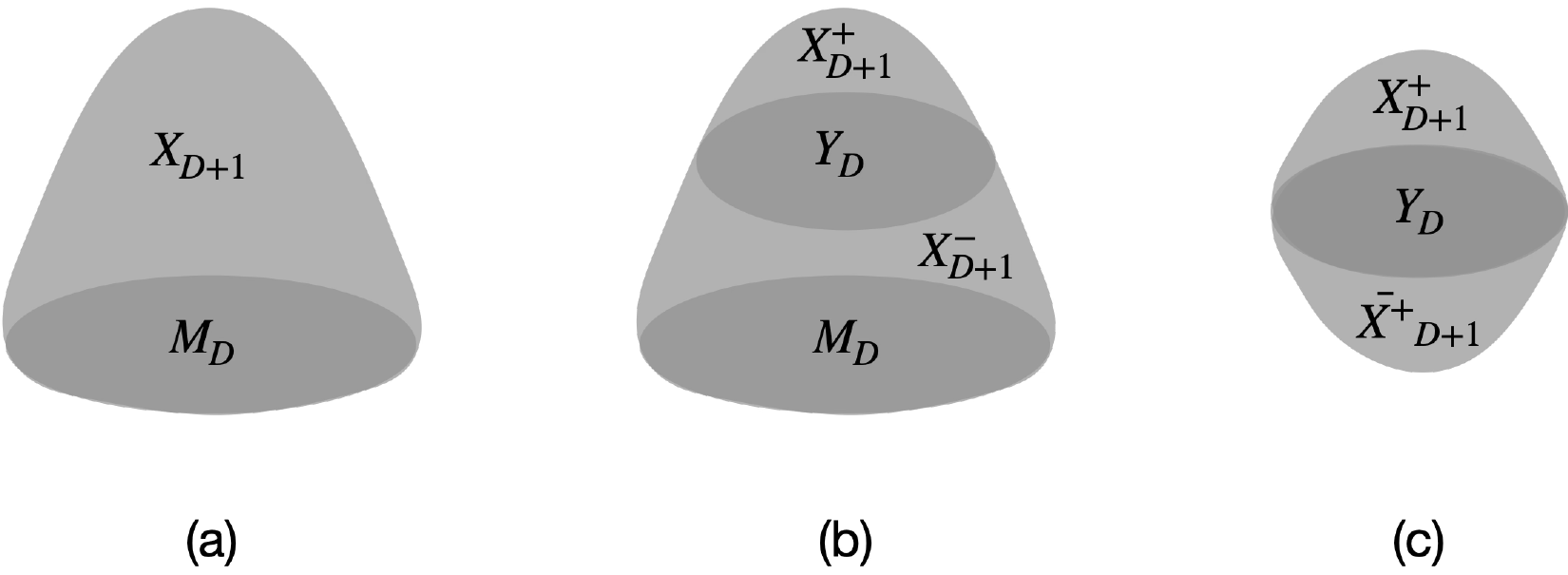}
    \caption{(a) Manifold $X_{D+1}$, on which a SymTFT is defined, with boundary $M_{D}$. (b) Cutting the bulk manifold into two parts $X_{D+1}^+$ and $X_{D+1}^-$ with a interface $Y_D$. (c) Gluing $X_{D+1}^+$ with its orientation reversal along $Y_{D}$.}
    \label{fig:bu-dimension}
\end{figure}

To rigorously establish (and refine) the relation \eqref{eq:dimequalsqqrtdim}, consider a QFT $\mathcal{T}_2(M_D)$ living on the boundary of a SymTFT worldvolume $X_{D+1}$ such that $\partial X_{D+1}=M_D$. This can be seen as an S-entanglement between $\mathcal{T}_2(M_D)$ and a QFT with a trivial worldvolume $\mathcal{T}_1(\varnothing)$ which is equivalent to a trivial theory. Holographically, this can be seen as an entanglement between the AdS$_{D+1}$ theory dual to $\mathcal{T}_2(M_D)$ and a baby universe dual to the trivial theory $\mathcal{T}_1(\varnothing)$. Let us now cut $X_{D+1}$ into two pieces $X^+_{D+1}$ and $X^-_{D+1}$ along some $Y_{D}$ such that $\partial X^-_{D+1}=\overline{Y}_{D}\cup M_D$ and $\partial X^+_{D+1}=Y_{D}$ (see Figure \ref{fig:bu-dimension} (a) and (b)).
Similar to what we have seen many times throughout this work, the SymTFT entangled state is a maximally entangled wavefunction
\begin{equation}
    \frac{1}{\sqrt{\mathrm{dim}(\mathcal{H}_{\mathcal{S}}(Y_D))}}\sum_{k}\ket{k}_1\otimes\ket{k}_2
\end{equation}
where $\ket{k}$ is a basis of states for $\mathcal{H}_{\mathcal{S}}(Y_D)$. Taking the partial trace $\mathrm{Tr}_2$ of this wavefunction yields the SymTFT density matrix $\rho=\frac{1}{k}\sum_{k}\ket{k}_1\bra{k}_1=\frac{1}{k}\mathbf{1}$. Following the discussion around \eqref{eq:Zsquared}, this allows us to compute the partition function for the partial trace of the S-entangled system to the trivial theory as
\begin{equation}\label{eq:Zssqaured2}
    |Z_{\mathcal{T}_1}(\varnothing)|^2=\mathrm{Tr}(\rho \ket{\psi_{X^+}}\bra{\psi_{X^+}})
\end{equation}
where $\ket{\psi_{X^+}}$ is defined\footnote{In the language of the Atiyah-Segal axioms, a manifold with a boundary is associated with a map $\mathcal{H}_{\mathcal{S}}(\varnothing)=\mathbb{C}\rightarrow \mathcal{H}_{\mathcal{S}}(Y_D)$ which selects out a state in $\mathcal{H}_{\mathcal{S}}(Y_D)$. Imposing a boundary condition $b$ with associated boundary state $\ket{b}$ to $Y_D$ is required to assign a partition function to $X^+$ which would be $\langle \psi_{X^+}|b\rangle$. Equation \eqref{eq:Zssqaured2} is then a generalization of this formula to a density matrix of boundary states. } as the state that the SymTFT $\mathcal{S}$ assigns to the manifold with boundary $X^+_{D+1}$. Note that in \eqref{eq:Zssqaured2} we have suppressed the notation in $Z_{\mathcal{T}_1}(\varnothing)$ which labels the information on the S-entanglement and partial tracing. This appears on the right hand side of the equation from the choice of $X^+_{D+1}$ and SymTFT theory. Since the trivial theory can be seen as holographically dual to a baby universe theory let us denote $|Z_{BU}|^2:=|Z_{\mathcal{T}_1}(\varnothing)|^2$. From our choice of $\rho$ we have
\begin{equation}\label{eq:z2bu}
    |Z|^2_{BU}=\langle \psi_{X^+_{D+1}}|\psi_{X^+_{D+1}} \rangle=Z_\mathcal{S}(X^+_{D+1}\cup_{Y}\overline{X^+}_{D+1})
\end{equation}
where $Z_\mathcal{S}(X^+_{D+1}\cup_{Y}\overline{X^+}_{D+1})$ is the SymTFT partition function evaluated on the $(D+1)$-dimensional manifold obtained from gluing $X_{D+1}^+$ and its orientated reversed partner $\overline{X^+}_{D+1}$ along $Y_D$ (see Figure \ref{fig:bu-dimension} (c) )which by construction is equal to the norm of $\ket{\psi_{X^{+}_{D+1}}}$. The argument relating the baby universe and SymTFT partition functions in \eqref{eq:z2bu} can be ``categorified" to relation between Hilbert spaces, category of boundary conditions, etc., by reducing the dimensionality of the manifolds involved by $1$, $2$, etc. At the level of Hilbert spaces we have 
\begin{equation}\label{eq:HBU}
    \mathcal{H}_{BU}\otimes \mathcal{H}^\dagger_{BU}=\mathcal{H}_S(X^{+}_D\cup_Y \overline{X^+}_{D})
\end{equation}
where now $X_D$ is $D$-dimensional and $Y$ is $(D-1)$-dimensional and which reproduces the intuited relation \eqref{eq:dimequalsqqrtdim}. 


\subsection{Generalized Ensemble Averaging Through Symmetry-Enriched $\alpha$-States}

The above derivation based on $(-1)$-form symmetry entanglement generalizes naturally to higher-form symmetries. For example, one can simply substitute the $(-1)$-form symmetry with the 1-form symmetry of 4D $\mathfrak{g}=\mathfrak{su}(N)$ theory, which we intensively discussed in Section 2, build the reduced density matrix for a maximally entangled state
\begin{equation}
\begin{split}
	&|\psi \rangle =\frac{1}{\sqrt{N}}\sum_{a=0}^{N-1}|a\rangle _1 \otimes  |a\rangle _2\\
	\Rightarrow &\rho_1=\text{Tr}_2\rho_\psi=\frac{1}{N}\sum_{a=0}^{N-1}|a\rangle  \langle a|.
\end{split}
\end{equation}
Consider a relative QFT state
\begin{equation}
	\mathcal{T}=\sum_{a=0}^{N-1}Z[B^{(2)}_a]|a\rangle,
\end{equation}
where $Z[B^{(2)}_a]$ is the partition function for $G=SU(N)$ theory coupled to the background $B^{(2)}_a$ field for the 1-form symmetry, associated to the topological boundary state $|a\rangle$ of the 5D SymTFT. Similar to Section \ref{sec:SymTFTentanglement} we leave implicit indices coming from basis elements of $H^2(M_4,U(1))$. With respect to this relative QFT state, the partition function corresponding to the mixed state $\rho_1$ is
\begin{equation}
	|Z_\text{entangled}|^2=\text{Tr}(\rho^\dagger_\mathcal{T}\rho_1)=\frac{1}{N}\sum_{a=0}^{N-1}|Z[B^{(2)}_a]|^2.
\end{equation}
This is likewise an ensemble-averaged observable, though the averaging is now over background configurations rather than over distinct local theories.

This motivates a more general construction that incorporates other generalized global symmetries into the conventional averaging built from the $(-1)$-form symmetry entanglement. Given that a $(-1)$-form symmetry entanglement resulting in an ensemble of $k$ theories (e.g., maximally entangled state for $\mathbb{Z}_k$ SymTFT) can be interpreted as $\alpha$-states in a $k$-dimensional baby universe Hilbert space, we introduce a notion of \emph{symmetry-enriched $\alpha$-state}, so that the ensemble averaging is not just about theories labeled by different parameters (coupling constants, gauge ranks etc.), but also over different symmetry background fields.

In simple cases such that different symmetries are independent to each other, namely the SymTFT for the QFT under consideration decomposes as
\begin{equation}
	S_{D+1}=\bigoplus_p S^{(p)}_{D+1}[B^{(p+1)}]\end{equation}
where $S_{D+1}^{(p)}$ is the SymTFT for the $p$-form symmetry, each of which has a separate field variable $B^{(p+1)}$ (as well as its canonical conjugate). An entangled state for two copies of QFTs associated to the SymTFT is then embedded in the Hilbert space
\begin{equation}
	\bigoplus_p \mathcal{H}^{(p)}_1\otimes \mathcal{H}^{(p)}_2.
\end{equation}
For illustration, consider the state
\begin{equation}
	|\psi \rangle =\sum_{\vec{a}^{(p)}}\sqrt{p_{\vec{a}^{(p)}}}|\vec{a}^{(p)}\rangle,
\end{equation}
where $\vec{a}^{(p)}=(a^{-1},a^{0}, a^{1},\cdots a^{D-1})$ labels eigenstates for all $p$-form symmetry sectors of the $(D+1)$-dimensional SymTFT under admitted polarizations.

Following the derivation of $(-1)$-form symmetry in the previous subsection, it is straightforward to obtain that for a relative QFT state
\begin{equation}
	|\mathcal{T}\rangle = Z[\vec{B}_a^{(p+1)}]|\vec{a}^{(p)}\rangle,
\end{equation}
where $Z[\vec{B}_a^{(p+1)}]$ is the partition function for the absolute theory with well-defined polarizations for all $p$-form symmetries with background fields $\vec{B}_a^{(p+1)}=(B_{a^{(-1)}}^{(0)},\cdots B_{a^{(D-1)}}^{(D)} )$, the resulting partition function for the entangled state $|\psi \rangle$
\begin{equation}
\begin{split}
	|Z_{\text{entangled}}|^2&=\text{Tr}(\rho^\dagger_\mathcal{T}\rho_1)=\text{Tr}(|\mathcal{T}\rangle^* \langle \mathcal{T}|^*~\text{Tr}_2(|\psi \rangle \langle \psi|))\\
	&=\sum_{\vec{a}^{(p)}}p_{\vec{a}^{(p)}}|Z[\vec{B}_a^{(p+1)}]|^2.
\end{split}
\end{equation}
This is an ensemble averaging governed by a joint probability $p_{\vec{a}^{(p)}}$ with $D$-dimensional random variables $\vec{a}^{(p)}=(a^{-1},a^{0}, a^{1},\cdots a^{D-1})$.

The conventional ensemble averaging over coupling constants can now be viewed as a first layer, i.e., focusing on the $p=-1$ sector. We can then interpret the $(-1)$-form symmetry as generating $\alpha$-states of the baby universe Hilbert space, and regard $p >-1$ sectors as enrichment on top of it. Grouping the $(-1)$-form label as $\alpha$ and the remaining as $\vec{b}$, the partition function becomes:
\begin{equation}
	|Z_{\text{entangled}}|^2 = \sum_{\alpha} \sum_{\vec{b}} p_\alpha(\vec{b}) \left| Z_\alpha[\vec{B}_b^{(q+1)}] \right|^2.
\end{equation}

Turning off all $\vec{B}_b^{(q+1)}$ fields, we recover the Marolf-Maxfield $\alpha$-state-driven ensemble averaging, while with nontrivial background $\vec{B}_a^{(q+1)}$, we are enriching each element theory $Z_\alpha$ labeled by the $\alpha$-state by symmetry background fields. Each $\alpha$-state now labels a family of theories rather than a single one, leading to a nested ensemble structure.

\subsection{Bulk Global Symmetries}

In this subsection we briefly comment on how the S-entanglement can lead to the appearance of global symmetries in the bulk.

It is widely believed that quantum gravity admits no exact global symmetries. In the standard AdS/CFT correspondence (without averaging), global symmetries of the CFT are realized by a SymTFT together with appropriate asymptotic boundary conditions, which occupies a topological sector of the bulk AdS theory.\footnote{Strictly speaking, the embedding of SymTFTs into the bulk works most transparently in situations with a semi-classical description. If the bulk theory is described by a perturbative string worldsheet, such as the tensionless limit of type IIB on AdS$_3$ discussed in Section \ref{subsec:noninver symentangle}, the SymTFT is better viewed as an organizational tool for symmetries of the CFT rather than as a literal target-space theory.} As a stand-alone quantum field theory, however, the SymTFT itself carries global symmetries. These must be either gauged or broken in the UV completion by gravity, so that no global symmetry survives in the full bulk theory. Often gauging alone cannot suffice due to the emergence of dual quantum symmetries. Ultimately one is forced to invoke additional breaking of such global symmetries in the bulk.

To be concrete, consider embedding a $(D+1)$-dimensional SymTFT localized near the AdS$_{D+1}$ conformal boundary (see Fig.~\ref{fig:adssliver}). For illustration, let us take the $\mathbb{Z}_N$ gauge theory
\begin{equation}
	S_{\text{TFT}}=\frac{N}{2\pi} \int_{D+1}b_{p+1}\wedge d c_{D-p}
\end{equation}
together with topological boundary conditions encode a $p$-form symmetry and/or a $(D-p-1)$-form symmetry of a $D$-dimensional QFT. The corresponding topological operators are
\begin{equation}
	U_m=e^{im\int_{\Sigma_{p+1}}b_{p+1}}, ~V_n=e^{in\int_{\Sigma_{D-p}}c_{D-p}},~m,n\in \{0,1,\cdots N-1 \}.
\end{equation}
As a $(D+1)$-dimensional QFT, this SymTFT therefore carries both a $(p+1)$-form $\mathbb{Z}_N^{(p+1)}$ symmetry and a $(D-p-1)$-form $\mathbb{Z}_N^{(D-p-1)}$ symmetry, with
\begin{equation}
	\begin{split}
		\mathbb{Z}_N^{(p+1)}&: \text{charged operators}~U_m,~\text{symmetry operators}~V_n,\\
		\mathbb{Z}_N^{(D-p-1)}&: \text{charged operators}~V_n,~\text{symmetry operators}~U_m	
	\end{split}.	
\end{equation}
For definiteness we focus on the $\mathbb{Z}_N^{(p+1)}$ symmetry and discuss its fate once the SymTFT is coupled to bulk gravity away from the conformal boundary. The case of $\mathbb{Z}_N^{(D-p-1)}$ symmetry works similarly.

\begin{figure}[h]
    \centering
    \includegraphics[width=0.6\linewidth]{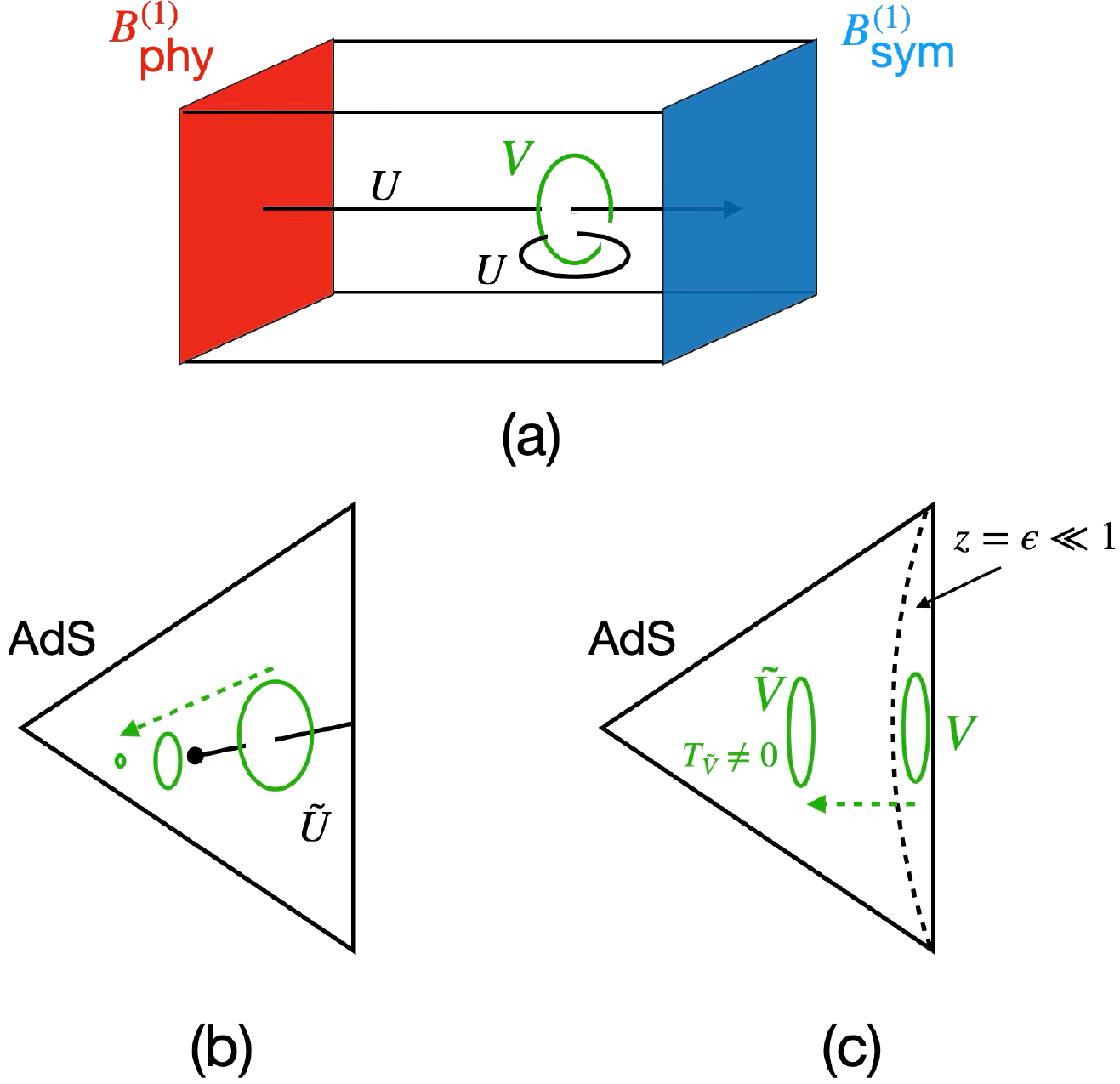}
    \caption{Two standard mechanisms by which bulk global symmetries fail in AdS/CFT.
(a) A SymTFT at the AdS boundary supports symmetry operators $V$ linking with charged operators $U$.
(b) If the bulk contains excitations (such as wrapped branes) that terminate the charged operator $\tilde U$, the linking defect $V$ becomes trivial.
(c) Even in the absence of such excitations, pulling the symmetry operator $V$ away from the boundary turns it into a tensionful object $\tilde V$, thereby destroying the bulk global symmetry.}
    \label{fig: symbreak standard}
\end{figure}

There are two complementary viewpoints for why this symmetry cannot persist deep in the AdS throat.
\begin{itemize}
	\item From a charged operator perspective, suppose the bulk contains a $p$-dimensional excitation on which the charged defect $U_m$ can end. In this case, the symmetry operator $V_n$, which previously linked with $U_m$, can now be freely deformed and shrunk to nothing due to its topological nature. See Figure \ref{fig: symbreak standard} (b) for an illustration.  If the mass of the $p$-dimensional object is $m_p$, the $\mathbb{Z}_N^{(p+1)}$ symmetry is effectively broken at a UV scale $\Lambda_{\text{UV}}\gtrsim m_p$. In string theory, such excitations are often provided by D$q$-branes wrapping $(q+1-p)$-cycles $\Sigma_{q+1-p}$, with a schematic mass
	\begin{equation}
		m_p \propto T_q \cdot \mathrm{vol}(\Sigma_{q+1-p})
		=\frac{1}{(2\pi)^q g_s (\alpha')^{(q+1)/2}}\cdot \mathrm{vol}(\Sigma_{q+1-p}).
	\end{equation}

	\item From the perspective of the $(D-p)$-dimensional symmetry operator itself, pulling it away from the conformal boundary into the bulk throat turns it into a tensionful object \cite{Apruzzi:2022rei, GarciaEtxebarria:2022vzq, Heckman:2022xgu, Heckman:2024oot}. See Figure \ref{fig: symbreak standard} (c) for a rough depiction. Denoting its tension by $T_{D-p-1}$, the symmetry is absent once the UV scale exceeds $\Lambda_{\text{UV}} \gtrsim T_{D-p-1}\cdot \mathrm{vol}(\Sigma_{D-p})$, where $\Sigma_{D-p}$ is the bulk cycle supporting the object. In string theory, this reflects the fact that symmetry operators at finite radial depth typically arise from tensionful ingredients such as wrapped D-branes or KK-monopoles.
\end{itemize}

\begin{figure}[t!]
    \centering
    \includegraphics[width=0.75\linewidth]{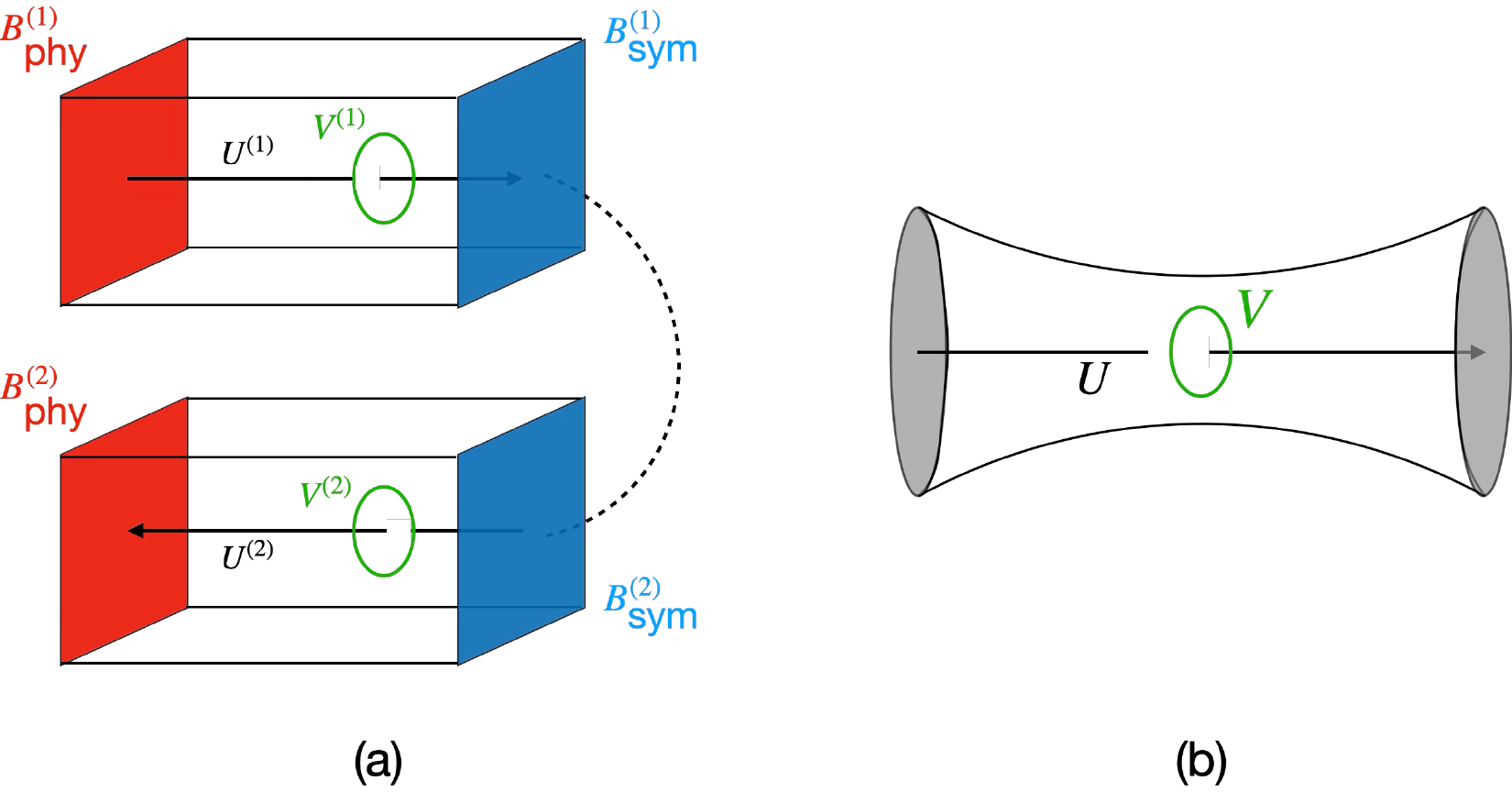}
    \caption{S-entanglement as a mechanism for bulk global symmetries.
(a) Two boundary theories $\mathcal{T}_D^{(1,2)}$ coupled via the S-entanglement admit teleportation of operators ($U^{(1)} \leftrightarrow U^{(2)}$, $V^{(1)} \leftrightarrow V^{(2)}$) across the entangled pair. The linking operator $V$ therefore persists without requiring tensionful bulk realizations.
(b) From the bulk perspective this corresponds to a defect threading a wormhole and linking the disconnected conformal boundaries of two asymptotic AdS regions, thereby generating a non-trivial topological sector that realizes a bulk global symmetry.}\label{fig:sympreserve entangle}
\end{figure}

The situation is qualitatively different in the entangled construction. Consider two $D$-dimensional theories $\mathcal{T}_D^{(1)}$ and $\mathcal{T}_D^{(2)}$ coupled through a common $(D+1)$-dimensional SymTFT, which creates an entangled state of the form illustrated in Fig.~\ref{fig:sympreserve entangle} (a).

In this setup an operator inserted in one sector can be transported through the SymTFT into the other sector, as shown in the figure by the teleportation of $U^{(1)}$ into $U^{(2)}$. The symmetry is therefore realized as a diagonal action across the two sectors rather than as an operator that must extend independently into the bulk. While the total bulk system is free of global symmetries, this mechanism allows the entangled pair to \textit{individually} avoid both of the obstacles to symmetry breaking that we discussed in the standard case: from the charged-operator viewpoint, a charged defect in one sector can be terminated by moving into the entangled partner rather than by the introduction of a bulk excitation; from the symmetry-operator viewpoint, the defect does not need to propagate into the bulk throat as a tensionful object, since its action can be represented in the partner sector.

Now suppose we trace out sector~2. The remaining state of sector~1 is described by the reduced density matrix
\begin{equation}
  \rho_1=\mathrm{Tr}_{2}\,|\Psi\rangle\langle\Psi| .
\end{equation}
As we discussed before, the resulting partition function reads
\begin{equation}
    Z^{(2)} = \frac{1}{N}\sum_{B_{p+1}}| Z(B_{p+1})|^2.
\end{equation}
which is an average over distinct background field sectors. This averaging indicates that the baby-universe Hilbert space is larger than one-dimensional (here $N$-dimensional). There hence are extra labels conserved, corresponding to unscreened charge in the bulk. As discussed in, e.g., \cite{McNamara:2020uza}, this inability of the bulk to interpolate between these superselection sectors is precisely the statement that there is a non-trivial cobordism class, leading to nontrivial topological operators generating global symmetries in the bulk \cite{McNamara:2019rup}.

From the bulk gravity point of view, the above process means that there exists a defect extending between the disconnected conformal boundaries of two asymptotic AdS regions of a wormhole geometry, as illustrated in Figure \ref{fig:sympreserve entangle} (b). The presence of this object implies that the bulk theory contains a non-trivial topological sector generated by operators linking with the defect. In contrast to the standard AdS/CFT case, where a symmetry operator pulled into the bulk necessarily becomes tensionful and thus destroys the global symmetry, here the linking operator is protected by the fact that the defect connects the two asymptotic boundaries. The entangled construction therefore provides a concrete mechanism by which bulk global symmetries can persist, realized as topological sectors supported by defects threading the wormhole.

It is also instructive to turn this logic around. Suppose instead that one knows enough about the UV complete gravitational theory to realize a bulk excitation on which $U$ terminates. In this case the linking defect $V$ necessarily trivializes: the putative bridge between the two sectors collapses, and the bulk global symmetry disappears. However, once the bridge is gone, the symmetry entanglement itself cannot be consistently defined, since there is no longer a defect that can mediate ensemble averaging across the two boundaries.

This is precisely in line with the general expectation that bulk global symmetries in quantum gravity can only arise through ensemble averaging rather than as exact symmetries of a single UV completion. The S-entanglement construction thus provides a concrete mechanism for how ensemble averaging can restore topological symmetry operators in the bulk.

\newpage

\section{Eternal AdS Black Holes}\label{sec:eternal}
We revisit Maldacena's proposal on the holographic description of eternal AdS black holes but now within the framework of SymTFT. In reviewing the original argument for this duality in \cite{Maldacena:2001kr}, we show that the usual entanglement present in the thermofield double state is independent from the amount of S-entanglement between the two CFT copies. In particular, the extent of the S-entanglement can be determined by various choices of boundary conditions involved for the SymTFT when the CFT$_D$ worldvolume is taken to be the (Euclidean signature) manifold $[0,\beta/2]\times S^{D-1}$. As we review below, the importance of the CFT$_D$ on this Euclidean manifold stems from the fact that it appears in an intermediate step in the duality argument of \cite{Maldacena:2001kr}.

We then show how the splittability property of objects threading the non-transversable wormhole, as conjectured in \cite{Harlow:2015lma}, can be violated when the S-entanglement is non-trivial. We explain how this is due to a lack of factorization of the total CFT data, such as certain spaces of genuine defect operators as well as the Hilbert itself, into the left and right copies. We then revisit the work of Marolf and Wall \cite{Marolf:2012xe} which, motivated by certain Alice/Bob experiments, proposed that eternal AdS black hole Hilbert spaces should take the form $\mathcal{H}_L\otimes \mathcal{H}_R\otimes V$. Here $\mathcal{H}_{L,R}$ are the Hilbert spaces for the left/right copies of the CFT$_D$ and $V$ is an extra ``superselection" sector which determines the outcome of the Alice/Bob experiments. In comparing to our work, we find that $V$ plays a similar role to the tensor product of the SymTFT Hilbert space, $\mathcal{H}_{\mathcal{S}}(M_{D})^{\otimes 2}$. We interpret entangled states in this Hilbert space as implying that the black hole has a single associated gauge charge, as opposed to a separate ones for the left and right boundaries (Statement 2 from the Introduction). Finally, we sketch how we expect these considerations to generalize beyond discrete bulk gauge symmetries.

\subsection{Maldacena's Argument Revisited}\label{ssec:IAS}

\begin{figure}
    \centering
\includegraphics[trim={0 3cm 0 2cm},clip,width=12cm,scale=0.8]{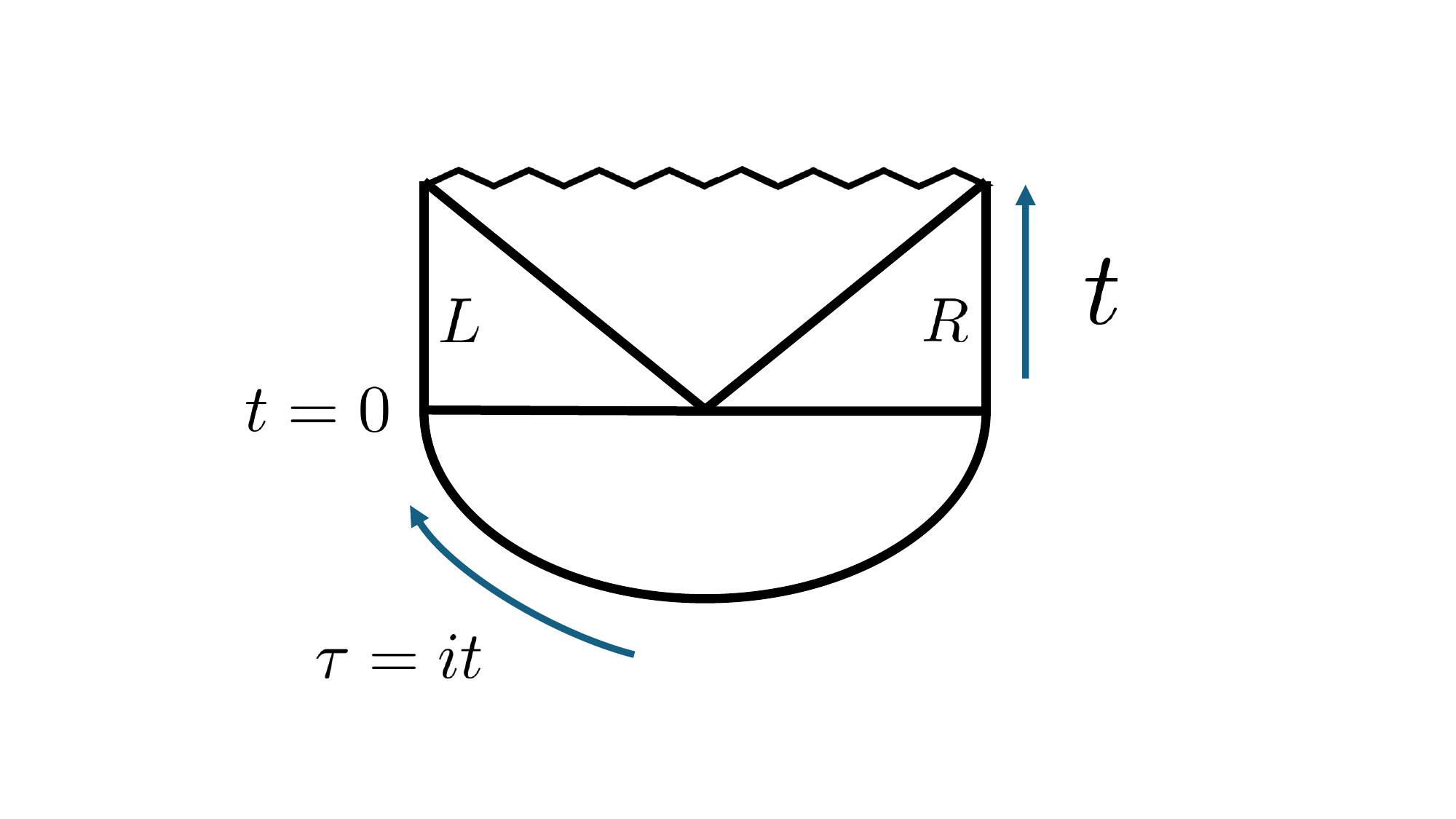}
    \caption{Hartle-Hawking preparation of the eternal AdS black hole for $t>0$. The top half of the figure is the Penrose diagram for the Lorentzian signature spacetime while the bottom half represents the $(\tau,r)$ coordinates of the Euclidean AdS black hole.}
    \label{fig:eternalBH}
\end{figure}

Given a CFT$_D$, $\mathcal{T}$, which we assume for the moment to be an \textit{absolute CFT}, dual to some QG theory on $\mathrm{AdS}_{D+1}\times M$, the main claim of \cite{Maldacena:2001kr} is that eternal (two-sided) black hole solutions of the low-energy gravitational theory are dual to two copies of $\mathcal{T}$ in the thermofield double state
\begin{equation}\label{eq:tfd}
    \ket{TFD}:=\frac{1}{Z_{\mathcal{T}}(\beta)}\sum_{n}e^{-\beta E_{n}/2} \ket{n}_{L}\otimes \ket{n}_R, \quad \ket{TFD}\in \mathcal{H}^{(L)}_{\mathcal{T}}\otimes \mathcal{H}^{(R)}_{\mathcal{T}}.
\end{equation}
Here $\mathcal{H}^{(L,R)}_{\mathcal{T}}$ are two identical copies of the Hilbert space of $\mathcal{T}$ on $S^{D-1}$ with ``left" and ``right" labels, $n$ runs over energy eigenstates, and $Z_{\mathcal{T}}(\beta)\equiv Z_{\mathcal{T}}(S^1_{\beta}\times S^{D-1})$. From the bulk perspective, the intuitive reason this state exists in the tensor product is because there are two asymptotically AdS boundaries (see Figure \ref{fig:eternalBH}) and performing a partial trace over, say $\mathcal{H}^{(R)}_{\mathcal{T}}$, reproduces the thermal density matrix $\rho=e^{-\beta H}$ which has the expected entropy for a large black hole with temperature $\beta^{-1}$ in an AdS space with a single asymptotic region.

The argument for this equivalence in \cite{Maldacena:2001kr} proceeds as follows. First consider the Euclidean $\mathrm{AdS}_{D+1}$-Schwarzschild black hole solution:
\begin{align}
    ds^2&=Fd\tau^2+F^{-1}dr^2+r^2d\Omega_{D-2}\label{eq:adsblackshield} \\
     F&:=1+\frac{r^2}{L^2}-\frac{\mu}{r^2}.
\end{align}
In \eqref{eq:adsblackshield}, the Euclidean time coordinate is periodic $\tau\sim \tau+\beta$, $\mu=\mu(\beta)$ parametrizes the mass of the black hole\footnote{In terms of the horizon length, $r^2_+=\frac{L^2}{2}(-1+\sqrt{1+4\mu/L^2})$, $\mu$ can be expressed in terms of the inverse temperature from the relation $\beta=\frac{2\pi L^2r_+}{2r^2_++L^2}$. Note that we are considering ``large" AdS black holes which have positive specific heat and satisfy $\sqrt{2}r_+>L$.}, $L$ is the AdS radius, and $d\Omega_{D-2}$ is the volume form for a unit $(D-2)$-sphere. Notice that the $L\rightarrow \infty$ limit reproduces the flatspace Schwarzschild black hole solution.


Next, consider the restriction of the Euclidean time coordinate to $\tau\in [0,\beta/2]$, then the asymptotic boundary of the Euclidean bulk spacetime is $I_{\beta/2}\times S^{D-1}$. From holography we expect this to be dual to a CFT on $I_{\beta/2}\times S^{D-1}$, but since $I_{\beta/2}\times S^{D-1}$ itself has a boundary, this procedure prepares the system in some particular state $\ket{\Psi}$. This state is in the vector space $\mathcal{H}^{(L)}_{\mathcal{T}}\otimes \mathcal{H}^{(R)}_{\mathcal{T}}$ whose two factors are due to the fact that $I_{\beta/2}\times S^{D-1}$ has two disconnected boundaries. With such a state in hand, we can evolve this system in Lorentzian time as in Figure \ref{fig:eternalBH}. In other words, this procedure gives a Hartle-Hawking preparation for the state $\ket{\Psi}$.

\begin{figure}[t]
    \centering
\includegraphics[trim={0 1cm 0 2cm},clip,width=12cm,scale=0.8]{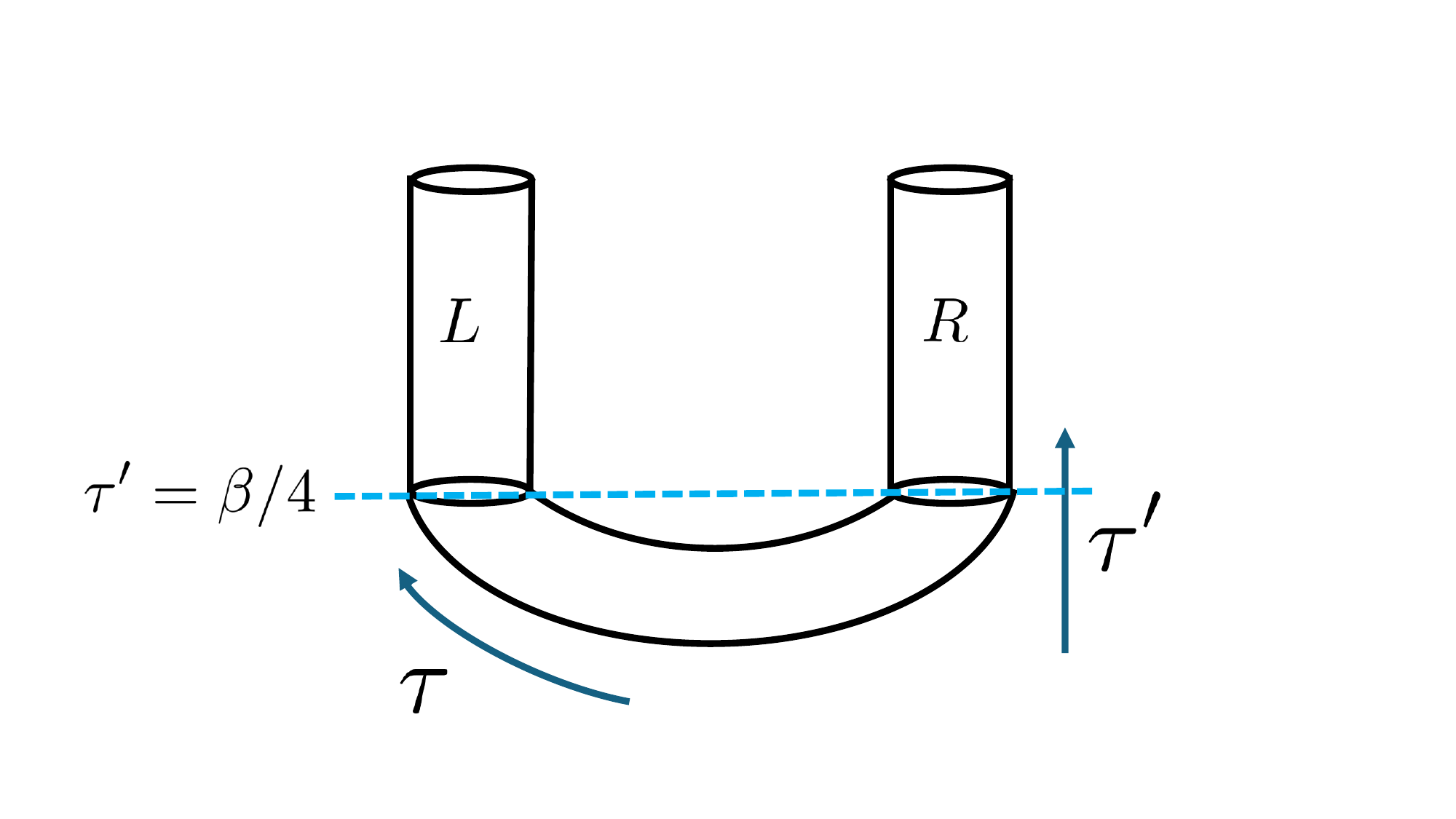}
    \caption{For a CFT on $I_{\beta/2}\times S^{D-1}$ one can consider the Euclidean time direction $\tau$ along $I_{\beta/2}$, or $\tau^\prime\equiv |\tau-\beta/4|$.}
    \label{fig:twoquantizations}
\end{figure}

\begin{figure}[h]
    \centering
\includegraphics[trim={0 2cm 0 1cm},clip,width=12cm]{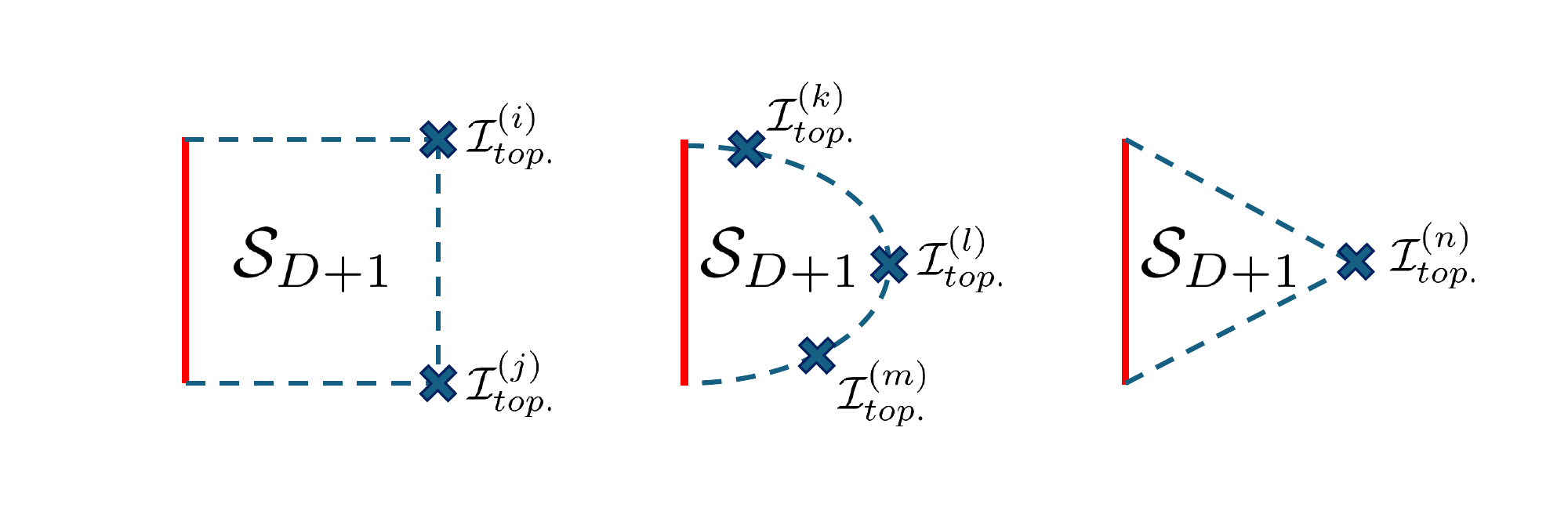}
    \caption{Several possible SymTFT configurations for a CFT on $I_{\beta/2}\times S^{D-1}$. The worldvolume of this CFT is depicted by the red interval. The blue dotted lines denotes topological boundaries for $\mathcal{S}_{D+1}$, while the blue crosses are topological interfaces.}
    \label{fig:plural}
\end{figure}

One can see that $\ket{\Psi}=\ket{TFD}$ as defined in \eqref{eq:tfd} from quantizing the Euclidean CFT on $I_{\beta/2}\times S^{D-1}$ in two different ways as illustrated in Figure \ref{fig:plural}. Quantizing along the $\tau$ direction means evolving states along the length of the cylinder, while quantizing along the $\tau^\prime\equiv |\tau-\beta/4|$ direction evolves states away from the middle of the cylinder. This implies that the inner produce of $\ket{\Psi}$ with some state $\ket{v_L}\otimes \ket{v_R}\in \mathcal{H}^{(L)}_{\mathcal{T}}\otimes \mathcal{H}^{(R)}_{\mathcal{T}}$ is given as
\begin{equation}\label{eq:psitfd}
    \big( \langle v_L|\otimes \bra{v_R}\big) \ket{\Psi} = \bra{v_L}e^{-\beta H/2}\ket{v_R^*}=\sum_{n}e^{-\beta E_n/2}\langle v_L | n\rangle \langle v_R | n\rangle
\end{equation}
where in the first equality follows from the equivalence of quantization directions in Figure \ref{fig:twoquantizations} and the second follows from inserting $1=\sum_{n}\ket{n}\bra{n}$. Equation \eqref{eq:psitfd} then exactly matches the matrix elements of $\ket{TFD}$.

Let us now replay this story while taking into account the SymTFT construction of $\mathcal{T}$ on $I_{\beta/2}\times S^{D-1}$. As pointed out in \cite{Cvetic:2024dzu}, there is a plurality of ways one can build a $(D+1)$-dimensional SymTFT sandwich when the $D$-dimensional field theory of interest is placed on a manifold with boundary\footnote{In \cite{Cvetic:2024dzu}, the suggested term for these general constructions was \textit{cheesesteaks} (a logical generalization of sandwiches). See also \cite{GarciaEtxebarria:2024jfv, Cordova:2024iti, Choi:2024tri, Bhardwaj:2024igy, Das:2024qdx} for recent works which also generalize the SymTFT construction to QFTs with boundaries.}. See Figure \ref{fig:plural} for a sampling of such possibilities which differ by the insertion of topological interfaces localized on the $D$-dimensional topological boundary of the SymTFT.
We define the corners where the topological boundaries and the CFT worldvolume meet (i.e. where the dotted blue lines hits the red intervals in Figure \ref{fig:plural}) in the simplest manner possible by just imposing the relevant boundary condition for the field shared between the CFT and the SymTFT, which are always discrete $p$-form gauge fields in our cases, at $\tau=0$ and $\tau=\beta/2$. Different choices of boundary conditions at corners can be obtained by placing different topological interfaces there. Furthermore, from the fact that $\mathcal{S}_{D+1}$ is topological, each of the SymTFT sandwiches in Figure \ref{fig:plural} can be deformed into each other. Therefore, one can simply restrict to the case of having only a single topological interface without loss of generality.

The SymTFT setup relevant for eternal AdS black holes then crucially relies on a choice of a topological interface which we label by $i$. These can be defined by matrices
\begin{equation}\label{eq:interfacedata2}
    M^{(\mathcal{B}_i)}\in \mathrm{Hom}(\mathcal{H}_{\mathcal{S}}(M^-_D), \mathcal{H}_{\mathcal{S}}(M^+_D)).
\end{equation}
where $M^+_D\cup M^-_D=M_D$ and $\partial M^{\pm}_D=M_{D-1}$ is the worldvolume of the interface. In other words, they transform SymTFT boundary states below the interface to SymTFT boundary states above the interface. Fusing topological interfaces then just amounts to finite dimensional matrix multiplication. This presentation is most natural if we quantize the SymTFT vertically in Figure \ref{fig:plural}, while if one quantizes in a radial direction away from the interface, then one can equivalently present it as a vector
\begin{equation}\label{eq:interfacedata}
    \ket{\mathcal{B}_i}\in \mathcal{H}_{\mathcal{S}}(M_D)^{\otimes 2}
\end{equation}
This is reminiscent of the fact that local operators on boundaries of string worldsheets are labeled by two Chan-Paton factors and follows the (extended) Atiyah-Segal axioms\footnote{For an introduction to extended TFTs see Section 1 of \cite{lurie2008classification}.} \cite{atiyah1988topological, segal1988definition}. One can go between these two descriptions by changing the basis vectors $\ket{I}_{L}\otimes \ket{J}_R\in \mathcal{H}_{\mathcal{S}}(M_D)^{\otimes 2}$ to basis matrices $\ket{I}_L\bra{J}_R\in \mathrm{Hom}(\mathcal{H}_{\mathcal{S}}(M^-_D), \mathcal{H}_{\mathcal{S}}(M^+_D))$.

Now, due to the topological nature of the SymTFT we can redraw the right-most diagram of Figure \ref{fig:plural}, into the suggestive form:
\begin{figure}[h]
    \centering
\includegraphics[trim={0 1cm 0 1cm},clip,width=12cm]{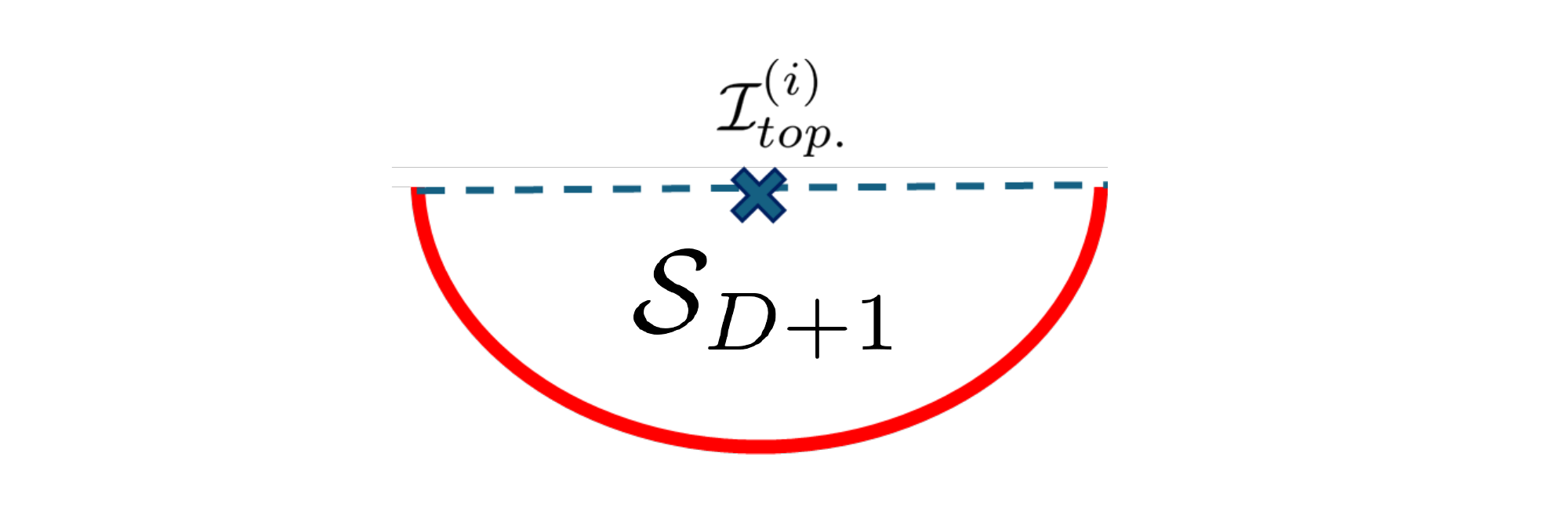}
    \caption{}
    \label{fig:sugform}
\end{figure}

\begin{figure}[t]
    \centering
\includegraphics[trim={0cm 2cm 0cm 1cm},clip,width=15cm,scale=2]{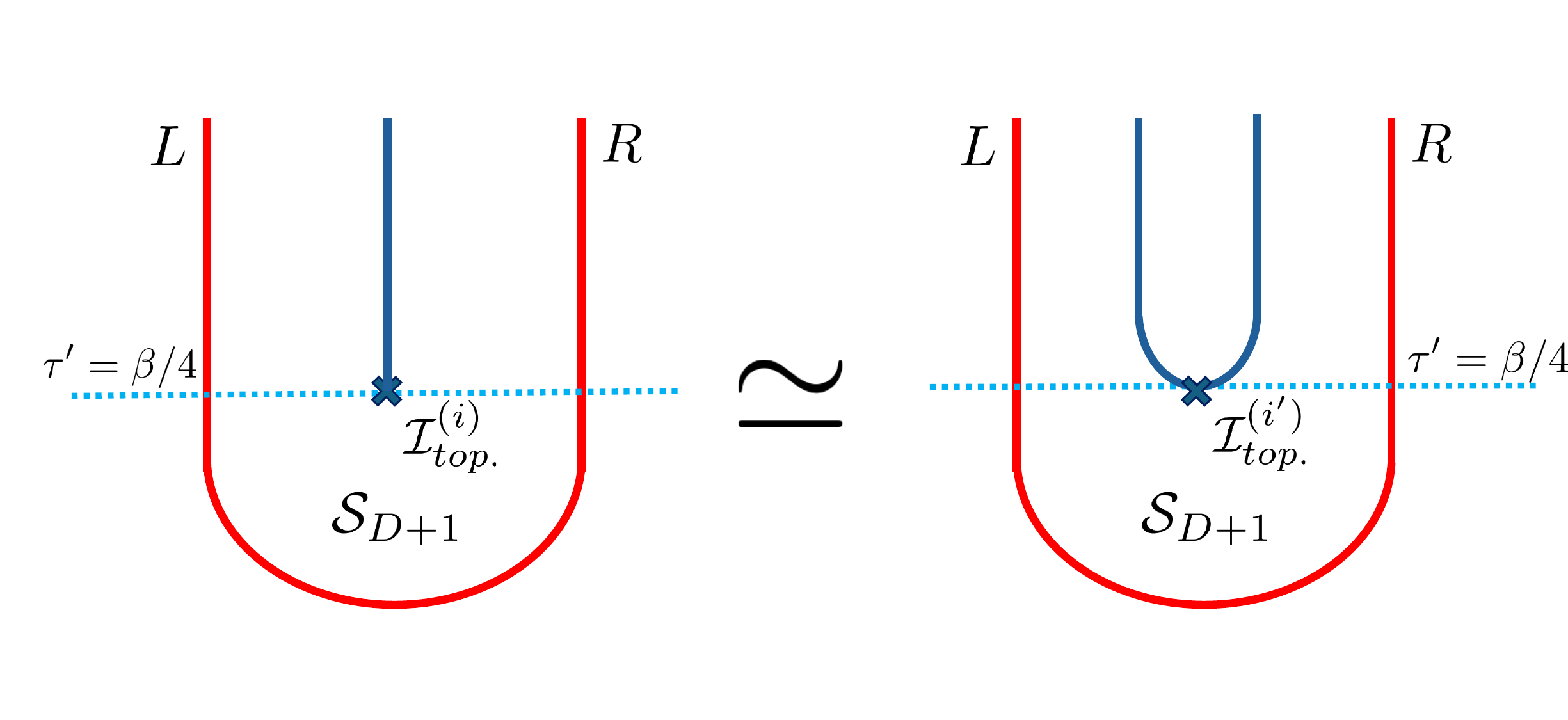}
    \caption{Two equivalent ways of (Lorentzian or Euclidean) time evolving the CFT-SymTFT system in Figure \ref{fig:sugform} beyond $\tau^\prime=\beta/4$. One the left, we evolve with a single SymTFT worldvolume for each time slice, albeit with an interface. While on the right, we evolve with two disconnected SymTFT worldvolumes with two (possibly entangled) topological boundaries. Note that the interface can change between the two presentations given, we focus on the left one in the text.}
    \label{fig:intervalsand}
\end{figure}

\noindent The cylinder endpoints are located at $\tau^\prime=\beta/4$, where from this location we can choose to time evolve the system into Lorentzian signature to produce to the CFT dual to the eternal AdS black hole, or choose to time evolve into Euclidean signature. Either way, the time evolution, as illustrated in Figure \ref{fig:intervalsand}, recovers the main picture of this paper where we have two CFTs whose SymTFT topological boundaries are specified by a state in $\mathcal{H}_{S}^{\otimes 2}$. Indeed, above some Euclidean time slice $\tau^\prime\geq \beta/4$ this setup is completely identical that of previous sections! As mentioned in the Introduction and shown in Section \ref{sec:SymTFTentanglement}, we have an equivalence between presenting the S-entanglement as a single connected SymTFT worldvolume (possibly with an interface) and a disconnected worldvolume (possibly with S-entanglement). This equivalence in the current setup is also illustrated in Figure \ref{fig:intervalsand}.

As we've shown in Section \ref{sec:SymTFTentanglement}, the SymTFT vector $\ket{\mathcal{B}}$ labels different absolute theories, it accordingly labels different Hilbert spaces and algebras of defect operators, e.g.
\begin{equation}
    \mathcal{H}_{\ket{\mathcal{B}}}, \; \;  \mathcal{V}^{(\textnormal{gen. line defects})}_{\ket{\mathcal{B}}}, \; \;  \mathcal{V}^{(\textnormal{non-gen. surface defects})}_{\ket{\mathcal{B}}}, \; \;  \mathrm{etc.}
\end{equation}
In the current context, this now includes a labeling on the thermofield double state $\ket{TFD,\mathcal{B}}\in \mathcal{H}_{\ket{\mathcal{B}}}$ whose form will be given in some examples below. In the case of a product SymTFT state $\ket{\mathcal{B}}=\ket{I}_L\otimes \ket{J}_R$ then we have that
\begin{equation}
    \mathcal{H}_{\ket{\mathcal{B}}}=\mathcal{H}^{(L)}_{\mathcal{T}_I}\otimes \mathcal{H}^{(R)}_{\mathcal{T}_J}
\end{equation}
where $\mathcal{H}_{\mathcal{T}_{I}}$ is the Hilbert space for the absolute CFTs $\mathcal{T}_{I}$ (and similarly for the polarization $J$). The algebras of defect operators also factorize for such a SymTFT state.



As we know from Section \ref{sec:SymTFTentanglement}, if $\ket{\mathcal{B}}$ implies a S-entanglement of $p$-form symmetries then the algebra of $p$-dimensional genuine defect operators of the combined system does not factorize
\begin{equation}
    \mathcal{V}^{(\textnormal{gen. $p$-defects})}_{\ket{\mathcal{B}}}\neq \mathcal{V}^{(\textnormal{gen. $p$-defects})}_{L}\otimes \mathcal{V}^{(\textnormal{gen. $p$-defects})}_{R}.
\end{equation}
For $p=0$, these are algebras of local operators which, by the CFT state-operator correspondence, is equivalent to the Hilbert space which implies that if $\ket{\mathcal{B}}$ displays 0-form S-entanglement, then the Hilbert space does not factorize
\begin{equation}\label{eq:Hspacenotfactorizing}
     \mathcal{H}_{\ket{\mathcal{B}}}\neq \mathcal{H}^{(L)}_{\mathcal{T}_I}\otimes \mathcal{H}^{(R)}_{\mathcal{T}_J}
\end{equation}
for some polarizations $I$ and $J$ for $\mathcal{T}$. Perhaps the simplest such case is when the interface $\mathcal{I}^{(i)}_{top.}$ is just the trivial identity interface: $\propto\sum_I\ket{I}_L\bra{I}_R$. In this case $\ket{\mathcal{B}}$ is just given by the maximally entangled state $\propto \sum_I \ket{I}_L\otimes \ket{I}_R$. This agrees with the intuitive fact that the left and right CFTs should have maximal S-entanglement if the left side of Figure \ref{fig:intervalsand} has no topological interface in the SymTFT sandwich.

While the thermofield double state, as traditionally defined in \eqref{eq:tfd}, is a state in the product of the left and right CFT Hilbert spaces, we now address how to construct such states when the Hilbert space of the AdS black hole does not factorize. For the sake of concreteness, let us assume that the SymTFT includes the $\mathbb{Z}_N$-valued gauge fields $b_1$ and $c_{D-1}$ and the usual BF term
\begin{equation}
   \frac{2\pi}{N}\int b_{1}\delta c_{D-1}  \subset S_{\mathcal{S}_{D+1}}.
\end{equation}
The topological Wilson line/surface operators for this SymTFT are
\begin{equation}
    U=\exp \left( \frac{2\pi i}{N} \int_{\sigma_1}b_1 \right), \quad V=\exp \left( \frac{2\pi i}{N} \int_{\sigma_{D-1}}c_{D-1} \right).
\end{equation}
Denote the boundary theory as $\mathcal{T}$, then recall that in the usual SymTFT sandwich construction it has a $\mathbb{Z}^{(0)}_N$ global symmetry if $b_1$ (resp. $c_{2}$) has
Dirichlet (resp. Neumann) boundary conditions along the topological boundary, and $V$ is the 0-form
topological symmetry operator. We now consider the $\mathcal{T}$ theory in the setup of Figure \ref{fig:intervalsand} to understand what the TFD states are when the topological boundary condition is entangled or not. The non-entangled state $\ket{00}$ is equivalent to considering the theory with the
polarization such that $\mathcal{T}_L$ and $\mathcal{T}_R$ each have their own  $\mathbb{Z}^{(0)}_N$ global symmetry and we do not have a background field turned on. Meanwhile the entangled state $\ket{\mathcal{B}}=\frac{1}{N}(\sum^{N-1}_{i=0}\ket{ii})$
is the polarization where the diagonal subgroup $(\mathbb{Z}_N)_{\text{diag}}\subset (\mathbb{Z}_N)_L\times (\mathbb{Z}_N)_R$ generated by $(+1,+1)$ is gauged. This means that the usual
thermofield double state ($\ket{TFD,00}$ in this notation) is transformed to one where we
gauge $(\mathbb{Z}_N)_{\text{diag}}$:
\begin{equation}
    \ket{TFD,00}\rightarrow \ket{TFD,\mathcal{B}}=\frac{1}{N}(\sum^{N-1}_{k=0}V^{k}_LV^{k}_R)\ket{TFD,00}.
\end{equation}
More explicitly, if we keep track of the $\mathbb{Z}_N$ charges $q_n$, we can write the usual thermofield double state \ref{eq:tfd} as
\begin{equation}
    \ket{TFD,00}=\frac{1}{Z_{\mathcal{T}}(\beta)}\sum_{n}e^{-\beta E_{n}/2} \ket{n,q_n}_{L}\otimes \ket{n,q_n}_R.
\end{equation}
One can see why the $\mathbb{Z}_N$ charges are the same for both the left and right states in the tensor product by returning to the derivation around \eqref{eq:psitfd} and deriving a selection rule:
\begin{align}
    \bra{v_L,q_L}e^{-\beta H/2}\ket{v^*_R,-q_R}&=e^{+2\pi i q_R/N}\bra{v_L,q_L}e^{-\beta H/2}V\ket{v^*_R,-q_R}\\
    &=e^{+2\pi i (q_R-q_L)/N}\bra{v_L,q_L}e^{-\beta H/2}\ket{v^*_R,-q_R} \\
    &\implies q_L\equiv q_R \; \mathrm{mod} \; N
\end{align}
Because the projection operator from the diagonal gauging, $\frac{1}{N}(\sum^{N-1}_{k=0}V^{k}_LV^{k}_R)$, requires the $\mathbb{Z}_N$ charges satisfy $q^{(L)}_n=-q^{(R)}_n$, only the zero-charge states remain
\begin{equation}
    \ket{TFD,\mathcal{B}}=\frac{1}{N}(\sum^{N-1}_{k=0}V^{k}_LV^{k}_R)\ket{TFD,00}=\frac{1}{Z_{\mathcal{T}}(\beta)}\sum_{\{n \;| \; q_n=0\}}e^{-\beta E_{n}/2} \ket{n,0}_{L}\otimes \ket{n,0}_R.
\end{equation}
This state is in the Hilbert space defined by projecting onto the invariant states of $\mathcal{H}^{(L)}_{\ket{0}}\otimes \mathcal{H}^{(R)}_{\ket{0}}$,
\begin{equation}
    \mathcal{H}_{\ket{\mathcal{B}}}=(\mathcal{H}^{(L)}_{\ket{0}}\otimes \mathcal{H}^{(R)}_{\ket{0}})/(\mathbb{Z}_N)_{\text{diag}},
\end{equation}
and the left- and right- $\mathbb{Z}^{(0)}_N$ symmetry operators are identified as
\begin{equation}
     V:=V_{L}\simeq  V^{-1}_R
\end{equation}
which is the symmetry operator for the remaining $\mathbb{Z}_N=[(\mathbb{Z}_N)_L\times (\mathbb{Z}_N)_R]/(\mathbb{Z}_N)_{\text{diag}}$ 0-form symmetry. Additionally, this gauging in the CFT implies that every genuine local operator will take the form
\begin{equation}
     \prod_i \mathcal{O}^L_i \prod_j \mathcal{O}^R_j, \; \textnormal{such that} \; \sum_i q^{(L)}_i(\mathcal{O}^L_i)=-\sum_j q^{(R)}_j(\mathcal{O}^R_j)
\end{equation}
which is the same answer one would get from applying the state-operator correspondence to $\mathcal{H}_{\ket{\mathcal{B}}}$. We can consider a non-trivial background field labeled by $a \; \mathrm{mod} \; N$ for the remaining $\mathbb{Z}^{(0)}_N$ global symmetry by taking the SymTFT state to be $\ket{\mathcal{B}}=\frac{1}{N}(\sum^{N-1}_{k=0}e^{2\pi i k a/N}\ket{kk})$ which correspondingly changes the projection operator to $\frac{1}{N}(\sum^{N-1}_{i=0}e^{2\pi i k a/N}V^k_LV^k_R)$. The thermofield double state in this case will be
\begin{equation}
\ket{TFD,\mathcal{B}}=\frac{1}{N}(\sum^{N-1}_{k=0}V^{k}_LV^{k}_R)\ket{TFD,00}=\frac{1}{Z_{\mathcal{T}}(\beta)}\sum_{\{n \; | \; q_n=a\}}e^{-\beta E_{n}/2} \ket{n,a}_{L}\otimes \ket{n,a}_R
\end{equation}
Additionally, notice that it was important that in this example the CFT is of dimension $D\neq 2$ because 0-form gauging because we would have to include twisted sector states (as the reader may be familiar with from string worldsheet orbifolding) which makes obtaining the SymTFT-entangled thermofield double states more complicated. We address this technicality in Appendix \ref{app:5point1extended}.


Alternatively, we can choose a SymTFT state $\ket{\mathcal{B}}$ that amounts to gauging by the ``anti-diagonal" subgroup of $(\mathbb{Z}_N)_L\times (\mathbb{Z}_N)_R$ generated by $(+1,-1)$. In this case, the thermofield double state is invariant under the projection operator $\frac{1}{N}(\sum^{N-1}_{k=0}V^{k}_LV^{-k}_R)$, while the Hilbert space will still be non-factorizable into left and right-pieces.

In summary, the analysis of this section shows that the definition of gauge charges of the eternal AdS black hole crucially relies on the degree of S-entanglement. This relation boils down to the following two points:
\begin{itemize}
    \item \textit{If an eternal AdS black hole has a single well-defined $\mathbb{Z}_N$ 0-form gauge charge under which objects in both causal regions outside of the black hole are charged, then $\ket{\mathcal{B}}$ must have a non-trivial S-entanglement.}
    \item \textit{If $\ket{\mathcal{B}}$ does not have S-entanglement, then the eternal AdS black hole has a separate $\mathbb{Z}_N$ gauge charges for each of its (left and right) horizons.}
\end{itemize}


\subsection{On Harlow's Splittability Argument}\label{ssec:MIT}

\begin{figure}[t]
    \centering
\includegraphics[trim={0cm 4cm 0cm 3cm},clip,width=15cm,scale=2]{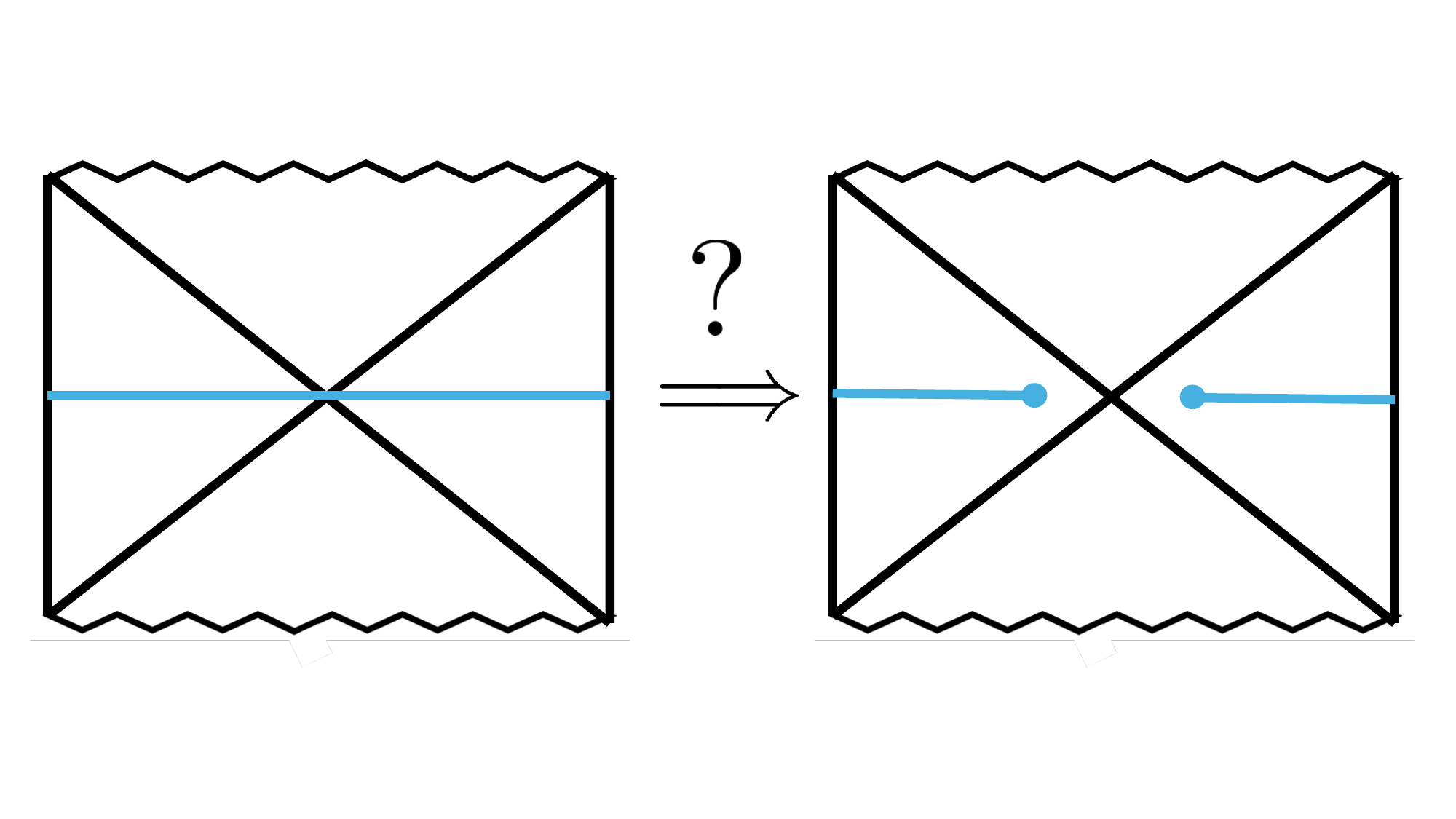}
    \caption{In blue, we have an object whose spatial worldvolume goes through the AdS black hole wormhole. Is this object necessarily splittable in the sense of the right-hand part of the figure?}
    \label{fig:2splitornot2split}
\end{figure}

We now comment on the ramifications of the previous section on the claims of \cite{Harlow:2015lma}\footnote{See also \cite{Guica:2015zpf} for a similar study of wormhole-threading bulk Wilson line operators.} that all objects threading the wormhole in these bulk geometries are splittable, see Figure \ref{fig:2splitornot2split}.


Briefly reviewing the physical reasoning \cite{Harlow:2015lma}, consider a Wilson line for a bulk $U(1)^{(0)}$ gauge theory with some charge $+q$ threading the wormhole as in the left side of Figure \ref{fig:2splitornot2split}. Such an insertion of a bulk Wilson line corresponds to some operator in quantum gravity theory on AdS, so assuming the factorization of the Hilbert space (as well as the algebra of all operators) into left and right pieces seems to imply that such an operator can be expressed in terms of a product of operators localized on the left and right exteriors of the black hole. This is further corroborated by the fact that the $U(1)^{(0)}$ gauge theory should have a complete spectrum of electrically charged particle states \cite{Polchinski:2003bq}. In particular, the creation operator for a particle states with electric charges $\pm q$ can furnish endpoints for the charge $+q$ which can cause it to split as in the right side of Figure \ref{fig:2splitornot2split}.

\begin{figure}
    \centering
\includegraphics[trim={0cm 2cm 0cm 0cm},clip,width=15cm,scale=2]{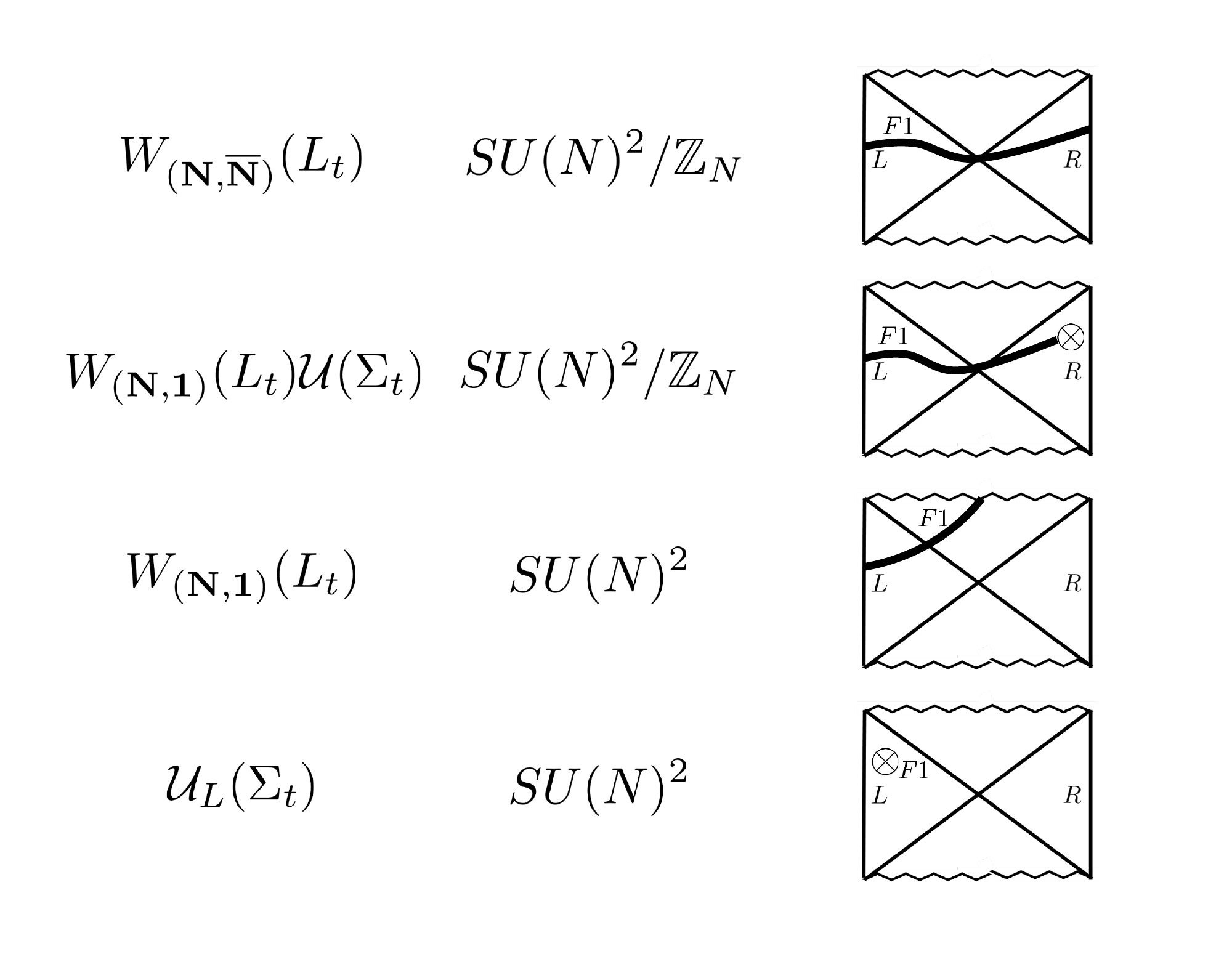}
    \caption{We give a sampling of objects (left column) involving Wilson lines/1-form symmetry operators present for 4D SYM theory with corresponding gauge group (middle column) and their bulk interpretations (right column). Note $\mathbb{Z}_N=\mathbb{Z}_N(1,-1)$ here. In the second row, $\Sigma_t$ is a timelike surface in the 4D gauge theory such that $\partial \Sigma_t=L_t$ and the symbol $\otimes$ means that the $F1$ string spatial worldvolume goes into non-radial spatial direction. In the third column, we interpret the Wilson line in the CFT as an $F1$ string which ends on the black hole. }
    \label{fig:3columns}
\end{figure}

Given that the presence of S-entanglement can cause a lack of factorization, let us understand how this alters the above conclusions. For concreteness our focus will be on the 4D $\mathcal{N}=4$ CFT case with maximal S-entanglement being between 1-form symmetries. In particular we take the gauge group to be $(SU(N)\times SU(N))/\mathbb{Z}_N$. We then place a timelike Wilson line in the representation $(\mathbf{N},\overline{\mathbf{N}})$ at $\mathbb{R}_t\times \{p_N\}\subset \mathbb{R}_t\times S^3$ where $p_N\in S^3$ denotes the north pole. From our holographic dictionary, this operator creates two copies of the $F1$-string configuration as in Figure \ref{fig:allowedbanned}, one for the left and right black hole exteriors which in the Penrose diagram appears as in the top line of \ref{fig:3columns}. These two sides of the $F1$ string are then connected in the bulk and charged under a common $(\mathbb{Z}_N)^{(1)}$ gauge group. This object is \textit{non-splittable} in the sense that the $(\mathbb{Z}_N)^{(1)}$ gauge charge in the left region must cancel the $(\mathbb{Z}_N)^{(1)}$ gauge charge in the right region. This charge cancellation is not possible in the imagined scenario in the right side of Figure \ref{fig:2splitornot2split} since if the configuration of an $F1$ string ending is valid in each separate left/right patch, then one is free to add multiple copies of this configuration on one side.

A subtlety of this construction is that such a Wilson line configuration will violate Gauss' law for the gauge theory since $S^3$ is closed, so we must include for instance a timelike Wilson line in the representation $(\overline{\mathbf{N}},\mathbf{N})$ at the south pole $p_S\in S^3$. This creates a wormhole threading $F1$ string with opposite orientation which can annihilate the $F1$ string at the north pole. This decay process is suppressed in the limit $\beta\rightarrow \infty$. For the sake of completeness we also detail in Figure \ref{fig:3columns}, interpretations for various Wilson lines/topological operator configurations when the gauge group is $SU(N)^2$ or $SU(N)^2/\mathbb{Z}_N$.

Notice that the key difference between the $U(1)$ Wilson line example of \cite{Harlow:2015lma} and our example is that in our case, the 1-branes threading the wormhole have a bulk gauge charge. The Wilson lines considered in \cite{Harlow:2015lma} are charged under an emergent $U(1)^{(1)}$ global symmetry at low energies which is broken at high energies due to the tower of electrically charge particles states required for the electric charge completeness for the $U(1)^{(0)}$ gauge symmetry. While this means that the non-splittable example of this section does not directly contradict the splittable example of \cite{Harlow:2015lma} (or Charge Completeness in general), it does illustrate how the S-entanglement can lead to configurations not considered in \cite{Harlow:2015lma} because this reference assumed factorizability.

Finally, note that for large finite $\beta$, the configuration of $F1$/anti-$F1$ strings mentioned above has long but finite lifetime so is ``splittable" in the sense that it can decay. However, the fact that this lifetime can be made arbitrarily large by increasing $\beta$ is in stark contrast with the $U(1)$ Wilson line example of \cite{Harlow:2015lma} because the lifetime for the Wilson line source to decay is proportional to the mass of the lightest $U(1)^{(0)}$ charged particle which is expected to be bounded above by the Planck mass \cite{Cordova:2022rer}.

\subsection{Comments on Marolf/Wall's ``Superselection Sectors"}\label{ssec:UCSBCambridge}

We now use the proposal of Section \ref{ssec:IAS} to understand a puzzle posed by Marolf and Wall in \cite{Marolf:2012xe} on the apparent inadequacy of the thermofield double state in describing the physics of eternal AdS black holes. We then compare our answer with their conjectured solution which involved what they referred to as ``superselection sectors" for the eternal AdS black hole Hilbert space.\footnote{Note that the reference \cite{Jafferis:2020ora} claims that some of the conceptual issues brought up by Marolf and Wall can be addressed in their formulation of bulk observers in AdS. Since we do not see a clear relation between their claims and our statements on the black hole charges, we leave an understanding of how our two works relate for the future.}

\begin{figure}
    \centering
\includegraphics[trim={0cm 4cm 0cm 3cm},clip,width=15cm,scale=2]{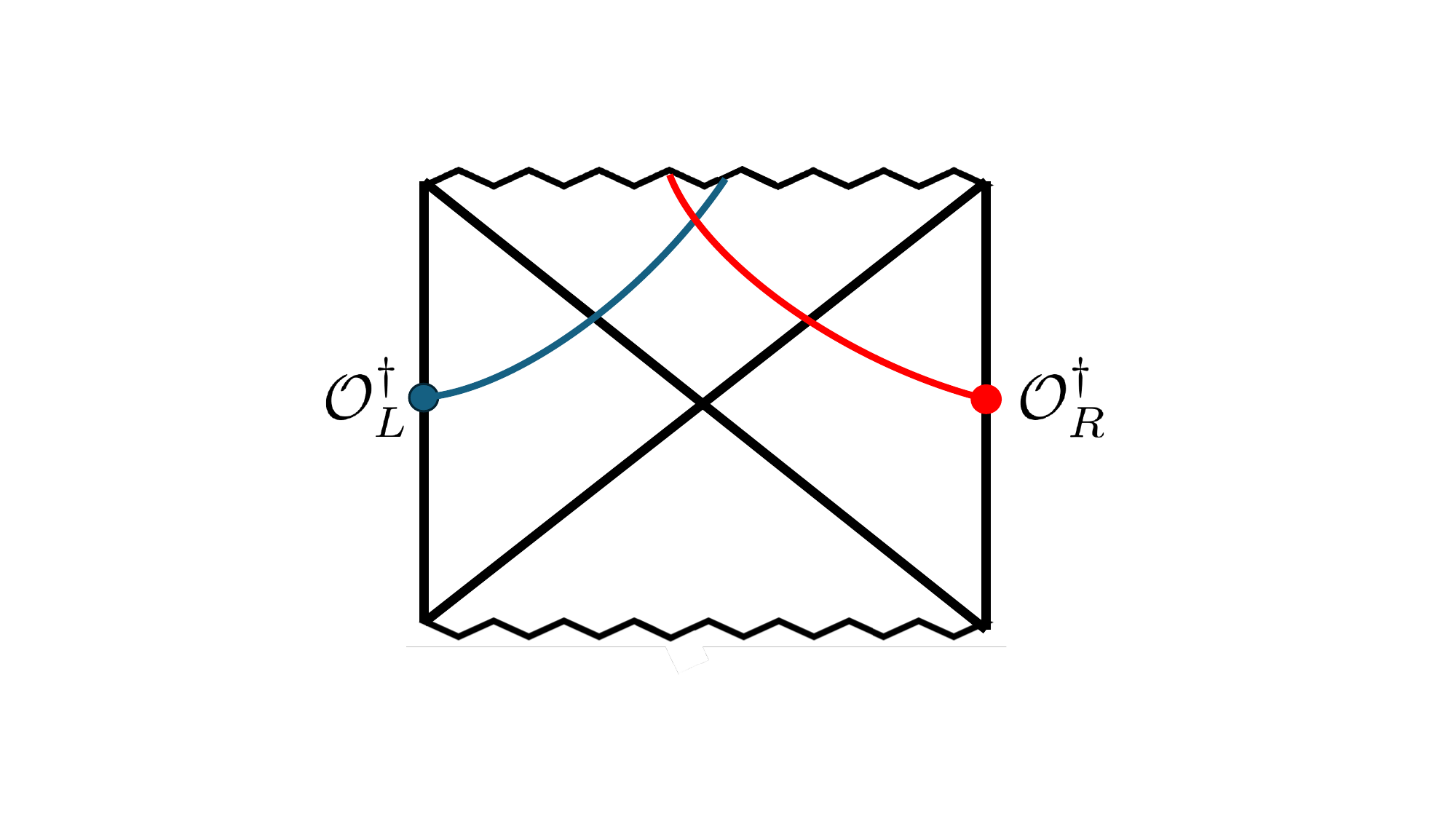}
    \caption{Two observers falling into the AdS black hole and meeting in the interior. One originates from the left asymptotic boundary and the other from the right. }
    \label{fig:LoisRyan}
\end{figure}

Reviewing Marolf and Wall's puzzle, they start by considering two in-falling observers, which we name ``Lois" and ``Ryan", that are respectively created by local CFT operators $\mathcal{O}^\dagger_L$ and $\mathcal{O}^\dagger_R$. The idea is that these observers are created near the left and right asymptotic boundaries, see Figure \ref{fig:LoisRyan}. In the original setting of Maldacena (i.e. with no S-entanglement) the Hilbert space factorizes into left and right pieces $\mathcal{H}=\mathcal{H}_L\otimes \mathcal{H}_R$, so $\mathcal{O}_{L,R}\in \mathcal{H}_{L,R}$. The authors of \cite{Marolf:2012xe} then claim that there should be a way to measure the interaction between Lois and Ryan inside the black hole, i.e. some observable whose vev in the Lois + Ryan + Eternal AdS black hole system is $\approx 1$ when they interact and $\approx 0$ when they do not. This was motivated by a the expectation that there should be a semiclassical understanding of the large black hole interior. More precisely, Marolf/Wall expect that there exists an operator $\mathcal{P}$ such that one simultaneously satisfies:
\begin{align}\label{eq:cond1}
    \mathrm{1.  } & \bra{TFD}\mathcal{O}_L \mathcal{O}_R \mathcal{P} \mathcal{O}^\dagger_R \mathcal{O}^\dagger_L \ket{TFD}\approx 1 \\
    \mathrm{2.  } & \bra{TFD}\mathcal{O}_L  \mathcal{P} \mathcal{O}^\dagger_L \ket{TFD}\approx \bra{TFD}\mathcal{O}_R \mathcal{P} \mathcal{O}^\dagger_R \ket{TFD} \approx 0\label{eq:cond2}
\end{align}
Due to the exact factorization of the Hilbert space, $\mathcal{P}=\mathcal{P}_L+\mathcal{P}_R$ where $\mathcal{P}_L$ commutes with all operators in $\mathcal{H}_R$, and similarly for $(L \leftrightarrow R)$. Therefore, satisfying the second condition will always make the correlation function in the first condition approximately vanish. The proposal of \cite{Marolf:2012xe} to address this discrepancy is that one should replace $\ket{TFD}$ in the above conditions by a new state they call $\ket{w2}$. They claim that both states should exist in an extended Hilbert space
\begin{equation}
    \mathcal{H}_{\overline{\mathbf{bulk}}}=\mathcal{H}_L\otimes \mathcal{H}_R\otimes V
\end{equation}
$\mathcal{H}_L\otimes \mathcal{H}_R\otimes V$ where $V$ is some auxiliary Hilbert space (the ``superselection sector" space) which is large enough such that $\ket{w2}\in \mathcal{H}_{\overline{\mathbf{bulk}}}$ with a non-trivial component in $V$ such that the two above conditions are satisfied.

Comparing to our proposal, we first specialize to the case that the two in-falling observers are charged under a $\mathbb{Z}^{(0)}_N$ bulk gauge symmetry. We consider an eternal AdS black hole with maximal entanglement for a $\mathbb{Z}^{0}_N$ symmetry, such as the AdS$_4$ black holes dual to two copies of CFTs possessing 0-form symmetries. In CFT language, we have a $\mathbb{Z}^{(0)}_N$ global symmetry obtained by gauging $(\mathbb{Z}^{(0)}_N)_L\times (\mathbb{Z}^{(0)}_N)_R$ by the $\mathbb{Z}_N$ subgroup generated by $(+1,-1)$. Let $q_{L,R}$ denote the charges under $(\mathbb{Z}^{(0)}_N)_{L,R}$, then genuine local operators will take the form $\mathcal{O}^{(q_L)}_L\mathcal{O}^{(q_R)}_R$ such that $q_L=q_R=: q$. Moreover, this gauging identifies the symmetry operators $\mathcal{U}_L\sim \mathcal{U}_R=:\mathcal{U}$. We can then calculate the following correlator
\begin{align}
    \bra{TFD,\mathcal{B}}\mathcal{O}^{(q)}_L\mathcal{O}^{(q)}_R  \mathcal{U}(\mathcal{O}^{(q)}_L)^\dagger (\mathcal{O}^{(q)}_R)^\dagger\ket{TFD,\mathcal{B}}=& e^{-2\pi i q/N} + O(\beta^{-1})
\end{align}
where $\ket{\mathcal{B}}=\frac{1}{N}\sum^{N-1}_{k=0}\ket{k}\otimes\ket{k}$ as in Section \ref{ssec:IAS}. The leading term comes from evaluating the correlator in the $\beta\rightarrow \infty$ where thermofield double state reduces the CFT vacuum. Similarly, one can compute that
\begin{align}
     \bra{TFD,\mathcal{B}}\mathcal{O}^{(q)}_L \mathcal{U}(\mathcal{O}^{(q)}_L)^\dagger \ket{TFD,\mathcal{B}}=&  \bra{TFD,\mathcal{B}}\mathcal{O}^{(q)}_R \mathcal{U}(\mathcal{O}^{(q)}_R)^\dagger \ket{TFD,\mathcal{B}} \\
     =& 1 + O(\beta^{-1})
\end{align}
due to the fact that these non-genuine local operators\footnote{Technically when one writes $\mathcal{O}^{(q)}_{L,R}$ one should include the dependence on the topological line operators attached to these local operators, e.g. $\mathcal{O}^{(q)}_{L}(x_0)\mathcal{V}(L)$ where $\partial L=x_0$.
} are not charged under $\mathcal{U}$. Consider now the operator:
\begin{equation}\label{eq:P}
    \mathcal{P}_q:=\frac{e^{2\pi i q/ N}}{1-e^{2\pi i q/N}}(\mathcal{U}-1).
\end{equation}
The above correlators of $\mathcal{U}$ implies that $\mathcal{P}_q$ satisfies the conditions \eqref{eq:cond1} and \eqref{eq:cond2}. We must caution that giving a more complete solution to Marolf/Wall's puzzle would involve not just generalizing our S-entanglement construction to continuous internal symmetries but also to spacetime symmetries: an observable measuring whether or not Lois and Ryan hit each other surely depends on their initial momenta. Indeed the operator $\mathcal{P}$ in \eqref{eq:P} only detects whether or not both Lois and Ryan exist in the spacetime or not; much more coarse information. We hope that our arguments here can serve as a stepping stone to understand how CFT data can measure such events past the horizon that one may expect semiclassically to be sensible information.

Finally, we note that the SymTFT Hilbert space $\mathcal{H}_{\mathcal{S}}(M_D)^{\otimes 2}$ played an identical role as the ``superselection sector" space $V$ in Marolf/Wall's conjectural answer to their puzzle. Indeed, if we chose $\ket{B}$ to be a direct product state, then it would have been impossible to construct an operator $\mathcal{P}$ satisfying \eqref{eq:cond1} and \eqref{eq:cond2}, and $\ket{TFD,\mathcal{B}}$ would have similar properties to the state Marolf/Wall denote as ``$\ket{TFD}$". Meanwhile the entangled $\ket{\mathcal{B}}$ used in this section behaves as their ``$\ket{w2}$".

\subsection{A Conjecture on Refining the Holographic Dual Proposal of an Eternal AdS Black Hole}\label{ssec:CERNVT}
In this section, we have so far been working in a top-down perspective to construct eternal AdS black holes with S-entanglement and understanding the physical consequences. While rigorous, our sacrifice is that we have had to restrict ourselves to consider S-entanglement for $p$-form discrete internal symmetries. In this subsection, we let ourselves speculate on how to extend our claims to continuous and spacetime symmetries.

As we saw from Section  \ref{ssec:IAS}, eternal AdS black holes only have a globally well-defined $\mathbb{Z}_N$ 0-form charge (as opposed to separately defined charges for the left and right boundaries) iff the S-entanglement was non-trivial. We suspect a similar conclusion holds also for all types of gauge charges as well. In particular, we first expect that there exists an embedding of the SymTFT of discrete symmetries, $\mathcal{S}_{D+1}$, into a SymTFT for all internal global symmetries of the CFT, $\mathcal{S}^{\mathrm{full,int}}_{D+1}$, which further embeds in a SymTFT construction for all global symmetries of the CFT:
\begin{equation}\label{eq:nest}
    \mathcal{S}_{D+1}\subset \mathcal{S}^{\mathrm{full,int}}_{D+1}\subset \mathcal{S}^{\mathrm{full}}_{D+1}.
\end{equation}
Proposals for SymTFT constructions for continuous internal symmetries have appeared recently \cite{Antinucci:2024zjp, Brennan:2024fgj, Bonetti:2024cjk} which involve TFTs with non-compact gauge groups, as well as the SymTh proposal of \cite{Apruzzi:2024htg} (see also \cite{Mignosa:2025cpg}) which replaces the TFT with a gapless (albeit scale-invariant) theory. See also \cite{Apruzzi:2025hvs} for work on developing a SymTFT construction for spacetime symmetries. In sketching such a nested sequence \eqref{eq:nest}, we remain agnostic as to whether the correct formulation of $\mathcal{S}^{\mathrm{full}}_{D+1}$ is truly topological or not, but we simply assume that a CFT of interest $\mathcal{T}_D$ is a valid boundary condition for $\mathcal{S}^{\mathrm{full}}_{D+1}$ and that all symmetry data such as topological operators and their linking with charged operators, is captured by $\mathcal{S}^{\mathrm{full}}_{D+1}$.

With these conservative assumptions, one can form S-entangled setups exactly analogous to Figure \ref{fig:twoversions} where we have $\mathcal{S}^{\mathrm{full}}_{D+1}$ placed on $M_{D}\times [0,1]$ with $\mathcal{T}_D$ at $M_{D}\times \{0\}$ and $\overline{\mathcal{T}_D}$ at $M_{D}\times \{1\}$. This follows from the fact that if the first boundary condition exists, then the conjugate boundary condition also exists. This is implies that for any complicated symmetry higher category $\mathcal{C}$, regardless of how complicated\footnote{For higher fusion categories, it is still an open question on how to rigorously treat the condition of unitarity. See \cite{Ferrer:2024vpn, Bartsch:2025drc} for recent work in this direction.}, there will always exist a notion of ``diagonal gauging" of $\mathcal{C}\otimes \mathcal{C}^\dagger$. See \cite{Robbins:2024tqf} for an example of defining such a gauging for the case when $\mathcal{C}=\mathrm{Rep}(G)$ which is non-trivial. Note that the existence of such a sandwich construction further implies that anomaly associated with this diagonal gauging cancels schematically as\footnote{For a similar recent discussion see \cite{Heckman:2025lmw}.}
\begin{equation}
\mathcal{A}_{\mathrm{diag}}=\mathcal{A}_\mathcal{C}+\mathcal{A}_{\mathcal{C}^\dagger}=\mathcal{A}_\mathcal{C}-\mathcal{A}_{\mathcal{C}}=0.
\end{equation}

We conjecture then that preparing a TFD state for the product CFT $\mathcal{T}\otimes \overline{T}_D$ with a S-entanglement associated with the diagonal gauging of their symmetries is dual to an eternal AdS black hole whose spacetime gauge charges, such as angular momentum, and internal gauge charges are globally well-defined in the sense that these charges are not separately defined with respect to the left and right event horizons. Such a construction should in principle constitute a full solution to the types of puzzles posed in  Marolf/Wall \cite{Marolf:2012xe}.


\section{Conclusions and Outlook}\label{sec:conc}



In this paper, we have introduced and explored the concept of SymTFT entanglement (S-entanglement for short) between QFTs. This provides a natural generalization of the familiar ``sandwich" construction of SymTFTs, clarifying how (non-)invertible subsymmetries can be gauged between two or more theories. From the holographic perspective, S-entanglement appears as a new entry in the AdS/CFT dictionary and is a necessary ingredient whenever multiple disconnected conformal boundaries are present. In particular, we have shown how baby universe Hilbert spaces (with more than one-dimension) and their associated ensemble averaging naturally arise from SymTFT entanglement and established a relation $\mathcal{H}_{BU}\otimes \mathcal{H}^\dagger_{BU}=\mathcal{H}_{\mathcal{S}}$ \eqref{eq:HBU}. Furthermore, in the case of two-sided AdS black holes, the SymTFT entanglement data can determine whether the Hilbert space factorizes or not, and also whether 0-form internal gauge charges of the black holes are independent between the left- and right-horizons or not.

We anticipate several avenues of future directions / generalizations of this work.

\begin{itemize}
    \item In this work we focused on discrete internal symmetries. A natural extension is to consider continuous and even spacetime symmetries.\footnote{A subtlety for spacetime symmetries, compared to internal ones, is the absence of a clear notion of topological operators generating them. This raises the puzzle of whether they admit a description purely in terms of SymTFTs or require a more general framework. See \cite{Apruzzi:2025hvs} for a recent work in this direction.} Appendix \ref{app:theta} illustrates how a $U(1)$ SymTFT can lead to ensemble averaging over $\theta$-angles in 4D holographic CFTs. We also sketched in Section \ref{ssec:CERNVT} how this might impact the physics of two-sided AdS black holes and clarify puzzles raised by Marolf and Wall \cite{Marolf:2012xe}.
    \item Another direction is to revisit stringy AdS$_3$ holography, discussed briefly in Section \ref{subsec:noninver symentangle}, in case where no semiclassical supergravity description is available near the conformal boundary. See \cite{Du:2024tlu, Gutperle:2024vyp, Knighton:2024noc} for recent work mapping bulk worldsheet data to CFT$_2$ symmetries. It would be interesting to explore S-entanglement for these string backgrounds because of the interesting resolution to the factorization puzzle in this setting from the bulk point of view \cite{Eberhardt:2020bgq}. A concrete question to ask then would be: how does one impose entangled boundary conditions from the point of view of the bulk tensionless string worldsheet in the presence of multiple conformal boundaries?

    \item Closely related, the AdS$_3$/CFT$_2$ correspondence has also been approached from bottom-up perspectives using 3D TFT path integrals. Interpreted as SymTFTs, these constructions capture the CFT data in rational models, where the vertex operator algebra is fully encoded in Verlinde lines and modular tensor categories. More generally, one may view the Virasoro TFT as the SymTFT governing 2D topological defects commuting with the Virasoro algebra. See \cite{Collier:2023fwi,Collier:2024mgv,Chen:2024unp,Hung:2024gma,Dymarsky:2024frx,Bao:2024ixc,Takahashi:2024ukk,Hung:2025vgs,Geng:2025efs,Hartman:2025cyj,Hartman:2025ula,Barbar:2025krh} for a partial list of recent work along these lines. It would be interesting to explore S-entanglement in this setting, as it may provide new insight into ensemble averaging phenomena in the bottom-up AdS$_3$/CFT$_2$ correspondence.
    \item In addition to considering S-entanglement as correlations between boundary conditions of AdS spacetimes, one can also ask what are the consequences of this for asymptotically Minkowski vacua. We addressed this briefly in Section \ref{ssec:HolInt} when considering the system of $N$ $D3$-branes before back reaction, but we expect that, in the context of celestial holography\footnote{See, e.g., \cite{Pasterski:2021raf} for a review of celestial holography.}, there should exist a construction which entangles two disconnected Minkowksi vacua in a similar manner as we have done for AdS vacua.
    \item In the context of ensemble averaging in AdS/CFT, there is an ongoing discussion of whether the averaging should be understood as taken over microscopically well-defined element theories (see, e.g., the Maloney–Witten construction of averaging over Narain moduli space \cite{Maloney:2020nni}), or instead as an averaging over more abstract axiomatic data of the CFT (see, e.g., the proposal to average over random three-point function data of large-$c$ CFTs \cite{Chandra:2022bqq}). The latter mechanism has in particular been motivated by the chaotic behavior of CFTs above the black hole threshold \cite{Schlenker:2022dyo}\footnote{See also \cite{Liu:2025cml,Gesteau:2025obm} for more recent discussion}. The ensemble averaging derived from S-entanglement in this work is closer in spirit to the former type, as it naturally produces mixtures of well-defined element theories. It would nevertheless be interesting to explore whether S-entanglement can also be adapted to mimic the latter mechanism.
    \item Finally, given that taking a partial trace of two (or more) S-entangled AdS vacua can lead to bulk global symmetries, it would be interesting to explore how other core Swampland principles are violated for these partial traced subsystems, given that so many Swampland conjectures rely on assuming the absence of global symmetries in quantum gravity theories. Such a direction would be in similar spirit to the work of \cite{Anastasi:2025puv}, which explored how Swampland principles can be violated for gravitationally coupled subsystems (in this case, end-of-the world branes) in the context of AdS/BCFT correspondence.
\end{itemize}




\acknowledgments

We thank Ibrahima Bah, Arkya Chatterjee, Elliott Gesteau, Monica Guica, Jonathan J. Heckman, Max H\"ubner, Da-Chuan Lu, Juan Maldacena, Jake McNamara, Ruben Monten, Miguel Montero, Sridip Pal, Kyriakos Papadodimas, and Zhengdi Sun for helpful discussions. ET thanks the Harvard Swampland Initiative and Uppsala Center for Geometry and Physics for their hospitality during which part of this work was carried out and presented. ET thanks the Aspen Center for Physics, which is supported by National Science Foundation grant PHY-2210452, for their hospitality during the completion of this work. XY thanks the Kavli Institute for the Physics and Mathematics of the Universe (IPMU) for the hospitality during the completion of this work. XY thanks the Southeastern Regional Mathematical String Theory Meeting for the hospitality during the completion of this work.
ET and XY thank the 2025 Simons Physics Summer Workshop for their hospitality during the completion of this work. The work of ET is supported in part by the ERC Starting
Grant QGuide-101042568 - StG 2021. The work of XY is partially supported by the NSF grant PHY-2310588.


\appendix






\section{SymTFT Entangled TFD states for Eternal AdS$_3$ Black Holes}\label{app:5point1extended}
We now address how to construct S-entangled thermofield double states when the CFT is two-dimensional. What makes this case special is that for a given $D=2$ CFT, $\mathcal{T}$ with non-anomalous global $0$-form symmetry $G^{(0)}$, gauging it results in a quantum symmetry which is also $0$-form symmetry. In general, this is a $(D-2)$-form symmetry $\mathrm{Rep}(G)^{(D-2)}$, and the fact that $D=2$, together with the state-operator correspondence, means that the Hilbert space of the gauged theory, $\mathcal{T}/G^{(0)}$, has sectors labeled by charges\footnote{These charges are classified by conjugacy classes of $G$.} of $\mathrm{Rep}(G)^{(0)}$. This is perhaps more familiar in the setting of orbifolds in string perturbation theory where we also have that the twisted sectors are labeled by charges of $\mathrm{Rep}(G)^{(0)}$. For a modern review of these statements see \cite{Bhardwaj:2017xup}.


For simplicity we will restrict to the case of $\mathcal{T}$ having a $G^{(0)}=\mathbb{Z}_N^{(0)}$ global symmetry where now $\mathrm{Rep}(G)^{(0)}\simeq \widehat{\mathbb{Z}}_N^{(0)}$ is the Pontryagin dual of the original group. We will let $q \; \mathrm{mod}\; N$ denote charges of the original group $\mathbb{Z}_N^{(0)}$ and will let $\hat{q} \; \mathrm{mod}\; N$ denote charges of the quantum symmetry $\widehat{\mathbb{Z}}_N^{(0)}$. With these preliminaries in hand we see that the algebra of genuine local operators of $\mathcal{T}$ decomposes as:
\begin{equation}\label{eq:vdecomp}
    \mathcal{V}^{\mathrm{gen.}}= \bigoplus_{q}\mathcal{V}^{\mathrm{gen.}}_q
\end{equation}
where each $\mathcal{V}^{\mathrm{gen.}}_q$ has charge $q\; \mathrm{mod} \; N$ under
$\mathbb{Z}_N^{(0)}$. Meanwhile non-genuine local operators are attached to topological
lines $V_{\hat{q}}$ which are symmetry operators for $\mathbb{Z}_N^{(0)}$. The reason why these have a $\hat{q}$ is label is because of the perfect pairing between finite abelian
groups and their Pontryagin duals: $\mathbb{Z}_N\times \widehat{\mathbb{Z}}_N\rightarrow
\mathbb{Q}/\mathbb{Z}$. More explicitly this means that the symmetry action on genuine local operators $\mathcal{O}_q\in \mathcal{V}^{\mathrm{gen.}}_q$ is given by\footnote{As the reader may have noticed this structure can be reproduced by a 3D SymTFT action $\propto N\int b_1\delta c_1$. }
\begin{equation}
    V_{\hat{q}}\mathcal{O}_q=\exp(2\pi i q \hat{q}/N)\mathcal{O}_q
\end{equation}
assuming $V_{\hat{q}}$ links once around $\mathcal{O}_q$. We can now define the total algebra of local operators as
\begin{equation}\label{eq:vtot}
    \mathcal{V}_\mathcal{T}^{\mathrm{total}}= \bigoplus_{\hat{q}}\mathcal{V}^{(\hat{q})}, \quad \; \;  \textnormal{where   } \mathcal{V}^{(\hat{0})}:=\mathcal{V}^{\mathrm{gen.}}.
\end{equation}
Note that each $\mathcal{V}^{(\hat{q})}$ for $\hat{q}\neq 0$ does not have a \textit{unique} decomposition into $q$-charge sectors as in \eqref{eq:vdecomp} because of the mixed 't Hooft anomaly between $\mathbb{Z}^{(0)}_N$ and $\widehat{\mathbb{Z}}^{(0)}_N$. Note that a natural interpretation of the non-genuine local operators of $\mathcal{T}$ are genuine local operators in the twisted sector of the orbifolded theory $\mathcal{T}/\mathbb{Z}_N$.  These are projected out of the spectrum of genuine local operators after gauging $\mathcal{T}/\mathbb{Z}_N$ by  $\widehat{\mathbb{Z}}_N$ which brings one back to $\mathcal{T}$. Also, by the state-operator correspondence one can similarly define a ``total Hilbert space" of $\mathcal{T}$ as:
\begin{equation}
\mathcal{H}_\mathcal{T}^{\mathrm{total}}=\bigoplus_{\hat{q}}\mathcal{H}^{(\hat{q})}
\end{equation}
where $\mathcal{H}^{(\hat{0})}=\oplus_q \mathcal{H}_q$ with $q\neq 0$ labeling the twisted sectors. The total Hilbert space
for $\mathcal{T}/\mathbb{Z}_N$ is identical to that of the $\mathcal{T}$ theory just as the set of local operators does not change: only which ones are labeled genuine or
non-genuine changes. However, the decomposition into subspaces does change. If we let $\mathcal{O}_{\hat{q}}$ denote a local operator of
charge $\hat{q}$ under $\widehat{\mathbb{Z}}^{(0)}_N$, then the symmetry operator is labeled by $q$ and has an action
\begin{align}
 U_{q}\mathcal{O}_{\hat{q}}=\exp(2\pi i q \hat{q}/N)\mathcal{O}_{\hat{q}}.
\end{align}
The total Hilbert space for the $\mathcal{T}/\mathbb{Z}_N$ theory then has the form
\begin{equation}
    \mathcal{H}^{\mathrm{total}}_{\mathcal{T}/\mathbb{Z}_N}=\bigoplus_{q}\mathcal{H}^{(q)}
\end{equation}
where $\mathcal{H}^{(q)}$ have a $U_q$ topological line defect inserted at a given point in the spatial $S^1$ and $\mathcal{H}^{(0)}=\oplus_{\hat{q}}\mathcal{H}^{(\hat{q})}$ is the decomposition of the usual Hilbert space into twisted sectors.

With this background in mind, let us apply it to the case of $\mathcal{T}:=\mathcal{T}_{L}\otimes \mathcal{T}_R$ and $\mathbb{Z}^{(0)}_N$ given by the diagonal subgroup of $(\mathbb{Z}^{(0)}_N)_L\times (\mathbb{Z}^{(0)}_N)_R$ generated by $(1,-1)\; \mathrm{mod}\; N$. The TFD states with the S-entanglement associated with gauging such a diagonal subgroup was studied in Section \ref{ssec:IAS}, and we saw there that the preparation of the thermofield double state followed from one without S-entanglement by acting with a projection operator $P\equiv \frac{1}{N}\sum^{N-1}_{k=0}V^k_LV^{-k}_R$. For $D=2$ however, the presence of twisted sector states modifies this answer to
\begin{equation}\label{eq:TFDdequals2}
    \ket{TFD, \mathcal{B}}=P\cdot \left( \frac{1}{Z_{\mathcal{T}}(\beta)}\sum_{\hat{q}, n^{(\hat{q})}}e^{-\beta E_{n^{(\hat{q})}}/2}\ket{n^{(+\hat{q})}}_L\otimes \ket{n^{(-\hat{q})}}_R\right).
\end{equation}
Explaining the notation, $\mathcal{B}$ denotes the SymTFT boundary conditions leading to such an S-entanglement,
while $\ket{n^{(+\hat{q})}}_L$ is
an energy eigenstate in the subspace $\mathcal{H}^{(\hat{q})}_L\subset \mathcal{H}^{\mathrm{total}}_{\mathcal{T}_L}$ which is
mapped to a non-genuine local operator attached to $V_{\hat{q}}$ after applying the state-operator correspondence to $\mathcal{T}_L$. The derivation of \eqref{eq:TFDdequals2}
follows from repeating the derivation of the TFD state reviewed in Section \ref{ssec:IAS} where now the key difference arises from the insertion of the complete basis of states (see below \eqref{eq:psitfd}) which includes twisted sector states.

\section{$U(1)$ SymTFT Entanglement and $\theta$-Angle Averaging}
\label{app:theta}

In this appendix, we sketch a generalization of the S-entanglement construction to the case of continuous symmetries, using the $U(1)$ SymTFT introduced in \cite{Brennan:2024fgj} and \cite{Antinucci:2024zjp}, whose string theory approach is proposed in \cite{Yu:2024jtk, Gagliano:2024off}. After building entangled states for $U(1)$ SymTFTs, we follow the discussion in Section \ref{sec:ohbaby} to build an ensemble averaging of holographic 4D (S)CFTs over the $\theta$-angle.

For a D-dimensional QFT with a $(p-1)$-form $U(1)$ symmetry, its associated SymTFT has the following action
\begin{equation}\label{eq: U(1) symtft}
    S=\frac{1}{2\pi}\int_{M_{D+1}} a_p \wedge df_{D-p},
\end{equation}
where $a_p$ is a $p$-form represents a $U(1)$ gauge field, while $f_{D-p}$ is a $R$ gauge field. The topological operators of this TFT are given by
\begin{equation}
\begin{split}
    &U_n(\gamma)=\exp \left( in\oint_{\gamma}a_p \right), ~n\in \mathbb{Z};\\
    &V_{\alpha}(\Gamma)=\exp \left( i\alpha \oint_{\Gamma}f_{D-p} \right), ~\alpha \in U(1).
\end{split}
\end{equation}
Under the canonical quantization on $M_D\times R_t$, we have $\gamma \in H_p(M_D, \mathbb{Z})$ and $\Gamma \in H_{D-p}(M_D, \mathbb{Z})$. The basic commutator then reads
\begin{equation}
    \left[ U_1(\gamma), V_1(\Gamma) \right]=e^{ \left( 2\pi i ~\text{Int} (\gamma, \Gamma) \right)},
\end{equation}
where $~\text{Int} (\gamma, \Gamma)$ is the intersection number between $\gamma$ and $\Gamma$ in $M_D$.

Since eventually we are interested in building entanglement and averaging for the $\theta$-angle, from now on we will set $p=0$ \cite{Yu:2024jtk}. The Hilbert space of this SymTFT\footnote{More precisely, the physical boundary in this case is not a single relative QFT but a deformation class of relative QFTs labeled by the circle-valued parameter \cite{Yu:2024jtk}.} can then be spanned by eigenstates of either $U_n$'s or $V_\alpha$'s. We will use $U_n$'s to build the polarization. The corresponding basis $\{ |\theta\rangle \}$ of the Hilbert space\footnote{There is an ongoing discussion in the community whether the $U(1)$ SymTFT is mathematically well-defined as a TFT. One subtlety is that the Hilbert space does not have finite dimensions but has a continuous basis, which does not obey the Atiyah-Segal axiom \cite{atiyah1988topological, segal1988definition} for the TFT.} is acted by $U_n$'s and $V_\alpha$'s as clock and shift operators:
\begin{equation}
\begin{split}
    &U_n|\theta \rangle =e^{i n \theta}|\theta \rangle,\\
    &V_\alpha |\theta \rangle = |\theta + \alpha \rangle.
\end{split}
\end{equation}
Given a relative QFT defined on the physical boundary of the SymTFT, its associated state in the Hilbert space reads
\begin{equation}
    | \mathcal{T}\rangle = \int [d\theta]Z_\mathcal{T}[\theta]|\theta \rangle.
\end{equation}
A specific QFT with the parameter value $\theta_*$ is picked by the symmetry boundary state $| \theta_* \rangle$:
\begin{equation}
    \langle \mathcal{T} | \theta_*\rangle=Z_\mathcal{T}[\theta_*].
\end{equation}

Now let us consider entangled states for this SymTFT and the resulting ensemble averaging. To be more explicit, we consider 4D QFTs associated to $D3$-branes probing singular Calabi-Yau 3-folds, which are cones over 5D Einstein-Sasaki manifolds. This includes 4D $\mathcal{N}=4$ SYM associated to the flat internal geometry $\mathbb{C}^3$, as well as $\mathcal{N}=1$ quiver SCFTs associated to $\mathbb{C}^3/\Gamma, \Gamma \subset SU(3)$ orbifolds or general toric local Calabi-Yau\footnote{There is a vast literature on the subject of 4D SCFTs on D-brane probes of singularities, see e.g., \cite{Douglas:1996sw, Franco:2005rj, Yamazaki:2008bt} for reviews.}. These 4D theories are holographically dual to AdS$_5\times X_5$ where $X_5$ is the 5D link geometry of the Calabi-Yau. This infinite class of 4D holographic QFTs is equipped with a $\theta$-angle, coming from the type IIB axion field $C_0$. The SymTFT for this $(-1)$-form symmetry is exactly the one in Eq.~(\ref{eq: U(1) symtft}), whose IIB approach is derived in \cite{Yu:2024jtk}.

Consider two copies of 4D relative QFTs within this infinite class, then we can write down an entangled state in the tensor product Hilbert space $\mathcal{H}_1 \otimes \mathcal{H}_2$ of the $U(1)$ SymTFT as
\begin{equation}
    \int [d\theta]~|\theta \rangle_1 \otimes |\theta \rangle_2.
\end{equation}
Following the discussion in Section \ref{sec:ohbaby}, we can then compute the ``partial trace'' over $\mathcal{H}_2$. The resulting mixed state is given by the following reduced density matrix,
\begin{equation}
    \rho_1= \int [d\theta]~|\theta \rangle \langle \theta|,
\end{equation}
which can be regarded as a continuous version of Eq. (\ref{eq: reduced density matrix discrete}). The ensemble averaging can then be realized by computing the partition function for the relative QFT state $|\mathcal{T}\rangle= \int [d\theta]Z_\mathcal{T}[\theta]|\theta \rangle$ with respect to this mixed state:
\begin{equation}
\begin{split}
    \text{Tr}\left(\rho_\mathcal{T}\rho_1 \right)&=\text{Tr}\left( |\mathcal{T}\rangle \langle \mathcal{T}| \rho_1 \right)\\
    &=\int d\theta ~|Z_\mathcal{T}[\theta]|^2.
\end{split}
\end{equation}
This is an ensemble averaging of 4D QFTs with random coupling constant $\theta$ under a uniform distribution. Interestingly, this can be interpreted as an one-dimensional slice of the $SL(2, \mathbb{Z})$ ensemble averaging over the complexified gauge coupling of 4D SYM, which is argued in \cite{Collier:2022emf} to be holographic dual to quantum AdS$_5\times S^5$. We leave the full treatment of top down/SymTFT approach to this $SL(2,\mathbb{Z})$ averaging for a future study.





\bibliographystyle{utphys}
\bibliography{entangle}
\end{document}